\documentclass[groupedaddress, showpacs, superscriptaddress]{revtex4-2}
\usepackage{amssymb,amsmath,amsthm,color,graphicx,times,graphicx}
\usepackage{hyperref}
\usepackage{subfigure}
\usepackage{times}
\usepackage{bbold}
\hypersetup{colorlinks,citecolor={cyan},linkcolor={magenta},urlcolor={blue}}
\usepackage{color}
\usepackage{academicons}
\newcommand{\orcid}[1]{\href{https://orcid.org/#1}{\textcolor[HTML]{A6CE39}{\aiOrcid}}}
\usepackage{array}
\usepackage{sidecap}
\usepackage[french]{babel}
\usepackage{array,multirow,makecell}
\usepackage[table]{xcolor}

\theoremstyle{plain}

\theoremstyle{definition}

\usepackage{orcidlink}
\makeatletter
\newsavebox{\@brx}
\newcommand{\llangle}[1][]{\savebox{\@brx}{\(\m@th{#1\langle}\)}%
  \mathopen{\copy\@brx\mkern2mu\kern-0.9\wd\@brx\usebox{\@brx}}}
\newcommand{\rrangle}[1][]{\savebox{\@brx}{\(\m@th{#1\rangle}\)}%
  \mathclose{\copy\@brx\mkern2mu\kern-0.9\wd\@brx\usebox{\@brx}}}
\makeatother

\begin{document}

\title{Monitored non-adiabatic and coherent-controlled quantum unital Otto heat engines: First four cumulants}
\author{Abdelkader El Makouri}\email{abdelkader\_elmakouri@um5.ac.ma}\affiliation{LPHE-Modeling and Simulation, Faculty of Sciences, Mohammed V University in Rabat, Morocco.} 
\author{Abdallah Slaoui \orcidlink{0000-0002-5284-3240}}\email{Corresponding author: abdallah.slaoui@um5s.net.ma}\affiliation{LPHE-Modeling and Simulation, Faculty of Sciences, Mohammed V University in Rabat, Morocco.}\affiliation{Centre of Physics and Mathematics, CPM, Faculty of Sciences, Mohammed V University in Rabat, Rabat, Morocco.}
\author{Rachid Ahl Laamara}\email{r.ahllaamara@um5r.ac.ma}\affiliation{LPHE-Modeling and Simulation, Faculty of Sciences, Mohammed V University in Rabat, Morocco.}\affiliation{Centre of Physics and Mathematics, CPM, Faculty of Sciences, Mohammed V University in Rabat, Rabat, Morocco.}

\begin{abstract}
Recently, measurement-based quantum thermal machines have drawn more attention in the field of quantum thermodynamics. However, the previous results on quantum Otto heat engines were either limited to special unital and non-unital channels in the bath stages, or a specific driving protocol at the work strokes and assuming the cycle being time-reversal symmetric i.e. $V^{\dagger}=U$ (or $V=U$). In this paper, we consider a single spin-1/2 quantum Otto heat engine, by first replacing one of the heat baths by an arbitrary unital channel and then we give the exact analytical expression of the characteristic function from which all the cumulants of heat and work emerge. We prove that under the effect of monitoring, $\nu_{2}>\nu_{1}$ is a necessary condition for positive work, either for a symmetric or asymmetric-driven Otto cycle. Furthermore, going beyond the average we show that the ratio of the fluctuations of work and heat is lower and upper-bounded when the system is working as a heat engine. However, differently from the previous results in the literature, we consider the third and fourth cumulants as well. It is shown that the ratio of the third (fourth) cumulants of work and heat is not upper-bounded by unity nor lower-bounded by the third (fourth) power of the efficiency, as is the case for the ratio of fluctuations. Finally, we consider applying a specific unital map that plays the role of a heat bath in a coherently superposed, manner and we show the role of the initial coherence of the control qubit on efficiency, on the average work and its relative fluctuations.\par 

$\mathbf{Keywords:}$ Monitored asymmetric quantum Otto heat engines, Unital qubit channels, Ratio of cumulants and coherently superposed channels.
\end{abstract}
\date{\today}

\maketitle
\section{Introduction}
The field of quantum thermodynamics (QT) \cite{AndersJ} can simply be defined as a new research area that is trying to develop and formulate a consistent theory of thermodynamics valid at the quantum scale. This new (actually it is old) field has started with the work of Scovil and Schulz-Duboiz on the thermodynamics of masers \cite{Scovil}. Masers can be considered the first prototype of quantum heat engines. Nowadays QT gains its progress from quantum resource theory to thermodynamics \cite{Nicole,Matteo}, fluctuations relations \cite{Esposito,Seifertt,Campisi1,Jarzynski,Crooks},   thermodynamic uncertainty relations  \cite{Horowitz1,Horowitz,Barato,Timparano} and so on. In this research area the quantum version of thermal machines either heat engines or refrigerators, occupy a very important place due to the role of their classical versions in our society. They were recently reported experimentally using different platforms, e.g. trapped ions \cite{Dawkins} and NMR setting \cite{Peterson}. For a general overview of other platforms, we recommend the recently written paper by Cangemi et al. \cite{Levy}. Quantum Otto cycle (QOC) \cite{Kieu,Rezek,Quan} is the most cycle studied in QT. This is because the heat and work exchanges are done in different strokes which makes their computation easy. This cycle is composed of two isochoric and two adiabatic quantum transformations. In previous works such as \cite{Rezek,Kieu} the source of heat is a thermal reservoir. However, it was shown that quantum measurement can as well provide heat to the system as long as its associated operators do not commute with the measured quantity \cite{JYi}. This idea of using measurement as a heat source is inspired by Maxwell-demon and Szilard-engine \cite{Maxwell}.

Usually in physics, we assume implicitly that events are causally ordered. More precisely, either the event A is in the past of B (A$\preceq$B), or B is in the past of A (B$\preceq$A), or in general the events may be space-like separated, i.e. there is no signaling between the events A and B. However, quantum mechanics allows for more general processes which have no definite causal order, and this is because quantum objects can be in a superposition of different states. This phenomenon would be useful when one wants to unify two important theories of the 20 century, i.e., quantum mechanics and general relativity, as was shown by Lucien Hardy in his works on indefinite causal order (ICO) and quantum gravity  \cite{Hardy,LHardy}. However, differently from ICO, where one focuses on the order in which the operations are applied \cite{Oreshkov1,Valiron}, e.g. $\mathcal{E}$ is before $\mathcal{F}$, $\mathcal{F}$ is before $\mathcal{E}$ or a superposition of the two orders, we can consider as well the case which quantum channel has been applied on the system e.g. $\mathcal{E}$ or $\mathcal{F}$ or in general it may be a superposition of the two. The latter case is called a coherent superposition (CS) of the two channels $\mathcal{E}$ and $\mathcal{F}$ \cite{Aharonov}. This is motivated by the fact that it was recently shown that coherently superposed channels (CSCs) as well provide an advantage e.g. in quantum communication \cite{Abbottr} and quantum metrology \cite{ChapeauCS,DXie,Xiaobin}, and even more as was shown in \cite{DXie} they improve the ICO in parameter estimation. Furthermore, CSCs have been not investigated in other areas such as QT to the authors' knowledge except \cite{RRubino,RRubino1}. 

As we said before, recently measurement-based quantum thermal machines \cite{Das,Elouard,Sagawa,Jordan,Chand,Buffoni,Lisboa,Daniel} draw more attention in the field of QT. However, the previous results on quantum heat engines were either limited to special unital \cite{Bresque,ShanheSu,Shanhe,Das,Buffoni} or non-unital channels \cite{Bezhadi,Pedro,Lisboa} in the bath stages, or a specific driving protocol \cite{Peterson} at the work strokes and under the assumption of the cycle being time-reversal symmetric i.e. $V^{\dagger}=U$, an exception are the results reported in Refs. \cite{BijayKumar,Poletti1,Prasanna1}. Even more, only counted works have considered quantities beyond averages, such as variance or in general higher cumulants, which have not been analyzed so far for quantum Otto heat engines (QOHEs). In this work, we try to fill these gaps. To do so, inspired by the QOHE studied in \cite{ShanheSu,Shanhe}, we consider the role of replacing one of the heat baths of the QOC with an arbitrary unital map on efficiency, heat, work, their relative fluctuations, skewness and kurtosis. In the second part of this paper, when considering applying the unital channel in a CS we limit ourselves to the unital map studied in Refs. \cite{Shanhe,ShanheSu}.

For example, in Ref \cite{SushantB} Saryal et al. have shown that in the adiabatic regime, the ratio of the $n$-th cumulants of work and heat (for quantum Otto heat engines with completely thermalizing reservoirs), is equal to the $n$-th power of the Otto efficiency and it is upper bounded by the $n$-th power of the Carnot efficiency. Translated into our notation this means that, $\llangle W^{n}\rrangle/\llangle Q_{M}^{n}\rrangle=\left(1-\nu_{1}/\nu_{2}\right)^{n} \leq \left(1-\beta_{h}/\beta_{c}\right)^{n}$. However, the authors did not give an interpretation of these bounds for $n\geq 3$ nor they have considered what would happen beyond the adiabatic regime. In Ref \cite{Gerry1}, Saryal et al. have proved that for autonomous coupled and continuous thermal machines in the linear response regime, the ratio of the $n$-th cumulants of the input and output currents are lower bounded by the $n$-th power of the efficiency and upper bounded by the $n$-th power of the Carnot efficiency. As a consequence of this interesting result, is that the engine's efficiency received a tighter bound than the Carnot bound. Furthermore, in Ref. \cite{Gerry2} the authors have also considered the ratio of the $n$-th cumulants of an ensemble of non-interacting quantum thermal machines. Suppose we have $i \in [1,N]$ individual machine, and that the ratio of $n$-th cumulants of every machine satisfy, $ \llangle W_{i}^{n}\rrangle/\llangle Q_{Mi}^{n}\rrangle \leq \left(1-\beta_{h}/\beta_{c}\right)^{n}$. Beyond the linear response limit, the authors have found that the ratio of the sum of the $n$-th cumulants of individual machines is bounded by the $n$-th power of the Carnot efficiency, i.e. $ \sum_{i=1}^{N}\llangle W_{i}^{n}\rrangle/\sum_{i=1}^{N}\llangle Q_{Mi}^{n}\rrangle \leq \left(1-\beta_{h}/\beta_{c}\right)^{n}$. For an ensemble of non-interacting quantum refrigerators, they have shown analytically and numerically under certain different assumptions, that the ratio of the sum of the $n$-th cumulants has a lower bound as well, given by the $n$-th power of the efficiency. The lower bound is saturated for example in the tight coupling limit, See \cite{Gerry1,Gerry2}. However, in the heat engine regime, only the upper bound was respected, but not the lower bound.

The main five messages to take from this paper can be summarised as follows:

\begin{enumerate}
\item First, considering a QOHE, we neither specify the driving protocol nor the unital channel replacing the thermal bath. Then we give the exact analytical expression of the characteristic function (CF) from which all the cumulants of heat and work would emerge. We show that considering arbitrary unital channel have a positive influence on heat absorbed, work and efficiency. Furthermore, we prove that under the effect of monitoring, changing the gap of the single spin-1/2 is a necessary condition for work extraction. This is true for either symmetric or asymmetric driven Otto cycle. We trace this back to the negative influence of projective measurement used to assess the fluctuations of thermodynamic quantities. Further, we show that for the asymmetric Otto cycle not only do the forward and the backward are needed to be treated on equal footing for meaningful bounds on relative fluctuations (RFs)\cite{BijayKumar}, but also help in restoring the fact that the Otto efficiency is still an upper bound.

\item We prove that when one of the heat baths in the Otto cycle is replaced by an arbitrary unital map, then the system cannot work as a refrigerator independently of the parameters. More precisely, the bath cannot be cooled. This is forbidden by the second law of thermodynamics in accordance with the Kelvin-Planck statement that heat only flows spontaneously from a hotter body to a cooler one. 

\item We show that the ratio of the fluctuations of $W$ and $Q_{M}$ are lower and upper bounded when the system is working as a heat engine and for some regions when it is working as an accelerator. This is proved to be true for the symmetric and asymmetric Otto cycle. We proved as well, that the square of the Otto efficiency provides a lower bound on the ratio of fluctuations for the symmetric Otto cycle. Then, numerically we show that the ratio of the third and the fourth cumulants can not be lower-bounded by the third (fourth) power of the efficiency as is the case for the relative fluctuations, nor they are bounded by unity. The reason behind this is purely quantum due to the driving as we show later on. Furthermore, the fact that we do not have always, $\llangle W^{n}\rrangle/\llangle Q_{M}^{n}\rrangle\geq\left(\llangle W\rrangle/\llangle Q_{M}\rrangle\right)^{n}$ for $n=3,4$ suggest that the ratio of third and fourth cumulants can not always provide a bound on the efficiency as it the case for the second cumulants.

\item We analyze in detail the effect of considering an arbitrary unital channel on the average work and its relative fluctuations. We show that the work average can be enhanced as well as its reliability (note that enhanced reliability is equivalent to decreasing relative fluctuations). When the inverse temperature is negative, we comment on the positive rule of non-adiabaticity.

\item Finally, when considering CSCs we consider the unital channel studied in \cite{Shanhe}. We show, that when the control qubit is projected in $|-\rangle_{c}$ Fourier basis has a positive influence on the work extracted and its reliability as well as efficiency, even though, the latter cannot exceed that of the Otto. When the control qubit is projected in the $|+\rangle_{c}$ Fourier basis we see the inverse of these conclusions. Note that efficiency enhancement in the non-adiabatic regime could be helpful when one is interested in both efficiency and power as in real-world applications.
\end{enumerate}

Our results put forward thoroughly and deeply the ones reported in Refs. \cite{Shanhe,ShanheSu}, by not only considering average quantities but as well the variance, skewness and kurtosis which were not considered there. Furthermore, inspired by the results reported in \cite{Rogerio,ShanheSu,Shanhe}.
we will express all the cumulants of work and heat in terms of only three transitions probabilities $\delta$, $\theta$ and $\zeta$ defined below, without specifying the unital channel providing heat to the system nor the driving protocol used to change the gap of the system and without the assumption that the cycle is time-reversal symmetric. In the main text below, we show the difference between our work and the one in \cite{ShanheSu,Shanhe}. Furthermore,  note that skewness and kurtosis have been taken into account here since they were not analyzed in the previous works on QOHEs in literature to the authors' knowledge. However, note that considering higher cumulants such as the third and fourth comes with more computational complexity, let alone higher than the fourth cumulant. For example, here we could not prove that the ratio of the third (fourth) cumulants is upper bounded by 1 and lower bounded by the third (fourth) power of the efficiency since their expressions are too complicated in terms of the parameters especially those of work. Fortunately, numerically we found that they do not obey the same bound as is the case for the ratio of fluctuations.

This paper is structured as follows:  In Sec.\ref{QOC}, we take the conventional QOC with two heat baths and replace one of them with an arbitrary unital qubit channel which would play the role of a hot bath. We give the exact analytical expression of the CF. Further, since we use the two-point measurement to assess fluctuations of work and heat we show that the latter has a negative effect on the work positive condition. More precisely, differently from \cite{Shanhe} we show that changing the frequency of the spin is a necessary condition for $\llangle W\rrangle>0$. The efficiency and the first four cumulants of work and heat are analyzed in detail. In Sec.\ref{QOCM}, first in Sec.\ref{CS} we define what we mean by CSCs, then we apply the unital channel considered in \cite{Shanhe} in a CS and show the effect of this on efficiency, work as well as its relative fluctuations. Finally, in Sec.\ref{Conclu}, we give a summary of our results. In the appendix section, we provide some technical details. Throughout the paper, we set $\hbar=k_{B}=1$.\par

\section{Monitored quantum unital Otto cycle}\label{QOC}
Before we dive into the main results of the paper let's remind the reader of the unmonitored QOC with either two heat baths or one heat bath and an arbitrary channel.
\subsection{Unmonitored (asymmetric) QOC}
\begin{figure}[hbtp]
\centering
\subfigure[]{
\includegraphics[scale=0.4]{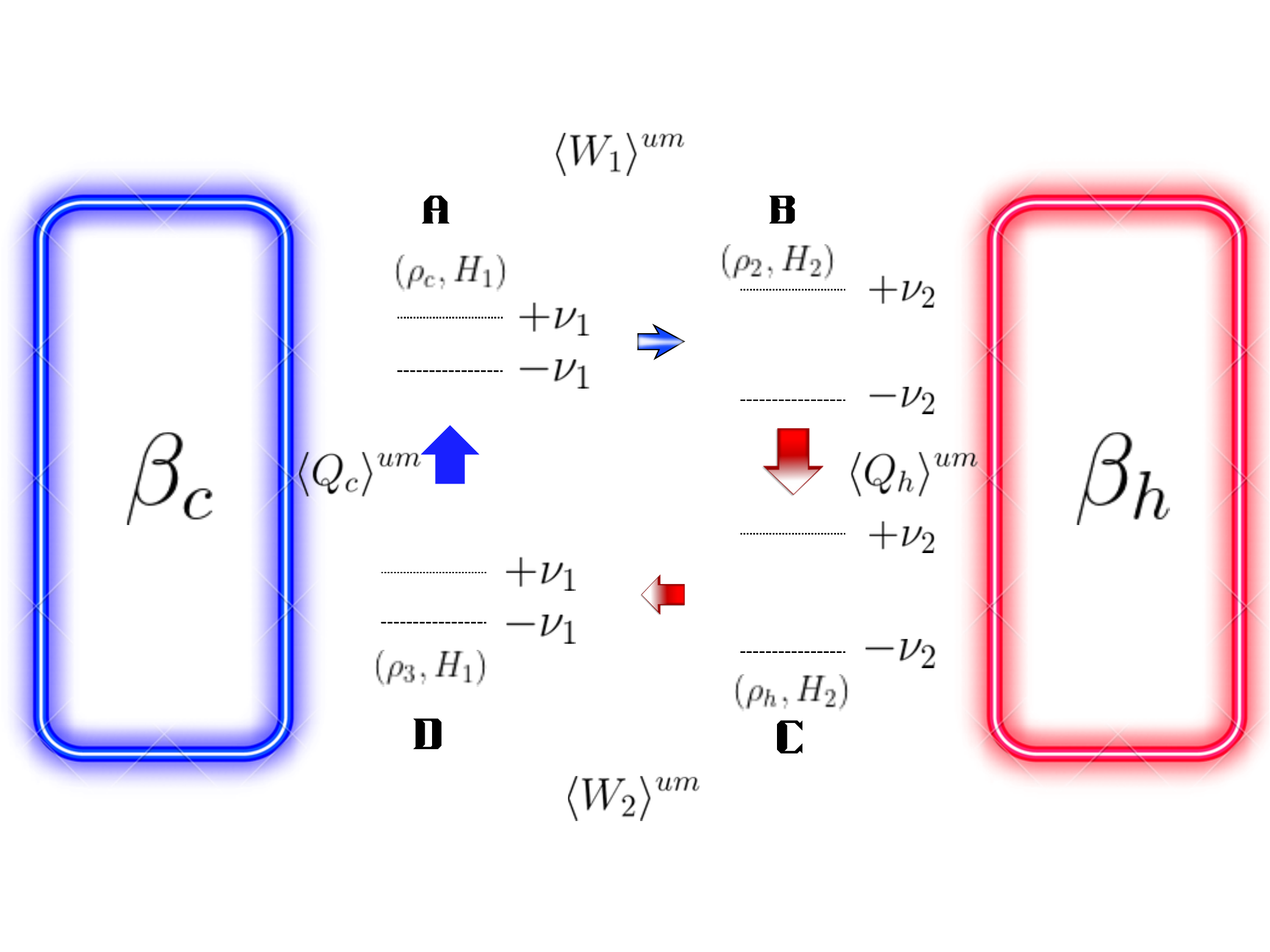}
}
\hfill
\subfigure[]{\includegraphics[scale=0.4]{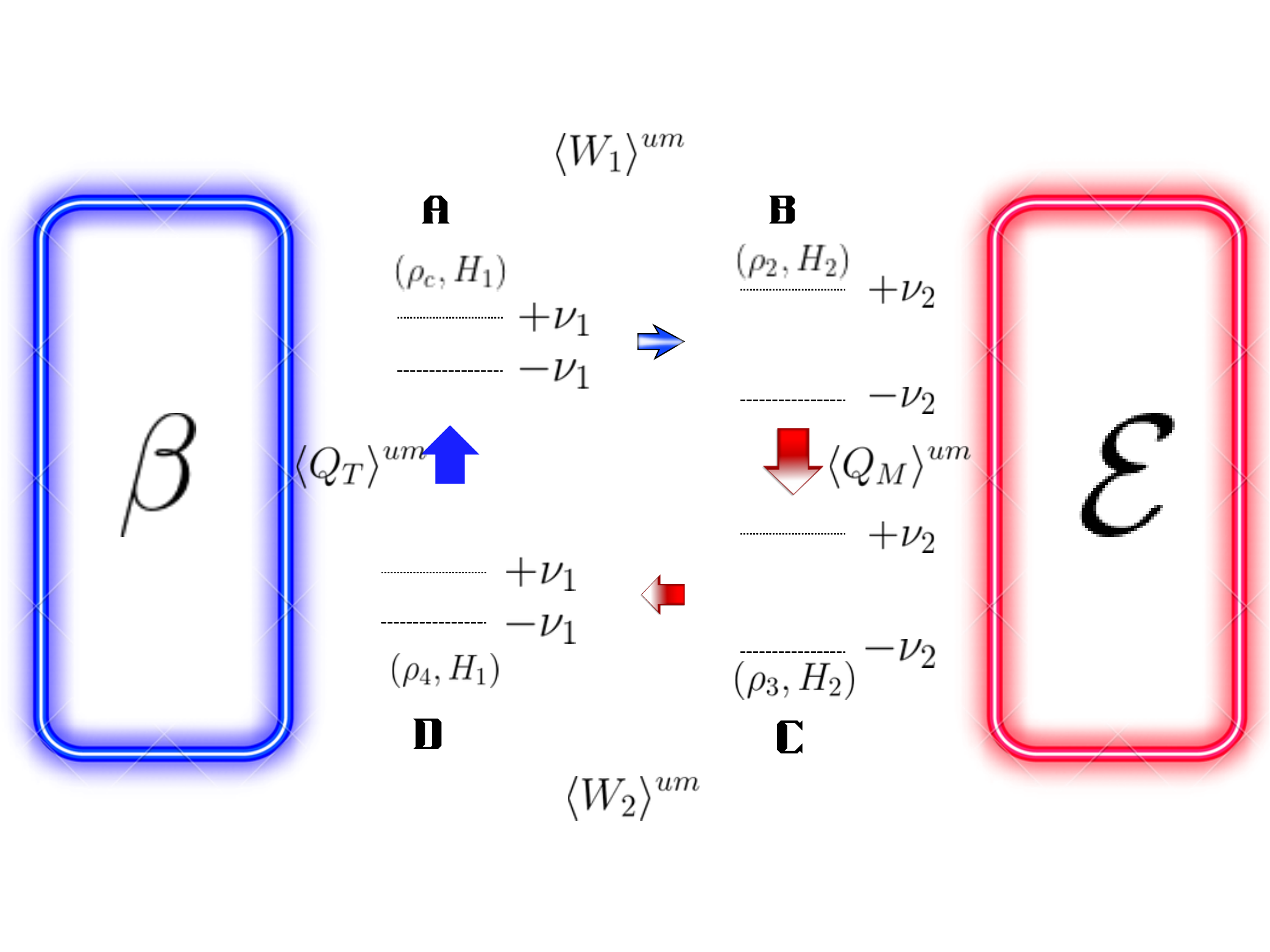}}
\caption{ (a) A two-level system undergoing the four strokes of the conventional QOC. $\mathbf{A}\rightarrow \mathbf{B}$ an adiabatic expansion $\nu_{1}\rightarrow\nu_{2}(>\nu_{1})$, $\mathbf{B}\rightarrow\mathbf{C}$ the system gets equilibrates with a hot bath, $\mathbf{C}\rightarrow \mathbf{D}$ an adiabatic compression $\nu_{2}\rightarrow\nu_{1}$ and finally $\mathbf{D}\rightarrow\mathbf{A}$ gets equilibrate with a cold bath, thus closing the cycle. In this case, we have two heat baths one at inverse temperature $\beta_{c}$ and the other at inverse temperature $\beta_{h}$. $\langle Q_{c}\rangle^{um}$ is the heat exchanged with the cold bath, $\langle Q_{h}\rangle^{um}$ is the heat exchanged with the hot bath and $\langle W_{1(2)}\rangle^{um}$ is the work done(extracted) on(from) the system during the expansion (compression) stroke. From the conservation of energy i.e. $\langle Q_{c}\rangle^{um}+\langle Q_{h}\rangle^{um}+\langle W_{1}\rangle^{um}+\langle W_{2}\rangle^{um}=0$, the total work is $\langle W\rangle^{um}=-(\langle W_{1}\rangle^{um}+\langle W_{2}\rangle^{um})=\langle Q_{c}\rangle^{um}+\langle Q_{h}\rangle^{um}$. (b) In this work instead of taking the hot bath to be a completely thermalizing channel, we will change it by a general quantum channel $\mathcal{E}$ described by $\sum_{i}K_{i}(.)K_{i}^{\dagger}$. Below when studying and analyzing the first four cumulants, we limit ourselves to the case when this channel is unital, i.e. $\mathcal{E}(\mathbb{1})=\mathbb{1}$. $\langle Q_{T}\rangle^{um}$ is the heat exchanged with the bath at inverse temperature $\beta$ and $\langle Q_{M}\rangle^{um}$ is the heat exchanged with the quantum channel $\mathcal{E}$.  In the same manner as before the total work is $\langle W\rangle^{um}=-(\langle W_{1}\rangle^{um}+\langle W_{2}\rangle^{um})=\langle Q_{M}\rangle^{um}+\langle Q_{T}\rangle^{um}$. We see below, that monitoring the system between strokes would affect the average quantities $\langle W\rangle^{um}$ and $\langle Q_{T}\rangle^{um}$ for this we used the subscript $um$(i.e. unmonitored) to distinguish between the average quantities computed when there is monitoring. Actually for (a) there is no difference in the averages when monitoring or not the system since the baths are considered to be completely thermalizing. This is not the case for (b).}\label{b0.55}
\hfill
\end{figure}

For a quantum system e.g. a two-level system, which we consider in our work, the steps of the conventional QOC with two thermal baths are given as follows (see Fig. \ref{b0.55}(a)): First, the system starts in thermal equilibrium with a cold bath at inverse temperature $\beta_{c}$. At the point $\mathbf{A}$, the state of the system and its Hamiltonian are given as follows: $\rho_{c}=e^{-\beta_{c} H_{1}}/Z_{c}$ and $H_{1}=\nu_{1}(|+\rangle_{11}\langle+|-|-\rangle_{11}\langle-|)$. $Z_{c}(=\mathrm{Tr}(e^{-\beta_{c} H_{1}}))$ is the partition function. Second, from $\mathbf{A}$ into $\mathbf{B}$, we isolate the system from the cold bath, and then an adiabatic expansion is applied by changing the gap from $\nu_{1}$ to $\nu_{2}$($>\nu_{1}$). This external driving is described by a unitary given by $U$. At the end of this process, point $\mathbf{B}$, the state and the Hamiltonian of the system are given as $\rho_{2}=U\rho_{c}U^{\dagger}$ and $H_{2}=\nu_{2}(|+\rangle_{22}\langle+|-|-\rangle_{22}\langle-|)$. In this unitary stroke an amount of work $\langle W_{1}\rangle^{um}=E_{2}-E_{1}=\mathrm{Tr}\left[\rho_{2}H_{2}\right]-\mathrm{Tr}\left[\rho_{c}H_{1}\right]$ is done on the system. Third, from $\mathbf{B}$ into $\mathbf{C}$, we thermalize again our system with a hot bath at inverse temperature $\beta_{h}$. At the end of this thermalization (point $\mathbf{C}$) the state of the system is $\rho_{h}=e^{-\beta_{h} H_{2}}/Z_{h}$, but the Hamiltonian stays fixed. In this thermalization stroke an amount of heat $\langle Q_{h}\rangle^{um}=E_{3}-E_{2}=\mathrm{Tr}\left[\rho_{h}H_{2}\right]-\mathrm{Tr}\left[\rho_{2}H_{2}\right]$ is exchange with the bath. Fourth, from point $\mathbf{C}$ into $\mathbf{D}$, an adiabatic compression is applied by getting back the gap of the system into its initial value, i.e. $\nu_{1}$. This process is described by another unitary given by $V$. At the end of this process the state and the Hamiltonian of the system are given as follows: $\rho_{3}=V\rho_{h}V^{\dagger}$ and $H_{1}$. Usually, people assume that $V^{\dagger}=U$(or $V=U$)-this equality is dependent on the expression of the Hamiltonian-which corresponds to the cycle being time-reversal symmetric, an exception \citep{Poletti1,BijayKumar,Prasanna1}. In this case, there is no difference between running the cycle forward or backward. In this second unitary process an amount of work $\langle W_{2}\rangle^{um}=E_{4}-E_{3}=\mathrm{Tr}\left[\rho_{3}H_{1}\right]-\mathrm{Tr}\left[\rho_{h}H_{2}\right]$ is extracted from the system. And finally, from point $\mathbf{D}$ into $\mathbf{A}$, we thermalize our system again with the cold bath, to close the cycle. In this latter thermalization step the system will exchange an amount of heat $\langle Q_{c}\rangle^{um}=E_{1}-E_{4}=\mathrm{Tr}\left[\rho_{c}H_{1}\right]-\mathrm{Tr}\left[\rho_{3}H_{1}\right]$ with the cold bath. $E_{1}=\mathrm{Tr}\left[\rho_{c}H_{1}\right]$, $E_{2}=\mathrm{Tr}\left[\rho_{2}H_{2}\right]$, $E_{3}=\mathrm{Tr}\left[\rho_{h}H_{2}\right]$ and $E_{4}=\mathrm{Tr}\left[\rho_{3}H_{1}\right]$ are the average energies at the points $\mathbf{A}$, $\mathbf{B}$, $\mathbf{C}$ and $\mathbf{D}$, respectively. The total work is given by, $\langle W\rangle^{um}=-(\langle W_{1}\rangle^{um}+\langle W_{2}\rangle^{um})$. One of the assumptions that we can relax is that instead of two heat baths as in figure \ref{b0.55} (a), one can replace one of them with a measurement channel as has been done e.g. if Refs. \cite{ShanheSu,Shanhe,JYi,Daniel,Das}. Furthermore, note that the common thing between these works is that the quantum channels were unital.\par

In this work, we go a step further and relax two assumptions. The first is that we do not assume that $V^{\dagger}=U$ nor do we specify the protocol used to change the gap $\nu_{1}\leftrightarrow\nu_{2}$, and second instead of considering a specific unital channel replacing the heat hot bath, we consider an arbitrary unital qubit channel. When $V^{\dagger}(\neq)=U$ the cycle is (a)symmetric. Here we take the convention that if the heat is entering the system it is positive and if it flowing out it is negative. Actually what it means here for the QOC to be asymmetric is that the degree of non-adiabaticity at the adiabatic expansion/compression stroke may not be equal. For example, adiabatic compression can be applied in a quasistatic manner but the expansion in a non-adiabatic or vice-versa and so on. This shows that an asymmetric-driven Otto cycle is more general than a symmetric case.\par

Thus differently from figure \ref{b0.55}(a), our quantum unital Otto cycle steps are given as follows (see figure \ref{b0.55}(b)): First the system starts in thermal equilibrium with a heat bath at inverse temperature $\beta$. At the point $\mathbf{A}$, the state and the Hamiltonian of the system are, respectively, $\rho_{1}=e^{-\beta H_{1}}/Z$ and $ H_{1}$. After this, from the point $\mathbf{A}$ into $\mathbf{B}$, a unitary transformation $U$ is applied on the system, and the state becomes $\rho_{2}=U\rho_{1}U^{\dagger}$ and the Hamiltonian is $H_{2}$. This is an adiabatic expansion. Then, contrary to figure \ref{b0.55}(a), from point $\mathbf{B}$ into $\mathbf{C}$, an arbitray quantum unital channel $\mathcal{E}$ defined as $\mathcal{E}(.)=\sum_{j}K_{j}(.)K_{j}^{\dagger}$, is applied to the system, and the state of the system becomes $\rho_{3}=\sum_{j}K_{j}\rho_{2} K_{j}^{\dagger}$. The Hamiltonian of the system is still $H_{2}$. Then from point $\mathbf{C}$ into $\mathbf{D}$ we apply another unitary $V$ to the state of the system. In this stroke, the gap of the system changed back to $\nu_{1}$, and the state of the system becomes $\rho_{4}=V\rho_{3}V^{\dagger}$ and the Hamiltonian $H_{1}$, but in this case, we do not assume that $V=U^{\dagger}$. Finally from point $\mathbf{D}$ into $\mathbf{A}$ we thermalize our system again with the bath at inverse temperature $\beta$, thus closing the cycle. In analogy to figure \ref{b0.55} (a), $E_{1}=\mathrm{Tr}\left[\rho_{c}H_{1}\right]$, $E_{2}=\mathrm{Tr}\left[\rho_{2}H_{2}\right]$, $E_{3}=\mathrm{Tr}\left[\rho_{3}H_{2}\right]$ and $E_{4}=\mathrm{Tr}\left[\rho_{4}H_{1}\right]$ are the average energies at the points $\mathbf{A}$, $\mathbf{B}$, $\mathbf{C}$ and $\mathbf{D}$, respectively, and the heat averages are: $\langle Q_{M}\rangle^{um}=\mathrm{Tr}\left[\rho_{3}H_{2}\right]-\mathrm{Tr}\left[\rho_{2}H_{2}\right]$ and $\langle Q_{T}\rangle^{um}=\mathrm{Tr}\left[\rho_{1}H_{1}\right]-\mathrm{Tr}\left[\rho_{4}H_{1}\right]$.\par

At this point, an important remark is needed. Note that, since between the steps that we just described there is no measurement, this cycle is called an unmonitored cycle. This is the reason behind the notation $um$. In this work, we will consider the monitored version of the cycle described in figure \ref{b0.55}(b), since we are interested in analyzing higher cumulants of work and heat as well.
\subsection{The CF of the Otto cycle based on arbitrary qubit channels}
Before we give the exact analytical expression of the characteristic function of the forward for the Otto cycle based on unital qubit channels, let's consider first computing it for arbitrary qubit channels. Let's first clarify an important issue. When $V^{\dagger}\neq U$, in this case, we have to take into consideration both the forward as well as backward cycle and treat them on an equal footing. The steps of the forward cycle are $\mathbf{A}\rightarrow \mathbf{B}\rightarrow \mathbf{C}\rightarrow \mathbf{D}$ and those of the backward cycle are $ \mathbf{D}\rightarrow \mathbf{C}\rightarrow \mathbf{B}\rightarrow \mathbf{A}$. Otherwise, one would obtain inconsistent results on the bounds of the RF of work and heat \cite{BijayKumar}. Below we show that taking into account the backward cycle is not only necessary to obtain meaningful bounds on fluctuations \cite{BijayKumar}, but as well one can restore the fact that Otto efficiency is always the maximum attainable efficiency.\par

Applying a projective measurements \cite{Campisi1} on the Hamiltonian along the cycle ($\mathbf{A}\rightarrow \mathbf{B}\rightarrow \mathbf{C}\rightarrow \mathbf{D}$), the stochastic quantities $W_{1}$, $Q_{M}$ and $W_{2}$ are given as follows,
\begin{equation}
\begin{split}
&
W_{1}=\nu_{m}-\nu_{n},
\\ &
Q_{M}=\nu_{k}-\nu_{m},
\\ &
W_{2}=\nu_{l}-\nu_{k}.
\end{split}
\end{equation}
Here, $\nu_{n}$, $\nu_{m}$, $\nu_{k}$ and $\nu_{l}$ are the measured energy at the four points $\mathbf{A}$, $\mathbf{B}$, $\mathbf{C}$ and $\mathbf{D}$ in figure \ref{b0.55}(b), respectively, and $n$, $m$, $k$ and $l$ are their corresponding quantum nuumbers.

By adopting the two-point measurement scheme \cite{Campisi1,Esposito}, the joint probability distribution (PD) of the stochastic work $W=-(W_{1}+W_{2})$ and the stochastic heat $Q_{M}$ exchanged with an arbitrary quantum qubit channel with Kraus operators $K_{j}$, of the forward cycle, is given as follows,
\begin{equation}
\begin{split}
P(W,Q_{M})_{F} & =\sum_{n,m,k,l}\frac{e^{-\beta \nu_{n}}}{Z}|_{2}\langle m|U|n\rangle_{1}|^{2}\sum_{j}|_{2}\langle k|K_{j}|m\rangle_{2}|^{2}
 |_{1}\langle l|V|k\rangle_{2}|^{2}
\delta(W+(\nu_{m}-\nu_{n}+\nu_{l}-\nu_{k}))
\delta(Q_{M}-(\nu_{k}-\nu_{m})).
\end{split}
\label{PWQ}
\end{equation}
To get the first, second (or in general higher cumulants) of work and heat we use the CF which follows from the PD, written as
\begin{equation}
\chi(\gamma_{W},\gamma_ {M})_{F}=\int P(W,Q_{M})_{F}e^{i\gamma_{W}W}e^{i\gamma_{M}Q_{M}}dWdQ_{M}.\label{Chi}
\end{equation}
This is when $W$ and $Q_{M}$ are continuous variables. Using the expression (\ref{PWQ}) then the CF becomes,
\begin{equation}
\chi(\gamma_{W},\gamma_ {M})_{F}  =\sum_{n,m,k,l}\frac{e^{-\beta \nu_{n}}}{Z}|_{2}\langle m|U|n\rangle_{1}|^{2}\sum_{j}|_{2}\langle k|K_{j}|m\rangle_{2}|^{2}
 |_{1}\langle l|V|k\rangle_{2}|^{2}
e^{-i\gamma_{W}(\nu_{m}-\nu_{n}+\nu_{l}-\nu_{k})}e^{i\gamma_{M}(\nu_{k}-\nu_{m})} .
\label{Chi}
\end{equation}
Here, $\gamma_{M}$ and $\gamma_{W}$ are the Fourier conjugate of $Q_{M}$ and W, respectively. The cumulants of heats and work of the forward cycle follow from $\chi(\gamma_{W},\gamma_ {M})_{F}$, by the equation
\begin{equation}
\llangle W^{n}Q_{M}^{m}\rrangle_{F}=\frac{\partial^{n}\partial^{m}\rm ln(\chi(\gamma_{W},\gamma_{M})_{F})}{\partial(i\gamma_{W})^{n}\partial(i\gamma_{M})^{m}}\bigg\rvert_{\gamma_{W},\gamma_{M}=0}.\label{WQNM}
\end{equation}
If we eliminate the logarithmic function then this expression would give us the central moments, which are denoted for an arbitrary stochastic variable $\phi$ as $\langle \phi^{n}\rangle=\left(\partial^{n}\rm \chi(\gamma_{\phi})_{F}/\partial(i\gamma_{\phi})^{n}\right)\big\rvert_{\gamma_{\phi}=0}$. Actually for the first three derivatives, i.e. $n=1,2$ and $3$, the central moments and cumulants are the same. However, from the fourth derivative, there is a difference between central moment and cumulant. $\llangle \phi\rrangle_{F}$, $\llangle \phi^{2}\rrangle_{F}$, $\llangle \phi^{3}\rrangle_{F}$ and $\llangle \phi^{4}\rrangle_{F}$ are, repectively, the first(average), the second(variance), the third and the fourth cumulants of $\phi$, in the forward cycle, where $\phi=W,Q_{M}$. The normalized quantities $\llangle \phi^{2}\rrangle_{}/\llangle \phi\rrangle^{2}$, $\llangle \phi^{3}\rrangle_{}/\llangle \phi^{2}\rrangle_{}^{3/2}$ and $\llangle \phi^{4}\rrangle_{}/\llangle \phi^{2}\rrangle_{}^{2}$ are, respectively, RFs, skewness and kurtosis of $\phi$. $\llangle \phi^{4}\rrangle/\llangle \phi^{2}\rrangle^{2}$ is kurtosis shifted by -3, which is called excess kurtosis. Here, we just call it kurtosis. Usually, people use the notation $\langle.\rangle$ to mean the average. In what follows we use $\llangle.\rrangle$ for $W$ and $Q_{M}$ to mean the average but for the efficiency we use the ordinary notation, i.e. $\langle.\rangle$. We hope this notation is not confusing. Till this point everything is clear. However, one should note that there will be a very important problem when taking into account the backward. This follows, from the fact that for general maps the adjoint map cannot be taken to be the reversal, since it may not give a positive probability, i.e. physical states. For this reason, below we limit ourselves to unital channels. 

Based on Table (\ref{Tbale1}) in appendix \ref{chi} and by defining $\delta=|_{2}\langle+|U|-\rangle_{1}|^{2}$ to be the probability of the system transitionning from the state $|-\rangle_{1}$ into the state $|+\rangle_{2}$ in the adiabatic expansion, $\theta=\sum_{j}|_{2}\langle-|K_{j}|+\rangle_{2}|^{2}$ the probability of the system in the state $|+\rangle_{2}$ being found in the state $|-\rangle_{2}$ after the unital map has been applied on it and finally $\zeta=|_{1}\langle+|V|-\rangle_{2}|^{2}$ the transition probability between the states $|-\rangle_{2}$ and $|+\rangle_{1}$ in the adiabatic compression, then plugging all this in the equation (\ref{Chi}) (see appendix \ref{chi}), one can obtain,
\begin{widetext}
\begin{equation}
\begin{split}
\chi(\gamma_{W},\gamma_{M})_{F}= & 
(1-\delta)(1-\zeta)\left(\frac{e^{\beta\nu_{1}}}{Z}(h-\theta)+\frac{e^{-\beta\nu_{1}}}{Z}(1-\theta)\right)+
(1-\delta)\zeta\left(\frac{e^{\beta\nu_{1}}}{Z} e^{-2i\gamma_{W}\nu_{1}}(h-\theta)+\frac{e^{-\beta\nu_{1}}}{Z} e^{2i\gamma_{W}\nu_{1}}(1-\theta)\right)
\\ &
+(1-\delta)\zeta\left(\frac{e^{\beta\nu_{1}}}{Z} (1-h+\theta)e^{2i\gamma_{W}\nu_{2}}e^{2i\gamma_{M}\nu_{2}}+\frac{e^{-\beta\nu_{1}}}{Z}e^{-2i\gamma_{W}\nu_{2}}e^{-2i\gamma_{M}\nu_{2}}\theta\right)
\\ &
+(1-\delta)(1-\zeta)\left(\frac{e^{\beta\nu_{1}}}{Z} (1-h+\theta)e^{2i\gamma_{W}(\nu_{2}-\nu_{1})}e^{2i\gamma_{M}\nu_{2}}+\frac{e^{-\beta\nu_{1}}}{Z}e^{-2i\gamma_{W}(\nu_{2}-\nu_{1})}e^{-2i\gamma_{M}\nu_{2}}\theta\right)
\\ &
+\delta(1-\zeta)\left(\frac{e^{\beta\nu_{1}}}{Z}e^{-2i\gamma_{W}\nu_{2}}e^{-2i\gamma_{M}\nu_{2}}\theta+\frac{e^{-\beta\nu_{1}}}{Z}e^{2i\gamma_{W}\nu_{2}}e^{2i\gamma_{M}\nu_{2}}(1-h+\theta)\right)
\\ &
+\delta\zeta\left(\frac{e^{\beta\nu_{1}}}{Z}e^{-2i\gamma_{W}(\nu_{1}+\nu_{2})}e^{-2i\gamma_{M}\nu_{2}}\theta+\frac{e^{-\beta\nu_{1}}}{Z}e^{2i\gamma_{W}(\nu_{1}+\nu_{2})}e^{2i\gamma_{M}\nu_{2}}(1-h+\theta)\right)
\\ &
+\delta\zeta\left(\frac{e^{\beta\nu_{1}}}{Z}(1-\theta)+\frac{e^{-\beta\nu_{1}}}{Z}(h-\theta)\right)+\delta(1-\zeta)\left(\frac{e^{\beta \nu_{1}}}{Z} e^{-2i\gamma_{W}\nu_{1}}(1-\theta)+\frac{e^{-\beta \nu_{1}}}{Z} e^{2i\gamma_{W}\nu_{1}}(h-\theta)\right),\label{Arbitrary}
\end{split}
\end{equation}
\end{widetext}
with, $h=\sum_{j} {}_{2}\langle-|K_{j}K_{j}^{\dagger}|-\rangle_{2}$. To obtain the CF of the backward cycle $\chi(\gamma_{M},\gamma_{W})_{B}$ it is not straightforward, since the $K_{j}^{\dagger}$ cannot describe a valid channel, i.e. a physical channel. Thus computing the CF of the backward, is difficult, for general qubit maps. Furthermore, below all thermodynamic quantities will be expressed in terms of $\delta$, $\theta$ and $\zeta$. And below, whenever we omit the subscript $F$ referring to the forward, it means that the forward and the backward are the same.
\subsection{$\chi(\gamma_{M},\gamma_{W})_{F}$ of the Otto cycle based on arbitrary unital qubit channels}
When the map $\mathcal{E}$ is unital,
completely positive and trace-preserving, so is its adjoint $\mathcal{E}^{\dagger}$. Therefore, the adjoint $\mathcal{E}^{\dagger}$ can be regarded as another time evolution. Thus the backward of,
\begin{equation}
\mathcal{E}(.)=\sum_{j}K_{j}(.)K_{j}^{\dagger},\hspace{1cm}{\rm is-defined-as}\hspace{1cm}\mathcal{E^{\dagger}}(.)=\sum_{j}K_{j}^{\dagger}(.)K_{j}.
\end{equation}
Furthermore, for a two-level system considered here, one should note that every unital map can be decomposed in terms of Pauli operators \cite{Nielsen}, for this reason, unital maps are also called Pauli channels, see \cite{Nielsen}. Thus,
\begin{equation}
\mathcal{E}(.)=\sum_{i=0}^{3}p_{i}\sigma_{i}(.)\sigma_{i}^{\dagger}.\label{PauliU}
\end{equation}
with, $\sum_{i=0}^{3}p_{i}=1$. The Pauli operators in the basis of the Hamiltonian $H_{2}$ are given as follows: $\sigma_{0}=|+\rangle_{22}\langle+|+|-\rangle_{22}\langle-|$, $\sigma_{1}=|+\rangle_{22}\langle-|+|-\rangle_{22}\langle+|$, $\sigma_{2}=i|-\rangle_{22}\langle+|-i|+\rangle_{22}\langle-|$ and $\sigma_{3}=|+\rangle_{22}\langle+|-|-\rangle_{22}\langle-|$. In this case, we have, $0\leq\theta=p_{1}+p_{2}\leq1$, which shows that the highest possible value of $\theta$ is 1. Below, we comment on this in detail. The backward does not mean that we reverse time, since in the lab the reverse process of the forward as well as is run in the forward. The reader should note that the adjoint of equation (\ref{PauliU}) is itself, i.e. $\mathcal{E}=\mathcal{E}^{\dagger}$.\par

When the map is unital one can easily show that $h=1$ (for unital maps we have $\sum_{j}K_{j}\mathbb{1}K_{j}^{\dagger}=\mathbb{1}$ and thus $h=1$), in this case, $\chi(\gamma_{M},\gamma_{W})_{F}$ get simplified into a more compact form, and we have,
\begin{widetext}
\begin{equation}
\begin{split}
\chi(\gamma_{W},\gamma_{M})_{F} & =(1-\theta)\left( 1+\left(\frac{\rm 2 cos((2\gamma_{W}+i\beta)\nu_{1})}{Z}-1\right)(\delta+\zeta-2\delta\zeta)\right)
\\ &
+\theta \left( \left(1-\delta\right) \left(\zeta \frac{\rm 2cos(2(\gamma_{W}+\gamma_{M})\nu_{2}-i\beta\nu_{1})}{Z}+(1-\zeta) \frac{\rm 2 cos(2(\gamma_{W}(\nu_{2}-\nu_{1})+\gamma_{M}\nu_{2})-i\beta\nu_{1})}{Z}\right)\right.
\\ &
+\left.\delta\left((1-\zeta) \frac{ 2 \cos(2(\gamma_{W}+\gamma_{M})\nu_{2}+i\beta\nu_{1})}{Z}+\zeta \frac{\rm 2cos(2((\nu_{1}+\nu_{2})\gamma_{W}+\gamma_{M}\nu_{2})+i\beta\nu_{1})}{Z}\right)\right).
\end{split}\label{chih}
\end{equation}
\end{widetext}
The backward CF follows from this equation, by the correspondence $\delta\leftrightarrow\zeta$. Note that $\theta_{F}=\theta_{B}=\theta$ since the channel is unital. The exact analytical equation (\ref{chih}) is the first main result of the paper. One can check easily that for $\gamma_{M}=\gamma_{W}=0$, we have $\chi(\gamma_{W},\gamma_{M})_{F}=1$, which is nothing but probability conservation.
\subsection{Exact analytical expressions of the first and second cumulants of work and heat}
From equations (\ref{WQNM}) and (\ref{chih}), one can derive the next cumulants for the forward cycle
\begin{widetext}
\begin{equation}
\llangle Q_{M}\rrangle_{F}=2(1-2\delta)\theta\nu_{2}\tanh(\beta\nu_{1}),
\end{equation}
\begin{equation}
\llangle Q_{M}^{2}\rrangle_{F}=-4\theta\nu_{2}^{2}(-1+(1-2\delta)^{2}\theta\tanh^{2}{(\beta\nu_{1})}),
\end{equation}
\begin{equation}
\llangle Q_{T}\rrangle=-2(\theta+(1-2\theta)(\delta+\zeta-2\delta\zeta))\nu_{1}\tanh(\beta\nu_{1}),
\end{equation}
\begin{equation}
\begin{split}
\llangle W\rrangle_{F} & =2((1-2\delta)\theta\nu_{2}-(\theta+(1-2\theta)(\delta+\zeta-2\delta\zeta))\nu_{1})\tanh(\beta\nu_{1}),
\end{split}
\end{equation}
\begin{equation}
\begin{split}
\llangle W^{2}\rrangle_{F} & = 4(\theta+\zeta-2\theta\zeta+\delta(-1+2\theta)(-1+2\zeta)) \nu_{1}^{2}+8\theta(-1+\delta+\zeta)\nu_{1}\nu_{2}+4\theta\nu_{2}^{2} 
\\ & 
-4((\theta+\zeta-2\theta\zeta+\delta(-1+2\theta)(-1+2 \zeta))\nu_{1}+(-1+2\delta)\theta\nu_{2})^{2}\tanh^{2}{(\beta\nu_{1})}.
\end{split}\label{Wvari}
\end{equation}
\end{widetext}

Note that when $\theta=0$ all the cumulants of $Q_{M}$ become null since in this case no energy will be exchanged between the system and the channel. The third and fourth cumulants are not reported here since they are cumbersome and not illuminating at all. Further, the average heat $\llangle Q_{T}\rrangle$ follows from the law of energy conservation. 

Before we proceed, some important remarks are needed to be mentioned; first, one can see clearly that when $\beta=0$ (high-temperature limit) the initial state will be maximally mixed, and one can show easily that $\llangle W\rrangle_{F}$, $\llangle Q_{M}\rrangle_{F}$ and $\llangle Q_{T}\rrangle$ are null. The same thing for backward quantities. This follows from the fact that the state along the cycle will always stay a maximally mixed state, thus no change in the average energy will be observed. This is true either for the monitored or unmonitored Otto cycle. This is because the maximally mixed state is a fixed point of the cycle. This shows that we have to focus on $\beta\neq0$. However, this does not mean necessarily that higher cumulants will be zero, such as the variance and fourth cumulant. One can show that all odd cumulants will be null since this is due to the fact the distribution of heat and work will be symmetric around the 0 value, which is nothing but the average value. For example when $\beta\rightarrow+\infty$ or $-\infty$, the initial state will be pure state (for $\beta\rightarrow +\infty(-\infty)$, the pure state will be the ground state, i.e. $|-\rangle_{1}$(exicted state $|+\rangle_{1}$)).\par

In appendix \ref{Proof}, we prove that for the considered measurement unital channel in \cite{Shanhe}, $\theta$ cannot exceed 1/2. However, in our case, we have $0\leq \theta\leq1$ as we saw before. As we show below, this can help in improving the work extracted, decreasing its relative fluctuations and enhancing the efficiency. It is already seen from the expression of $\llangle Q_{M}\rrangle$ that when $\delta<1/2$ and for $\beta>0$, the more we increase $\theta$ the more heat would be provided by the unital channel. The maximum value is reached when $\theta=1$.

At this point, we want to emphasize that for $p_{1}+p_{2}=1$ and when the state is classical i.e. there are no off-diagonal elements, e.g. in the adiabatic regime, the unital channel Eq. (\ref{PauliU}), will supply the system with energy in the form of work and not heat, since in this case, the channel will not change the entropy of the system. Therefore, even though the channel is a mixture of $\sigma_{1}$ and $\sigma_{2}$ Pauli operators, the diagonal elements will only be swapped, thus the entropy of an arbitrary state $\rho$ and $\mathcal{E}(\rho)$ is equal. In the case when $p_{1}=p_{2}=0$, $\llangle Q_{M}\rrangle=0$ as expected, since the Kraus operators (i.e. the identity and the $\sigma_{3}$ Pauli operator,) commute with the Hamiltonian. When the system state is non-diagonal, i.e. quantum state for example in the non-adiabatic regime, and when $p_{1}+p_{2}=1$, the bath can provide the system with energy in the form of heat, as long as $p_{1}$ and $p_{2}$ in Eq.(\ref{PauliU}) are both non vanished. Thus when the unital channel is a mixture of the $\sigma_{1}$ and $\sigma_{2}$ Pauli operators acting on a quantum state, then the channel will provide the system with heat, since in this case entropy will be changed. However, when either $p_{1}$ or $p_{2}$ is null in this case the unital channel will be unitary and thus will not change the entropy of the system. In this latter case, the energy exchanged would be work and not heat. This analysis shows that in the adiabatic regime and when $p_{1}+p_{2}=1$ the unital channel will exchange energy in the form of work (ordered energy) with the system. However, beyond that regime, it can be either heat or work. See appendix \ref{WHL} for more details.

Equation (\ref{Wvari}) which characterizes the variance of $W$, is difficult to be analyzed, e.g. if one wants to know when it becomes vanishing, in contrast to the variance of $Q_{M}$. This is because it depends on the parameters in a complicated way. When $\delta=\zeta=0$, we have
\begin{equation}
\llangle W^{2}\rrangle=4\theta(\nu_{2}-\nu_{1})^{2}(1-\theta\tanh^{2}(\beta\nu_{1})).
\end{equation}
In this case, we see that the fluctuations of work vanished in three cases: when $\nu_{1}=\nu_{2}$ or $\theta=\tanh^{2}(\beta\nu_{1})=1$ or when $\theta=0$. While the variance of $Q_{M}$ when $\delta=\zeta=0$ vanishes in two cases: $\theta=0$ or $\theta=\tanh^{2}(\beta\nu_{1})=1$. In this case, the distributions of $W$ and $Q_{M}$  become degenerate and localized around one fixed value with all other outcomes having zero probability. Therefore, the variance would be zero. For example when $\theta=0$ then the only possible value of $W$ and $Q_{M}$ is 0. The PD becomes $P(W,Q_{M})_{F}=\delta(W)\delta(Q_{M})$ and the CF is $\chi(W,Q_{M})_{F}=1$. In this case, the third and fourth cumulants as well become null. Even higher cumulants than the first four cumulants become zero since the characteristic function is constant.

Let's now give a quick comparison between the case when we have a completely thermalizing channel and an arbitrary unital channel. When the unital qubit map is completely unital it describes the thermalization of a system with a heat bath at infinite temperature i.e. $T=\infty$(or $\beta=0$). In this case, the two levels become populated by $1/2$. A heat bath at positive inverse temperature can make the population of the higher level with at most $1/2$. However, when considering arbitrary unital channel the higher level can become more populated than the ground state, especially when the Pauli operators $\sigma_{1}$ and $\sigma_{2}$ become more dominated than the $\sigma_{0}$ and $\sigma_{3}$ operators. Let's show this statement with an example. Suppose the initial state before the unital channel is given by,
\begin{equation}
\rho_{i}=\begin{pmatrix}
1-a &  0 \\
0 & a \\
\end{pmatrix}.\label{Wzw}
\end{equation}
With $0\leq 1-a\leq1/2$. After applying the unital channel Eq. (\ref{PauliU}) on Eq. (\ref{Wzw}), and with simple steps of calculations one obtains
\begin{equation}
\mathcal{E}(\rho_{i})=\begin{pmatrix}
1-(a+(1-2a)(p_{1}+p_{2})) &  0 \\
0 & a+(1-2a)(p_{1}+p_{2}) \\
\end{pmatrix}.
\end{equation}
In this case one can show that $ 1-a\leq1-(a+(1-2a)(p_{1}+p_{2}))\leq a$. The upper bound is reached when $p_{1}+p_{2}=1$, but as we said before in this case the energy difference is work. Therefore we see that the population range of e.g. the excited level populated with $1-a$ before applying the unital channel, using an arbitrary unital map can vary from $1-a$ to $a$, thus can exceed $1/2$. This shows the strength of considering arbitrary unital maps.
\subsection{Positive work condition and efficiency bound for symmetric and asymmetric driven Otto cycle: heat engine regime}
Here we focus on the analysis of the first cumulant of $W$ and $Q_{M}$. Now, consider $\nu_{2}=\nu_{1}$ and $\zeta=\delta$, in such a case we have
\begin{equation*}
\llangle Q_{M}\rrangle=2(1-2\delta)\theta\nu_{1}\tanh(\beta\nu_{1}),
\end{equation*}
\begin{equation*}
\llangle Q_{T}\rrangle=-2(\theta+2\delta(1-2\theta)(1-\delta))\nu_{1}\tanh(\beta\nu_{1}),
\end{equation*}
and,
\begin{equation*}
\begin{split}
\llangle W\rrangle=4\delta(\delta+\theta-2\delta\theta-1)\nu_{1}\tanh(\beta\nu_{1})\leq0.
\end{split}
\end{equation*}
The latter inequality follows from the fact that $0\leq\delta+\theta-2\delta\theta(=\delta(1-\theta)+\theta(1-\delta))\leq1$. For the asymmetric Otto cycle case i.e. $\delta\neq\zeta$ but we still have $\nu_{2}=\nu_{1}$, we obtain,
\begin{equation*}
\begin{split}
\llangle W\rrangle_{F}&=-2(\delta+\zeta-2\delta\zeta-2\theta\zeta(1-2\delta))\nu_{1}\tanh(\beta\nu_{1}),
\end{split}
\end{equation*}
and,
\begin{equation*}
\begin{split}
\llangle W\rrangle_{B}&=-2(\delta+\zeta-2\delta\zeta-2\theta\delta(1-2\zeta))\nu_{1}\tanh(\beta\nu_{1}).
\end{split}
\end{equation*}
Numerical analysis showed that these two expressions separately can be either positive or negative, but taking both of the two quantities into consideration one obtains,
\begin{equation*}
\begin{split}
\llangle W\rrangle_{F}+\llangle W\rrangle_{B}&=-4(\delta+\zeta-2\delta\zeta-\theta\zeta+\delta\theta(-1+4\zeta))\nu_{1}\tanh(\beta\nu_{1}).
\end{split}
\end{equation*}
Simple manipulations further give,
\begin{equation}
\begin{split}
\llangle W\rrangle_{F}+\llangle W\rrangle_{B}=-4((\delta+\zeta-2\delta\zeta)(1-\theta)+2\delta\theta\zeta)\nu_{1}\tanh(\beta\nu_{1})(\leq0).
\end{split}\label{Manipula}
\end{equation}
The latter inequality follows from the fact that $(\delta+\zeta-2\delta\zeta)(1-\theta)$ and $2\delta\theta\zeta$ are always positive or null. Therefore for $\nu_{2}=\nu_{1}$, work cannot be extracted and the system can only work as either an accelerator ($A$) for $\delta<1/2$ or a heater ($H$) for $\delta>1/2$. Thus the condition $\nu_{2}\neq\nu_{1}$ is necessary for the system to work as a heat engine, independently of the cycle being symmetric or asymmetric. This is the main second result of the paper. We trace this to the the negative effect of the projective measurement on the average energy $E_{4}=\mathrm{Tr}\left[V\sum_{j}K_{j}U\rho_{1}U^{\dagger}K_{j}^{\dagger}V^{\dagger}H_{2}\right]$ as we explain further below. This is why the engine in \cite{Shanhe} could extract work even when $\nu_{2}=\nu_{1}$, i.e. even without changing the frequency of the spin system. Therefore, our results and theirs are not in contradiction with each other as one may think from the first sight. Note that for the asymmetric cycle we have taken both the forward and backward work averages for a meaningful study, since as was shown in \cite{BijayKumar} for one to obtain meaningful bounds on fluctuations one needs to take both the forward as well the backward into account.

\begin{figure}[hbtp]
\centering
\includegraphics[scale=0.9]{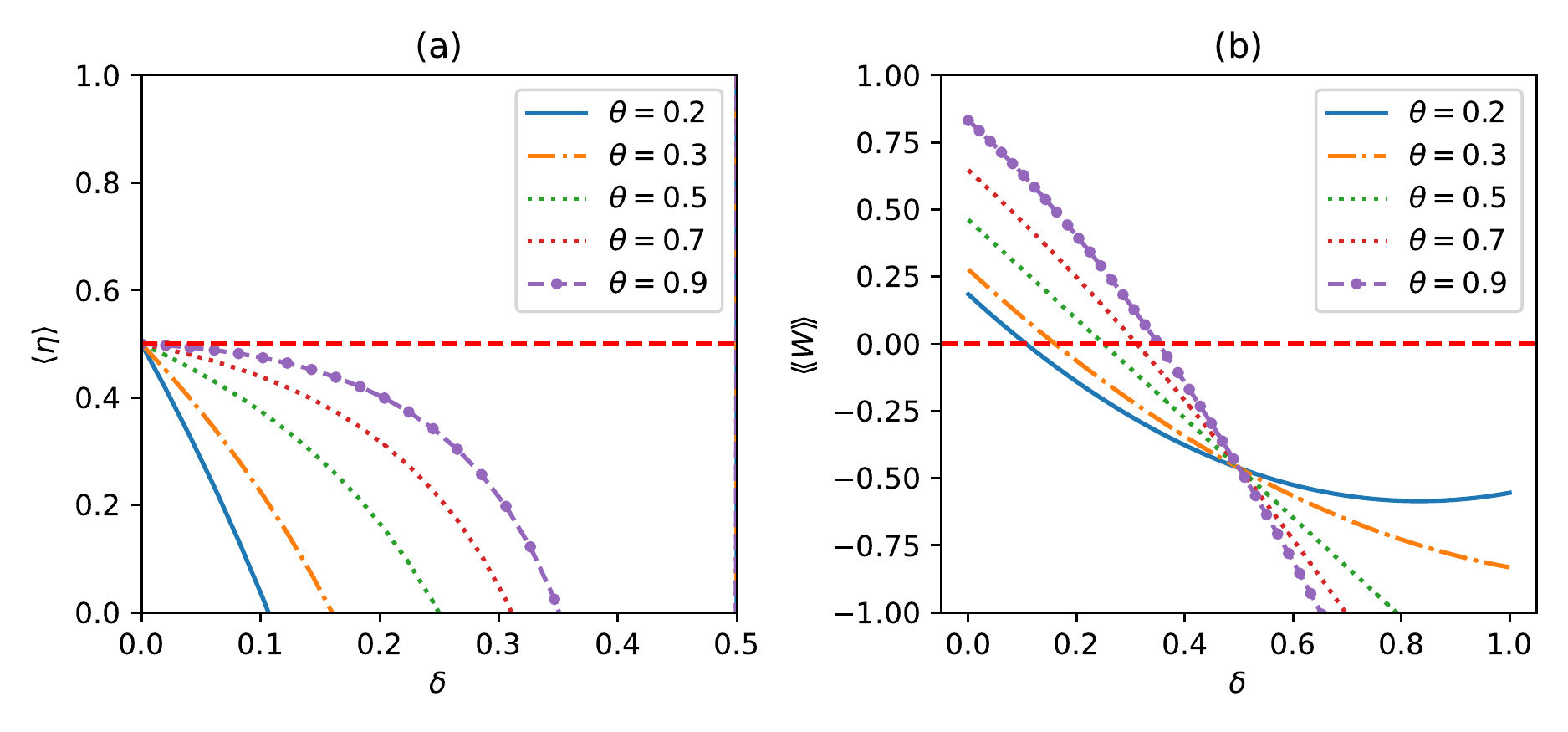}
\caption{(a) Plot of efficiency $\langle\eta\rangle$ as a function of $\delta$ with $\zeta=\delta$, $\beta=0.5$, $\nu_{1}=1$ and $\nu_{2}=2$. The horizontal dashed red line is the Otto efficiency which is equal to 0.5. We plot $\langle\eta\rangle$ only for $\delta\in[0,0.5]$, since beyond this interval $\llangle Q_{M}\rrangle$ becomes negative. One should note that differently from Ref. \cite{Shanhe} we see that the fact that $\theta$ can be higher than 1/2 enhances the efficiency. (b) Plot of extracted work $\llangle W\rrangle$ as a function of $\delta$ with $\zeta=\delta$ with the same value of $\beta$, $\nu_{1}$ and $\nu_{2}$ as in figure (a). For the considered parameters we see that considering an arbitrary unital channel could increase the value of $\delta$ above which work becomes negative. Even further, we see that the more we increase $\theta$ the more the extracted work gets enhanced. However, note that increasing $\delta$ diminishes the work extracted. This shows that non-adiabatic transitions are detrimental, but we show these transitions can be harnessed positively under an important circumstance.}
\label{WEff}
\end{figure}

For the system to work as a heat engine ($E$) when $\delta=\zeta$, the conditions $\llangle Q_{M}\rrangle>0$, $\llangle Q_{T}\rrangle<0$ and $\llangle W\rrangle>0$ has to be staified. For $\llangle Q_{M}\rrangle$ to be positive we have to ensure $\delta<1/2$. $\llangle Q_{T}\rrangle$ as we show below, is always negative or equal to zero. Thus for $\llangle W\rrangle>0$, we have to have, $|\llangle Q_{T}\rrangle|<\llangle Q_{M}\rrangle$. However, in terms of the parameters, for $\llangle W\rrangle>0$ we have the next condition on $\nu_{2}$,
\begin{equation}
\nu_{2}>\left(\frac{\theta+2\delta(1-2\theta)(1-\delta)}{\theta(1-2\delta)}\right) \nu_{1}(\geq\nu_{1} \ \mathrm{for} \ \delta<1/2).\label{Hey}
\end{equation}
This because $\theta+2\delta(1-2\theta)(1-\delta)-\theta(1-2\delta)(=2\delta(1-\delta-\theta+2\delta\theta))\geq0$. For an accelerator the conditions are: $\llangle Q_{M}\rrangle>0$, $\llangle Q_{T}\rrangle<0$ and $\llangle W\rrangle<0$. For a heater the conditions are: $\llangle Q_{M}\rrangle<0$, $\llangle Q_{T}\rrangle<0$ and $\llangle W\rrangle<0$. To satisfy these latter conditions we need to have $\delta>1/2$ with all other parameters are nonzero.

Now let's consider the case when $\zeta\neq\delta$. For the forward cycle to work as a heat engine, the condition $\delta <1/2$ and $\nu_{2}>\frac{\theta+(1-2\theta)(\delta+\zeta-2\delta\zeta)}{\theta(1-2\delta)}\nu_{1}$ are necessary. For the backward cycle, we have to ensure as well that $\zeta <1/2$ and $\nu_{2}>\frac{\theta+(1-2\theta)(\delta+\zeta-2\delta\zeta)}{\theta(1-2\zeta)}\nu_{1}$. Thus for the forward and backward to work as a heat engine we need to ensure that $\delta$ and $\zeta<1/2$ and $\nu_{2}>\mathbf{max}\{\frac{\theta+(1-2\theta)(\delta+\zeta-2\delta\zeta)}{\theta(1-2\delta)}\nu_{1},\frac{\theta+(1-2\theta)(\delta+\zeta-2\delta\zeta)}{\theta(1-2\zeta)}\nu_{1}\}$. Of course we may have other combinations such that the overall cycle is a heat engine. For example, the forward could be a heat engine and the backward an accelerator or vice versa and so on. In the latter case, when one imposes the positivity of work only on the sum i.e. $\left(\llangle W\rrangle_{F}+\llangle W\rrangle_{B}\right)$ to be $>0$ we have the next condition on $\nu_{2}$,
\begin{equation}
\nu_{2}\geq\left(\frac{\theta+(1-2\theta)(\delta+\zeta-2\delta\zeta)}{\theta(1-\delta-\zeta)}\right)\nu_{1}(>\nu_{1} \ \mathrm{for} \ \delta+\zeta <1).\label{KamKam}
\end{equation}
This is when we impose the positivity on only the sum of $\llangle W\rrangle_{F}$ and $\llangle W\rrangle_{B}$ and not on them individually.

Now let's look at the efficiency. For $\zeta=\delta$, the efficiency is given by,
\begin{equation}
\langle \eta\rangle=\frac{\llangle W\rrangle}{\llangle Q_{M}\rrangle}=1-\frac{\nu_{1}}{\nu_{2}}\frac{\theta+2\delta(1-2\theta)(1-\delta)}{(1-2\delta)\theta}.\label{symmm}
\end{equation}
Since $\theta+2\delta(1-2\theta)(1-\delta)-(1-2\delta)\theta=2\delta(1-(\delta+\theta-2\delta\theta))\geq0$, it follows that $-\frac{\nu_{1}}{\nu_{2}}\frac{\theta+2\delta(1-2\theta)(1-\delta)}{(1-2\delta)\theta}\leq-\frac{\nu_{1}}{\nu_{2}}$. Therefore the efficiency is limited by the one of the Otto, i.e. $1-\nu_{1}/\nu_{2}$. This latter bound is reached only when the adiabatic parameter $\delta$ is null. Analogously when $\delta\neq\zeta$ the efficiency expression \cite{BijayKumar} is given by,
\begin{equation}
\langle \eta\rangle=\frac{\llangle W\rrangle_{F}+\llangle W\rrangle_{B}}{\llangle Q_{M}\rrangle_{F}+\llangle Q_{M}\rrangle_{B}}=1-\frac{\nu_{1}}{\nu_{2}}\frac{(\theta+\zeta-2\theta\zeta+\delta(-1+ 2\theta)(-1+2\zeta))}{(1-\delta-\zeta)\theta}.\label{PropreEff}
\end{equation}
Further we have, $((\theta+\zeta-2\theta\zeta+\delta(-1+ 2\theta)(-1+2\zeta)))-((1-\delta-\zeta)\theta)=((\delta+\zeta-2\delta\zeta)(1-\theta)+2\delta\theta\zeta)\geq0$. From the latter inequality one can show that, 
\begin{equation}
\langle \eta\rangle=\frac{\llangle W\rrangle_{F}+\llangle W\rrangle_{B}}{\llangle Q_{M}\rrangle_{F}+\llangle Q_{M}\rrangle_{B}}\leq1-\frac{\nu_{1}}{\nu_{2}}.\label{PropE}
\end{equation}
Therefore for the symmetric and asymmetric Otto cycle, the efficiency is always less or equal to $1-\nu_{1}/\nu_{2}$. We should mention that for the asymmetric cycle, if one does not take into consideration the backward, the efficiency can be greater than that of the Otto, i.e. $1-\nu_{1}/\nu_{2}$. This means that not only one should treat the backward and the forward cycle on equal footing for meaningful bounds on RF as was stated in Ref. \cite{BijayKumar}, but also to restore the Otto efficiency being the upper bound of the efficiency as our result (Eq.(\ref{PropE})) shows.

In figure (\ref{WEff}) we plot the efficiency $\langle \eta\rangle$ (left figure) and work extracted $\llangle W\rrangle$ (right figure) as a function of the adiabatic parameter $\delta$ for the symmetric Otto cycle $\zeta=\delta$ and for different values of $\theta$. It is seen that work and efficiency get enhanced as we increase $\theta$. Further, it is found that efficiency becomes less perturbed by $\delta$, when we increase $\theta$.
\subsection{Monitored (our engine) and unmonitored engine (Ref. \cite{Shanhe})}\label{NegativeQc}
Let's now comment on why our heat engine needs the condition (\ref{Hey}) to output work, in contrast to what has been shown in Ref. \cite{Shanhe}. As we said before, in \cite{Shanhe}, the authors did not consider monitoring the working medium between the strokes. However, in our case we consider it. Thus we have that the average work and heats are not equal. Note that if one considers a completely unital map or a in general a completely thermalizing map then the two approaches give the same averages and fluctuations of average energies. Therefore monitoring or not monitoring the state of the system will not change the story. However, since the unital map is arbitrary, which may not erase the coherence generated by the unitary $U$, the monitored and unmonitored heat engines would be different. Our analysis is in agreement with what has been shown in Ref. \cite{JuzarSong}, that projective measurement may be detrimental to efficiency, work and its reliability and power. In Ref. \cite{PrasanaI}, the authors have considered a two-stroke Otto heat engine, where the hot bath being replaced by a non-selective quantum measurement. They have shown that the invariant state of the cycle is dependent on whether the cycle is being monitored or not, and consequently the work average and its fluctuations are different. More precisely, it was found that depending on the parameters, the work for the unmonitored cycle can be either greater or less than that of the monitored cycle. The same thing was found for work reliability. From our study we conclude that monitoring can have a negative influence on the performance of the heat engine, and this is seen above as the engine cannot output work under the condition $\nu_{1}=\nu_{2}$, either in the symmetric or the asymmetric Otto cycle. 

Let's now show a comparison between the monitored and unmonitored Otto cycle cases in terms of average energy at the point $\mathbf{D}$ in figure \ref{b0.55}(b). One can easily show that $E_{1}$, $E_{2}$ and $E_{3}$, the average energies at the points $\mathbf{A}$, $\mathbf{B}$ and $\mathbf{C}$ in figure \ref{b0.55}(b), respectively, will not be affected by using projective measurement to assess the statistics of thermodynamic quantities. Actually the state $\rho_{3}$ of the unmonitored Otto cycle can be written as follows,
\begin{equation}
\begin{split}
\rho_{3}&=\sum_{i=0}^{3}p_{i}\sigma_{i}\rho_{2}\sigma_{i}
\\ &
=\left(|+\rangle_{22}\langle+|+|-\rangle_{22}\langle-|\right)\left(\sum_{i=0}^{3}p_{i}\sigma_{i}\rho_{2}\sigma_{i}\right)\left(|+\rangle_{22}\langle+|+|-\rangle_{22}\langle-|\right)
\\ &
=\rho_{3}^{deph}+\sum_{i=0}^{3}p_{i}\sigma_{i}({}_{2}\langle+|\rho_{2}|-\rangle_{2} |+\rangle{}_{2}{}_{2}\langle-|+{}_{2}\langle-|\rho_{2}|+\rangle{}_{2} |-\rangle_{2}{}_{2}\langle+|)\sigma_{i}\label{Wow}.
\end{split}
\end{equation}
Note that we have employed Eq.(\ref{PauliU}). $\rho_{2}=U\rho_{1}U^{\dagger}$ and $\rho_{3}^{deph}(={}_{2}\langle+|\rho_{3}|+\rangle_{2}|+\rangle_{22}\langle+|+{}_{2}\langle-|\rho_{3}|-\rangle_{2}|-\rangle_{22}\langle-|)$ is the dephased state $\rho_{3}$. Let's now forget about the second term. Similarly to Eq.(\ref{Wow}) the state $\rho_{4}$ of the unmonitored Otto cycle is,
\begin{widetext}
\begin{equation}
\rho_{4}=\rho_{4}^{deph}+V\sum_{i}\sigma_{i}p_{i}(\langle+_{2}|\rho_{2}|-_{2}\rangle |+\rangle_{22}\langle-|+\langle-_{2}|\rho_{2}|+_{2}\rangle |-\rangle_{22}\langle+|)\sigma_{i}V^{+}\label{444}.
\end{equation}
\end{widetext}

One can show that the second term in Eq.(\ref{Wow}) will not contribute to the average energy $E_{3}$. This can be explained by the fact that the Kraus operators of an arbitrary unital channel, which are Pauli operators, do not couple the populations and coherences, thus the coherence created in the first stroke by $U$ can only affect $E_{4}$ and not $E_{3}$, thus the heat provided by the unital channel $\llangle Q_{M}\rrangle$ is coherence independent. What this means is that the difference between the average energies of the monitored and unmonitored will only be in $E_{4}$. Furthermore, note that the second term in Eq.(\ref{444}) will not contribute to the average energy  $E_{4}$ when $\rho_{2}$ is diagonal in the basis of the Hamiltonian $H_{2}$, since ${}_{2}\langle+|\rho_{2}|-\rangle_{2}={}_{2}\langle-|\rho_{2}|+\rangle_{2}=0$. In this case, there is no difference between the monitored and unmonitored heat engines from an energetic point of view.  Thus when the arbitrary unital channel is not completely unital, the unitary $V$ may couple the populations and coherences such that the latter would contribute to the average work $\llangle W\rrangle$ and heat exchanged with the bath $\llangle Q_{T}\rrangle$.

In appendix \ref{LZM}, for the Landau-Zener model, we give a detailed comparison between the work extracted and efficiency for the monitored and unmonitored single qubit.
\subsection{Negative temperature: Can the system work as a refrigerator in this case?}
Now let's first prove that $\llangle Q_{T}\rrangle$ is always negative. We have, $0\leq\delta+\zeta-2\delta\zeta=\delta(1-\zeta)+\zeta(1-\delta)\leq1$, since $0\leq\left( (1-\delta) \ \rm and, \ (1-\zeta)\right) \leq 1$. The term, $(\theta+\zeta-2\theta\zeta+\delta(1-2\zeta)(1-2\theta))$ can be rewritten as follows: $(\theta+\zeta-2\theta\zeta+\delta(1-2\zeta)(1-2\theta))=\theta+(1-2\theta)(\delta+\zeta-2\delta\zeta)$. For $0\leq \theta\leq1/2$ this term is clearly positive or null, since $\delta+\zeta-2\delta\zeta$ and $1-2\theta$ are greater than or equal to zero. To show this for $1/2<\theta\leq1$ we have, 
$(\theta+\zeta-2\theta\zeta+\delta(1-2\zeta)(1-2\theta))=\theta-(2\theta-1)(\delta+\zeta-2\delta\zeta)$. $(2\theta-1)(\delta+\zeta-2\delta\zeta)$ is $\geq0$. The biggest value that $(2\theta-1)(\delta+\zeta-2\delta\zeta)$ can take, independtly of $\theta$, is when $\delta+\zeta-2\delta\zeta=1$, in this case we have, $\theta-\max_{0\leq\delta,\zeta\leq1}(\zeta-2\theta\zeta+\delta(1-2\zeta)(1-2\theta))=\theta-(2\theta-1)=1-\theta$, which is clearly always $\geq0$. Thus, $\llangle Q_{T}\rrangle$ is always $\leq0$ as in Ref. \cite{Shanhe}. Therefore, heat will only flow out from the system when it will be in interaction with the heat bath, thus this bath will play the role of a heat-cold bath and the unital channel the role of hot heat bath. However, in contrast to them, our proof is valid for arbitrarily monitored Otto heat engines based on unital qubit channels, not just the specific measurement unital channel considered there. This is another main result of our work.

The notion of inverse temperature bath was introduced in Refs. \cite{Ramsey,Purcell}. Above, and in the previous work in Ref. \cite{Shanhe} it was shown that $\llangle Q_{T}\rrangle$ is always $\leq0$ but this is true only under the assumption that the temperature of the heat reservoir is positive. However, if it is negative then in this case the $\llangle Q_{T}\rrangle$ can be positive, since the sign of $\tanh(\beta\nu_{1})$ gets flipped. When $\llangle Q_{T}\rrangle$ is negative this means that the average energy of the system at the point $\mathbf{D}$ is bigger than the one when the system gets equilibrated with its environment i.e. point $\mathbf{A}$, see Fig.\ref{b0.55}(b). However, if the temperature is negative then in this case $E_{4}$ will be smaller than $E_{1}$ thus heat will flow from the environment into the system. However, in this case, the system is not a refrigerator as a heat bath with a negative temperature is hotter than any heat bath with a positive temperature, thus the bath will act as a hot heat bath \cite{Marco}. To summarise the operation modes of the system are:
\begin{equation*}
\mathrm{for} \ \beta>0:
\begin{split}
\\ &
\mathrm{E}: \llangle Q_{T}\rrangle< 0, \hspace{1cm}\llangle Q_{M}\rrangle> 0, \hspace{1cm}\ \rm and \hspace{1cm} \ \llangle W\rrangle> 0,
\\ &
\mathrm{A}: \llangle Q_{T}\rrangle> 0, \hspace{1cm}\llangle Q_{M}\rrangle> 0, \hspace{1cm}\ \rm and \hspace{1cm} \ \llangle W\rrangle< 0,
\\ &
\mathrm{H}: \llangle Q_{T}\rrangle< 0, \hspace{1cm}\llangle Q_{M}\rrangle< 0, \hspace{1cm}\ \rm and \hspace{1cm} \ \llangle W\rrangle< 0.
\end{split}\label{23}
\end{equation*}
and
\begin{equation*}
\mathrm{for} \ \beta<0:
\begin{split}
\\ &
\mathrm{A}: \llangle Q_{T}\rrangle> 0, \hspace{1cm}\llangle Q_{M}\rrangle< 0, \hspace{1cm}\ \rm and \hspace{1cm} \ \llangle W\rrangle< 0,
\\ &
\mathrm{E}: \llangle Q_{T}\rrangle> 0, \hspace{1cm}\llangle Q_{M}\rrangle< 0, \hspace{1cm}\ \rm and \hspace{1cm} \ \llangle W\rrangle> 0,
\\ &
\mathrm{E'}: \llangle Q_{T}\rrangle> 0, \hspace{1cm}\llangle Q_{M}\rrangle> 0, \hspace{1cm}\ \rm and \hspace{1cm} \ \llangle W\rrangle> 0.
\end{split}
\end{equation*}

For $\beta$ negative, we only have to reverse the sign of work and heats given in the above equation. Therefore when $\beta\rightarrow-\beta$ a heat engine becomes an accelerator, an accelerator becomes a heat engine and a heater becomes a heat engine $E'$, but the latter comes with unit efficiency since the system will absorb heat from both baths and transform it into work, thus $\langle\eta\rangle=\llangle W\rrangle/\left(\llangle Q_{T}\rrangle+\llangle Q_{M}\rrangle\right)=1$. This latter case is possible because of two reasons: the first one is the non-adiabaticity and the second is the negativity of the temperature of the bath. Since a heater is not possible in the adiabatic case, i.e. a nonequilibrium thermal machine. However, note that if the cost of preparing baths at negative temperatures-nonequilibrium baths-is taken into account, then there will be no inconsistency with the laws of thermodynamics, see Refs. \cite{Struchtrup,Warren} and the efficiency would be less than 1. The fact that the efficiency of a two-level Otto cycle with two heat baths, one of them with a positive temperature and the other with a negative temperature-being enhanced in the non-adiabatic regime, which seems counterintuitive, was also noted in Ref. \cite{Rogerio}.\par

Thus refrigeration is not allowed even when the temperature is negative. Furthermore, note that the heater is not possible as well, when $\beta$ is negative, since a heater demands the next conditions: $\llangle Q_{T}\rrangle< 0, \llangle Q_{M}\rrangle< 0 \ \rm and \ \llangle W\rrangle< 0$, which is not possible, since when $\beta<0$, $\llangle Q_{T}\rrangle$ will always be $\geq0$. We have demonstrated in equation (\ref{Manipula}) that a positive work condition is not possible when $\nu_{2}=\nu_{1}$ and when the inverse temperature is positive. However, as we see in this section, when the inverse temperature is negative we may have, $\llangle W\rrangle>0$ either for symmetric or asymmetric Otto cycle.\par

Until now we only considered the first cumulants of work and heat. Let's now take a step further towards higher cumulants such as variance, skewness and kurtosis which are defined as measures of fluctuations, asymmetry and tailedness and peakedness of a probability distribution, respectively.
\subsection{A lower and upper bound on the ratio of the fluctuations of $W$ and $Q_{M}$}
In this section, we prove rigorously that the ratio of the fluctuations of $W$ and $Q_{M}$ is lower bounded by the square of the efficiency and upper bounded by 1. We show when these bounds are respected either for the symmetric or the asymmetric Otto cycle. Furthermore, for the symmetric Otto cycle, we prove that the square of the Otto efficiency gives a tighter lower bound on the ratio of fluctuations than the one derived from the square of the efficiency of the engine.
\subsubsection{Symmetric Otto cycle}
The authors of Ref. \cite{BijayKumar} have shown that for a symmetric and asymmetric Otto cycle based on two heat baths, the ratio $\llangle W^{2}\rrangle/\llangle Q_{M}^{2}\rrangle$ is lower and upper bounded when the system is working as a heat engine. Since our Otto cycle is different from theirs we want to know if this is still true here as well. Let's consider $\llangle W^{2}\rrangle/\llangle W\rrangle^{2}-\llangle Q_{M}^{2}\rrangle/\llangle Q_{M}\rrangle^{2}$. We now only consider the case $\zeta=\delta$. After messy but straightforward steps of calculations which are not important to be included here, we obtain,
\begin{widetext}
\begin{equation}
\begin{split}
\frac{\llangle W^{2}\rrangle}{\llangle W\rrangle^{2}}-\frac{\llangle Q_{M}^{2}\rrangle}{\llangle Q_{M}\rrangle^{2}}&=\frac{2(1-\delta)\delta\nu_{1}(-(\theta+2\delta(1-\delta)(1-2\theta))\nu_{1}+2\theta(1-2\delta)\nu_{2})\coth^{2}(\beta\nu_{1})}{\theta(1-2\delta)^{2}((\theta+2\delta(1-\delta)(1-2\theta))\nu_{1}-\theta(1-2\delta)\nu_{2})^{2}}
\\ &
=\frac{(1-\delta)\delta\nu_{1}(\langle\langle W\rangle\rangle+2(1-2\delta)\nu_{2}\tanh(\beta\nu_{1}))\coth^{3}(\beta\nu_{1})}{\theta(1-2\delta)^{2}((\theta+2\delta(1-\delta)(1-2\theta))\nu_{1}-\theta(1-2\delta)\nu_{2})^{2}}.
\end{split}\label{wqh}
\end{equation}
\end{widetext}
This equation can be either positive, equal to zero or negative. The denominator is always $\geq0$. For equation (\ref{wqh}) to be $\geq0$ we have to ensure that,
\begin{equation}
\nu_{2}\geq\frac{(\theta+2\delta(1-\delta)(1-2\theta))}{2(1-2\delta)\theta}\nu_{1}.\label{condnu}
\end{equation}
When this latter condition is met we have,
\begin{equation}
\frac{\llangle W^{2}\rrangle}{\llangle W\rrangle^{2}}\geq\frac{\llangle Q_{M}^{2}\rrangle}{\llangle Q_{M}\rrangle^{2}}.\label{mqcqhw}
\end{equation}
Furthermore, note that the lower bound $\frac{(\theta+2\delta(1-\delta)(1-2\theta))}{2(1-2\delta)\theta}\nu_{1}$ on $\nu_{2}$ is half of the necessary condition needed for the system to work as a heat engine. Thus this condition does not ensure that the system is working as a heat engine. This shows that $\llangle W^{2}\rrangle/\llangle W\rrangle^{2}\geq\llangle Q_{M}^{2}\rrangle/\llangle Q_{M}\rrangle^{2}$ can still be valid even when the system is not working as a heat engine, such as an accelerator as we found numerically, to be shown later on. However, when the system is working as a heater, then we have,
\begin{equation}
\frac{\llangle W^{2}\rrangle}{\llangle W\rrangle^{2}}<\frac{\llangle Q_{M}^{2}\rrangle}{\llangle Q_{M}\rrangle^{2}}.
\end{equation}
This is because for the system to work as a heater we need $\delta>1/2$, and under this constraint, one can show that $2(-(\theta+2\delta(1-\delta)(1-2\theta))\nu_{1}+2\theta(1-2\delta)\nu_{2})\left(=\left(\llangle W\rrangle+2(1-2\delta)\nu_{2}\tanh(\beta\nu_{1})\right)/\tanh(\beta\nu_{1})\right)<0$, since $\llangle W\rrangle<0$. Therefore, we conclude that $\llangle W^{2}\rrangle/\llangle W\rrangle^{2}\geq\llangle Q_{M}^{2}\rrangle/\llangle Q_{M}\rrangle^{2}$ is always respected (violated) when the system as a heat engine (heater), but it can or cannot be violated when it is an accelerator. The violation depends on the parameters. Straightforwardly from Eq.(\ref{mqcqhw}) it follows that,
\begin{equation}
\frac{\llangle W^{2}\rrangle}{\llangle Q_{M}^{2}\rrangle}\geq\frac{\llangle W\rrangle^{2}}{\llangle Q_{M}\rrangle^{2}}=\langle\eta\rangle^{2}.
\end{equation}
This equation tells us that when equation (\ref{condnu}) is satisfied, the ratio of the fluctuations of $W$ and $Q_{M}$ is always greater or equal to the square of efficiency. In the heat engine region, the lower bound is reached in the adiabatic limit. Now let's see if the ratio of the relative fluctuations of $W$ and $Q_{M}$ has any upper bound as in Ref. \cite{BijayKumar}. In the latter reference, it was shown that for a symmetric and asymmetric QOHE with two completely thermalizing baths that $\llangle W^{2}\rrangle/\llangle Q_{M}^{2}\rrangle<1$. Let's now investigate under which condition $1-\llangle W^{2}\rrangle/\llangle Q_{M}^{2}\rrangle$ is $\geq0$. To do this, this inequality can be modified into checking if $\llangle Q_{M}^{2}\rrangle-\llangle W^{2}\rrangle\geq0$. We have
\begin{widetext}
\begin{equation}
\begin{split}
\llangle Q_{M}^{2}\rrangle-\llangle W^{2}\rrangle&=
4\nu_{1}(2\delta^{2}(-1+2\theta)\nu_{1}+\theta(\nu_{1} -2\nu_{2})+2\delta(\nu_{1}-2\theta\nu_{1}+2\theta\nu_{2}))(-1+(\theta+2(-1+\delta)\delta(-1+2\theta)) \tanh^{2}(\beta\nu_{1}))
\\ &
=4\nu_{1}(\llangle W\rrangle+(\theta+2(-1 +\delta)\delta(-1+2\theta))\nu_{1}\tanh(\beta\nu_{1}))(1-(\theta+2(-1+\delta)\delta(-1+2\theta)) \tanh^{2}(\beta\nu_{1}))\coth(\beta\nu_{1}).
\end{split}\label{34R}
\end{equation}
\end{widetext}
The term $(-1+(\theta+2(-1+\delta)\delta(-1+2\theta)) \tanh^{2}{(\beta\nu_{1}}))$ is $\leq0$, since $(\theta+2(-1+\delta)\delta(-1+2\theta)) \tanh^{2}{(\beta\nu_{1}})$ is $\leq$ 1, thus for $\llangle Q_{M}^{2}\rrangle-\llangle W^{2}\rrangle$ to be $\geq0$ one has to ensure that, $(2\delta^{2}(-1+2\theta)\nu_{1}+\theta(\nu_{1} -2\nu_{2})+2\delta(\nu_{1}-2\theta\nu_{1}+2\theta\nu_{2}))$ is $\leq0$ which is equivalent into, $\nu_{2}\geq\frac{(\theta+2\delta(1-\delta)(1-2\theta))}{2(1-2\delta)\theta}\nu_{1}$, which is nothing, but the condition given in Eq. (\ref{condnu}). Therefore, under the condition (\ref{condnu}), we have
\begin{equation}
\langle\eta\rangle^{2}=\frac{\llangle W\rrangle^{2}}{\llangle Q_{M}\rrangle^{2}}\leq\frac{\llangle W^{2}\rrangle}{\llangle Q_{M}^{2}\rrangle}\leq1.\label{UppLo}
\end{equation}
This equation is the first main result of this section. It shows that the ratio of the fluctuations of $W$ and $Q_{M}$ is lower and upper-bounded. Therefore, similarly to the previous analysis on the lower bound, we have $\llangle W^{2}\rrangle/\llangle Q_{M}^{2}\rrangle<1$ is always respected when the system as a heat engine, since $\llangle Q_{M}^{2}\rrangle-\llangle W^{2}\rrangle>0$ from the heat engine conditions, but it can or cannot be violated when it is an accelerator. The violation depends on the parameters, and to be precise whenever equation (\ref{condnu}) is violated then $\llangle W^{2}\rangle/\langle Q_{M}^{2}\rangle\geq1$. Under the conditions of a heat engine ($\delta<1/2$ and $\nu_{2}>((\theta+2\delta(1-2\theta)(1-\delta))/(\theta(1-2\delta)))\nu_{1}$), equation (\ref{34R}) is always positive, which means that the upper bound in equation (\ref{UppLo}) is not achieved, it is reached only when the system is working as an accelerator depending on the values of the parameters.

Let's now give an interpretation to equation (\ref{UppLo}). In the adiabatic regime, the ratio of the fluctuations of $W$ and $Q_{M}$ saturates to its lower bound which is the square of the Otto efficiency. However, when $0<\delta<1/2$ then the more we increase $\delta$, but not exceeding 1/2, the more the ratio of fluctuations increases until the upper bound 1 is violated in the accelerator regime depending on the parameters. In the heat engine region, it means that the fluctuations of $W$ are always less than those of $Q_{M}$, but the lower bound on the ratio of the fluctuations cannot be arbitrary. However, when the system is working as an accelerator then the fluctuations of $W$ can be either less, greater or equal to those of $Q_{M}$.

Note that equation can be interpreted also as a witness of the operational regime. More precisley when the ratio of fluctuations does not respect Eq. (\ref{UppLo}), then this indicates that the system now starts working as a useless machine, such as a heater or an accelerator.

Actually one can show that the square of the Otto efficiency provide a tighter lower bound on the ratio of fluctuations of $W$ and $Q_{M}$ than the square of the efficiency of the engine. After messy but strauthforward algebra one can find that,
\begin{equation}
\begin{split}
\llangle W^{2}\rrangle\nu_{2}^{2}-\llangle Q_{M}^{2}\rrangle\left(\nu_{2}-\nu_{1}\right)^{2}
& =8\delta\nu_{1}\nu_{2}^{2}\left((1-\delta)\nu_{1}+2\delta\theta\nu_{1}+2\theta(\nu_{2}-\nu_{1})\right).
\\ &
+16\delta\nu_{1}\nu_{2}^{2}\left(1-\delta-\theta+2\delta\theta\right)\left((1-2\delta)\theta\nu_{2}-(1-\delta)(\delta+\theta-2\delta\theta)\nu_{1}\right)\tanh^{2}\left(\beta\nu_{1}\right).
\end{split}\label{Tight}
\end{equation}
For $\nu_{2}\geq\nu_{1}$, one can easily see that $8\delta\nu_{1}\nu_{2}^{2}\left((1-\delta)\nu_{1}+2\delta\theta\nu_{1}+2\theta(\nu_{2}-\nu_{1})\right)\geq0$. On the other hand for $(1-2\delta)\theta\nu_{2}-(1-\delta)(\delta+\theta-2\delta\theta)\nu_{1}$ to be greater or equal to zero, this translates into 
\begin{equation}
(1-2\delta)\theta\nu_{2}\geq(1-\delta)(\delta+\theta-2\delta\theta)\nu_{1}.\label{HOPM}
\end{equation}
For a positive work we already have proved that we need $(1-2\delta)\theta\nu_{2}\geq(\theta+2\delta(1-2\theta)(1-\delta))\nu_{1}$, see Eq. (\ref{Hey}). We have,
\begin{equation}
(\theta+2\delta(1-2\theta)(1-\delta))\nu_{1}-(1-\delta)(\delta+\theta-2\delta\theta)\nu_{1}=\delta(1 - \delta-\theta+2\delta\theta)\nu_{1}\geq0.
\end{equation}
Since the necessary condition for positive work is always greater or equal to the one necessary (Eq. \ref{HOPM}) for the term in the second line in Eq. (\ref{Tight}) to be $\geq0$, this means that even when the positive work condition is not satisfied still the second tem in Eq. (\ref{Tight}) can be $\geq0$. Therefore under the condition Eq. (\ref{HOPM}) we have,
\begin{equation}
\frac{\llangle W^{2}\rrangle}{\llangle Q_{M}^{2}\rrangle}\geq\left(1-\nu_{1}/\nu_{2}\right)^{2}.
\end{equation}
We already have proved that efficiency is limited by the Otto efficiency, see Eq. (\ref{KamKam}). Therefore we have,
\begin{equation}
\langle\eta\rangle^{2}=\frac{\llangle W\rrangle^{2}}{\llangle Q_{M}\rrangle^{2}}\leq\left(1-\nu_{1}/\nu_{2}\right)^{2}\leq\frac{\llangle W^{2}\rrangle}{\llangle Q_{M}^{2}\rrangle}<1.\label{UppLoM8}
\end{equation}
This is another cnetral result of this section. Unfortunately, below, we could not generalize this result to the asymmetric Otto cyclc. From Eq. (\ref{UppLoM8}) we have,
\begin{equation}
\langle\eta\rangle\leq\left(1-\nu_{1}/\nu_{2}\right)\leq\sqrt{\frac{\llangle W^{2}\rrangle}{\llangle Q_{M}^{2}\rrangle}}<1.\label{UppLoM8}
\end{equation}
This means that the square root of the ratio of fluctuations can also provide an upper bound on the efficiency, however, it is not tighter than the Otto efficiency. In Ref. \cite{Gerry2,Gerry1}, the authors have found for autonomous continuous thermal machines, that the square root of the ratio of the second cumulants provide a tighter bound on the efficiency than the Carnot efficiency. A result that we see does not hold here for driven discrete quantum heat engines.

\subsubsection{Asymmetric Otto cycle}
Using the Mathematica software, one can show that for the case when $\zeta\neq\delta$ we have
\begin{equation}
\begin{split}
&\frac{2(\llangle W^{2}\rrangle_{F}+\llangle W^{2}\rrangle_{B})}{(\llangle W\rrangle_{F}+\llangle W\rrangle_{B})^{2}}-\frac{2(\llangle Q_{M}^{2}\rrangle_{F}+\llangle Q_{M}^{2}\rrangle_{B})}{(\llangle Q_{M}\rrangle_{F}+\llangle Q_{M}\rrangle_{B})^{2}}
\\&
=\frac{\nu_{1}(-(\theta+\zeta-2\theta\zeta+\delta(-1+2\theta) (-1+2\zeta))\nu_{1}+2\theta(1-\delta-\zeta)\nu_{2})}{\theta(-1+\delta+\zeta)^2} 
\\ &
\times \frac{-(\theta(\delta-\zeta)^{2}(\theta+\zeta-2\theta\zeta+\delta(-1+2\theta)(-1+2\zeta))+(\delta^{2}\theta+\delta(-1+2\zeta-2\theta\zeta)+\zeta(-1+\theta\zeta)) \coth^{2}(\beta\nu_{1}))}{((\theta+\zeta-2\theta\zeta+\delta(-1+2\theta)(-1+2 \zeta))\nu_{1}+\theta(-1+\delta+\zeta)\nu_{2})^{2}}.\label{Asssm}
\end{split}
\end{equation}
Note that we used the properly symmetrized expression of the relative fluctuations introduced in \cite{BijayKumar}. One can show that $-(\theta(\delta-\zeta)^{2}(\theta+\zeta-2\theta\zeta+\delta(-1+2\theta)(-1+2 \zeta))+(\delta^{2}\theta+\delta(-1+2\zeta-2\theta\zeta)+\zeta(-1+\theta\zeta))\coth^{2}(\beta\nu_{1}))$ is $\geq0$. Let's now prove this latter inequality.

We have, $\theta(\delta-\zeta)^{2}(\theta+\zeta-2\theta\zeta+\delta(-1+2\theta)(-1+2 \zeta))\geq0$ and $-(\delta^{2}\theta+\delta(-1+2\zeta-2\theta\zeta)+\zeta(-1+\theta\zeta))\coth^{2}(\beta\nu_{1})=-(\theta(\delta-\zeta)^{2}-(\delta+\zeta-2\delta\zeta))\coth^{2}(\beta\nu_{1})$. One can show that $\theta(\delta-\zeta)^{2}-(\delta+\zeta-2\delta\zeta)\leq0$. This latter inequality follows from a simple maximazation oevr $\theta$, since both $(\delta-\zeta)^{2}$ and $(\delta+\zeta-2\delta\zeta)$ are $\geq0$. From this one can conclude that $-(\delta^{2}\theta+\delta(-1+2\zeta-2\theta\zeta)+\zeta(-1+\theta\zeta))\coth^{2}(\beta\nu_{1})=-(\theta(\delta-\zeta)^{2}-(\delta+\zeta-2\delta\zeta))\coth^{2}(\beta\nu_{1})\geq0$. The terms $\theta(-1+\delta+\zeta)^2$ and $((\theta+\zeta-2\theta\zeta+\delta(-1+2\theta)(-1+2 \zeta))\nu_{1}+\theta(-1+\delta+\zeta)\nu_{2})^{2}$ are $\geq0$.

Now let's define the parameter $A$ to be given as follows,
\begin{equation}
A=-(\delta^{2}\theta+\delta(-1+2\zeta-2\theta\zeta)+\zeta(-1+\theta\zeta))\coth^{2}(\beta\nu_{1})-\theta(\delta-\zeta)^{2}(\theta+\zeta-2\theta\zeta+\delta(-1+2\theta)(-1+2 \zeta)).\label{AAAA}
\end{equation}
In such a case our task for proving that $A\geq0$, reduce to checking that,
\begin{equation}
-(\delta^{2}\theta+\delta(-1+2\zeta-2\theta\zeta)+\zeta(-1+\theta\zeta))\coth^{2}(\beta\nu_{1})\geq\theta(\delta-\zeta)^{2}(\theta+\zeta-2\theta\zeta+\delta(-1+2\theta)(-1+2 \zeta)).\label{Wallah}
\end{equation}
Actually to prove the latter inequality it is sufficient to prove it for the lowest value of $\coth^{2}(\beta\nu_{1})$ which is one. In this case (i.e. $\coth^{2}(\beta\nu_{1})=1$) Eq. (\ref{AAAA}) reduces into,
\begin{equation}
A=-(\theta(\delta-\zeta)^{2}-(\delta+\zeta-2\delta\zeta))-\theta(\delta-\zeta)^{2}(\theta+\zeta-2\theta\zeta+\delta(-1+2\theta)(-1+2 \zeta)).
\end{equation}
From this, after a simple algebra, we have
\begin{equation}
A=(\delta+\zeta-2\delta\zeta)-\theta(\delta-\zeta)^{2}(1+\theta+(1-2\theta)(\delta+\zeta-2\delta\zeta)).\label{AZ+}
\end{equation}
One can show that $(\delta+\zeta-2\delta\zeta)$ and $\theta(\delta-\zeta)^{2}(1+\theta+(1-2\theta)(\delta+\zeta-2\delta\zeta)$ are always positive or equal to zero.  Further the highest value of $\theta(\delta-\zeta)^{2}(1+\theta+(1-2\theta)(\delta+\zeta-2\delta\zeta)$ for arbitrary $\delta$ and $\zeta$ is when $\theta=1$. This letter statement can be checked easily by using the first and the second derivative with respect to $\theta$. In this case Eq. (\ref{AZ+}) becomes
\begin{equation}
\begin{split}
A & =\delta(1-\delta)+\zeta(1-\zeta)-(\delta-\zeta)^{2}(1-\delta-\zeta+2\delta\zeta). 
\end{split}
\end{equation}
After simple steps of calculations, one can find that,
\begin{equation}
\begin{split}
A =(1-2\delta)(\zeta^{3}+\delta(1-\zeta^{2}))+(1-2\zeta)(\delta^{3}+\zeta(1-\delta^{2}))+2\delta\zeta. 
\end{split}
\end{equation}
This equation is more simple to be analyzed by looking at it. Taking into account the fact that for a heat engine to be possible in the forward as well as in the backward cycle, $\delta$ and $\zeta$ need to be less than 1/2. Imposing this condition is only necessary but not sufficient as we have seen before. Since this leaves the possibility of an accelerator as well. From this constraint, it is straightforward to show that $A\geq0$. Therefore for $2(\llangle W^{2}\rrangle_{F}+\llangle W^{2}\rrangle_{B})/(\llangle W\rrangle_{F}+\llangle W\rrangle_{B})^{2}\geq 2(\llangle Q_{M}^{2}\rrangle_{F}+\llangle Q_{M}^{2}\rrangle_{B})/(\llangle Q_{M}\rrangle_{F}+\llangle Q_{M}\rrangle_{B})^{2}$ one need to ensure that,
\begin{equation}
\nu_{2}\geq\frac{\theta+(1-2\theta)(\delta+\zeta-2\delta\zeta)}{2(1-\delta-\zeta)\theta}\nu_{1}.\label{ConNu2}
\end{equation}
Actually this lower bound $\frac{\theta+(1-2\theta)(\delta+\zeta-2\delta\zeta)}{2(1-\delta-\zeta)\theta}\nu_{1}$ on $\nu_{2}$ is half of the necessary condition for $(\llangle W\rrangle_{F}+\llangle W\rrangle_{B})\geq0$, see equation (\ref{KamKam}). Analogously to the symmetric Otto cycle, let's now look at $(\llangle Q_{M}^{2}\rrangle_{F}+\llangle Q_{M}^{2}\rrangle_{B})-(\llangle W^{2}\rrangle_{F}+\llangle W^{2}\rrangle_{B})$. The explicit expression of the latter difference is given as,
\begin{equation}
\begin{split}
&(\llangle Q_{M}^{2}\rrangle_{F}+\llangle Q_{M}^{2}\rrangle_{B})-(\llangle W^{2}\rrangle_{F}+\llangle W^{2}\rrangle_{B})=
\\ &
=-8\nu_{1}((\theta+\zeta-2\theta\zeta+\delta(-1+2\theta)(-1+2\zeta))\nu_{1}+2\theta(-1+\delta+\zeta)\nu_{2})
(1-(\theta+\zeta-2\theta\zeta+\delta(-1+2\theta) (-1+2\zeta))\tanh^2(\beta\nu_{1}))
\\ &
=4\nu_{1}(\llangle W\rrangle_{F}+\llangle W\rrangle_{B}+2(\theta+(1-2\theta)(\delta+\zeta-2\delta\zeta))\nu_{1}\tanh(\beta\nu_{1}))
\\ &
\times (1-(\theta+(1-2\theta)(\delta+\zeta-2\delta\zeta))\tanh^2(\beta\nu_{1}))\coth(\beta\nu_{1}).
\end{split}
\end{equation}
Since $(1-(\theta+\zeta-2\theta\zeta+\delta(-1+2\theta) (-1+2\zeta))\tanh^2(\beta\nu_{1}))\geq0$, one can show that the condition (\ref{ConNu2}) is necessary in this case as well for $(\llangle Q_{M}^{2}\rrangle_{F}+\llangle Q_{M}^{2}\rrangle_{B})-(\llangle W^{2}\rrangle_{F}+\llangle W^{2}\rrangle_{B})$ to be $\geq0$. Therefore, when equation (\ref{ConNu2}) is satisfied we have
\begin{equation}
\langle \eta\rangle^{2}=\left(\frac{\llangle W\rrangle_{F}+\llangle W\rrangle_{B}}{\llangle Q_{M}\rrangle_{F}+\llangle Q_{M}\rrangle_{B}}\right)^{2}\leq\frac{\llangle W^{2}\rrangle_{F}+\llangle W^{2}\rrangle_{B}}{\llangle Q_{M}^{2}\rrangle_{F}+\llangle Q_{M}^{2}\rrangle_{B}} \leq1. \label{UppLo2}
\end{equation}
This shows that the ratio of the fluctuations of $W$ and $Q_{M}$ is still lower and upper-bounded even in the asymmetric case. Equation (\ref{UppLo2}) is a generalization of equation (\ref{UppLo}) to the asymmetric driven Otto cycle. Furthermore, note that the upper bound value 1, is not achieved when the system is working as a heat engine. This follows from imposing the heat engine conditions on the forward and the backward. These conditions are: $\llangle W\rrangle_{F(B)}>0$ and $\delta$($\zeta$)$<1/2$. In that case one can easily show that $(1-(\theta+(1-2\theta)(\delta+\zeta-2\delta\zeta))\tanh^2(\beta\nu_{1}))\coth(\beta\nu_{1})$ and $4\nu_{1}(\llangle W\rrangle_{F}+\llangle W\rrangle_{B}+2(\theta+(1-2\theta)(\delta+\zeta-2\delta\zeta))\nu_{1}\tanh(\beta\nu_{1}))$ are $>0$. From Eq. (\ref{UppLo2}), and analogously to Eq. (\ref{UppLoM8}) we have,
\begin{equation}
\langle \eta\rangle\leq \sqrt{\left(\frac{\llangle W^{2}\rrangle_{F}+\llangle W^{2}\rrangle_{B}}{\llangle Q_{M}^{2}\rrangle_{F}+\llangle Q_{M}^{2}\rrangle_{B}}\right)} <1\label{UppLo3},
\end{equation}
which is less than the upper bound 1 allowed by the law of energy conservation. Even though we could not prove it, we believe that the square root of the ratio of fluctuations of the asymmetric Otto cycle does not provide a tighter bound on the efficiency of the engine than the Otto efficiency. 
\subsection{The ratio of the third and fourth cumulants of $W$ and $Q_{M}$}
We have proved before, that the efficiency of the asymmetric Otto cycle is limited by the one of the Otto if we take into account both the forward and backward and treat them on an equal footing. Then we have proved the ratio of fluctuations is lower and upper bounded. In Ref \cite{Gerry1} Saryal et al. and in Ref \cite{Gerry2} Gerry et al. have proved that for continuous thermal machines, the ratio of the $n$-th cumulants has a lower and upper bound, where the lower bound is defined by the $n$-th power of the efficiency and the upper bound by the $n$-th power of the Carnot efficiency. The lower bound was shown to be saturated in the tight coupling limit. Therefore, analogously to the bounds on the ratio of the second cumulants for quantum unital Otto heat engines, let's ask if the ratio of the third (fourth) cumulants can be lower bounded by third (power) power of the efficiency and upper bounded by 1, as it is the case for autonomous continuous thermal machines?

The standardized third and fourth cumulants are respectively, the skewness and kurtosis. Skewness is a measure of the asymmetry of a distribution. When it is negative (positive) it means that the left tail (right tail) of a distribution is longer than the one on the right (left) side. For kurtosis (or excess kurtosis) it is a measure of the tailedness and peakedness of a distribution. These two quantities have to do with the shape of the distribution. In the quasistatic limit, the ratio of the third and fourth cumulants of heat and work are given as follows,
\begin{equation}
\frac{\llangle W^{3}\rrangle}{\llangle Q_{M}^{3}\rrangle}=\left(1-\frac{\nu_{1}}{\nu_{2}}\right)^{3}
\\  \
\rm{and}, \
\\
\frac{\llangle W^{4}\rrangle}{\llangle Q_{M}^{4}\rrangle}=\left(1-\frac{\nu_{1}}{\nu_{2}}\right)^{4}.\label{3434}
\end{equation}
This is still valid even for higher cumulants in the adiabatic regime. Furthermore, what these two relations tell us, is that the sign of the third and fourth cumulants of heat and work are the same in the quasistatic limit, and that the third (fourth) cumulant of $Q_{M}$ is greater than that of $W$. However, beyond this regime, their signs can be different, which makes this regime more interesting to be investigated. We will return to this point in a moment.

\begin{figure}[hbtp]
\centering
\includegraphics[scale=0.8]{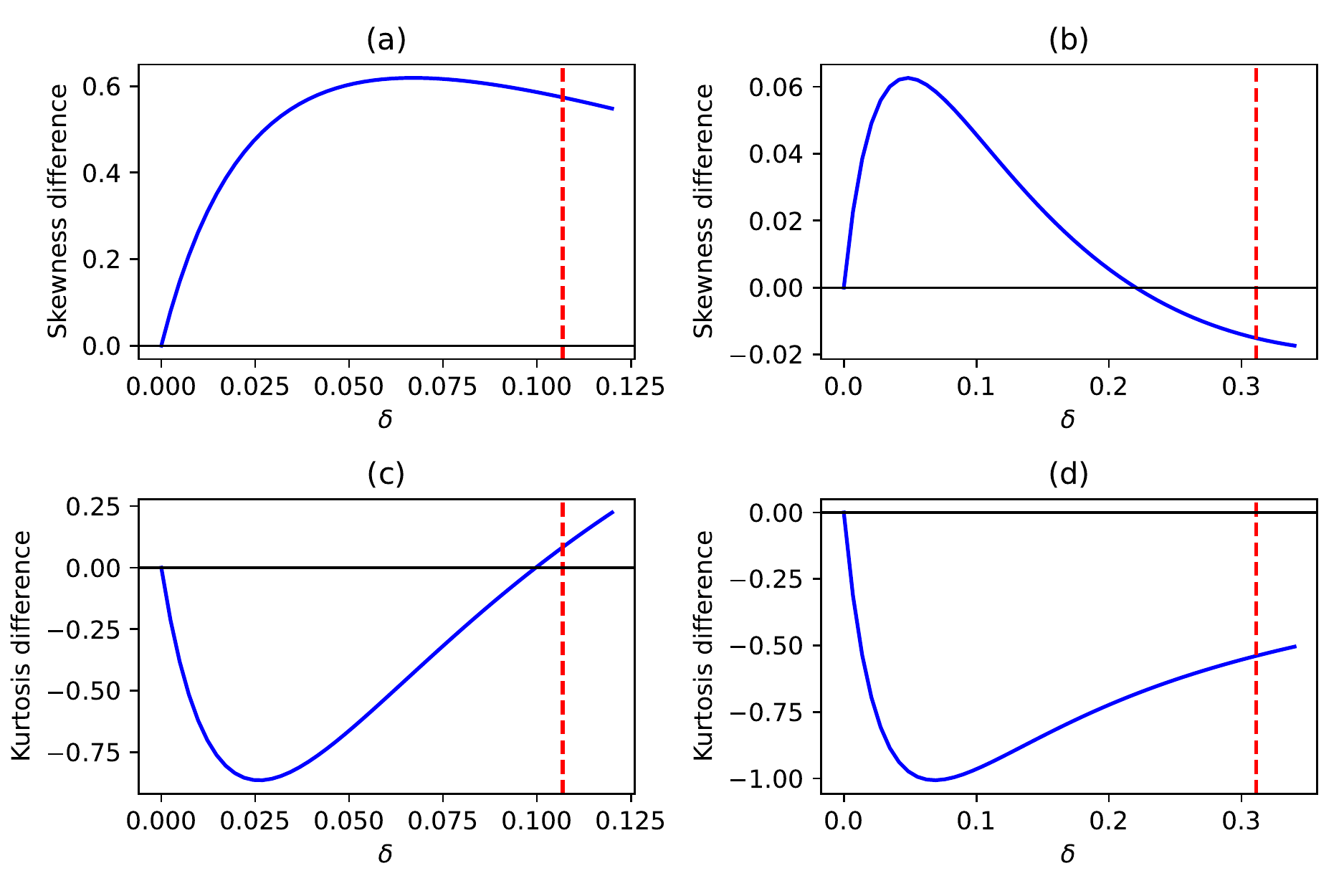}
\caption{Plot of $\llangle Q_{M}^{3}\rrangle/\llangle Q_{M}^{2}\rrangle^{3/2}-\llangle W^{3}\rrangle/\llangle W^{2}\rrangle^{3/2}$ (top figures) and $\llangle Q_{M}^{4}\rrangle/\llangle Q_{M}^{2}\rrangle^{2}-\llangle W^{4}\rrangle/\llangle W^{2}\rrangle^{2}$ (bottom figures). The parametrs are: $\beta=0.7$, $\nu_{1}=1$, $\nu_{2}=2$ and $\zeta=\delta$. In (a) and (c) $\theta=0.2$, and in (b) and (d) $\theta=0.7$. The dashed red vertical line in (a) and (c) corresponds to $\delta=0.10685$ and in (b) and (d)) corresponds to $\delta=0.31125$. In the left (right) figures when $\delta\geq0.10685$ $(\geq0.31125)$ the system stops working as a heat engine for the considered parameters. (a) For $\delta\leq0.10685$ the skewness order of $W$ and $Q_{M}$ is $\llangle Q_{M}^{3}\rrangle/\llangle Q_{M}^{2}\rrangle^{3/2}\geq \llangle W^{3}\rrangle/\llangle W^{2}\rrangle^{3/2}$ not $\llangle W^{3}\rrangle/\llangle W^{2}\rrangle^{3/2}\geq\llangle Q_{M}^{3}\rrangle/\llangle Q_{M}^{2}\rrangle^{3/2}$ as one may expect by analogy to the order of the RFs. (b) For $\delta\leq0.10685$: $\llangle Q_{M}^{3}\rrangle/\llangle Q_{M}^{2}\rrangle^{3/2}\geq \llangle W^{3}\rrangle/\llangle W^{2}\rrangle^{3/2}$ then $\llangle W^{3}\rrangle/\llangle W^{2}\rrangle^{3/2}\geq \llangle Q_{M}^{3}\rrangle/\llangle Q_{M}^{2}\rrangle^{3/2}$.  (c) The kurtosis order for $\delta\leq0.10685$ is: $\llangle W^{4}\rrangle/\llangle W^{2}\rrangle^{2}\geq \llangle Q_{M}^{4}\rrangle/\llangle Q_{M}^{2}\rrangle^{2}$ and then $\llangle Q_{M}^{4}\rrangle/\llangle Q_{M}^{2}\rrangle^{2}\geq \llangle W^{4}\rrangle/\llangle W^{2}\rrangle^{2}$. (d) The kurtosis order is always: $\llangle W^{4}\rrangle/\llangle W^{2}\rrangle^{2}\geq \llangle Q_{M}^{4}\rrangle/\llangle Q_{M}^{2}\rrangle^{2}$. }\label{0.20.7}
\end{figure}

In figure (\ref{0.20.7}) we plot the skewness difference $\llangle Q_{M}^{3}\rrangle/\llangle Q_{M}^{2}\rrangle^{3/2}-\llangle W^{3}\rrangle/\llangle W^{2}\rrangle^{3/2}$ and kurtosis difference $\llangle Q_{M}^{4}\rrangle/\llangle Q_{M}^{2}\rrangle^{2}-\llangle W^{4}\rrangle/\llangle W^{2}\rrangle^{2}$ as a function of $\delta$ for $\theta=0.2$ and $\theta=0.7$. We found that the order of skewness and kurtosis of $W$ and $Q_{M}$ can be arbitrary and not like their relative fluctuations when the system is working as a heat engine (see equation (\ref{mqcqhw})).

Note that in the numerical simulation, we do not plot the cumulants for the case when $\beta<0$, since in this regime one can show that the sign of odd cumulants such as average and skewness get flipped but that of even cumulants such as kurtosis preserve their sign. This means that the order of the relative fluctuations of $W$ and $Q_{M}$ we have proved before will stay the same, but the order of the skewness will be flipped.

\begin{figure}[hbtp]
\centering
\includegraphics[scale=0.75]{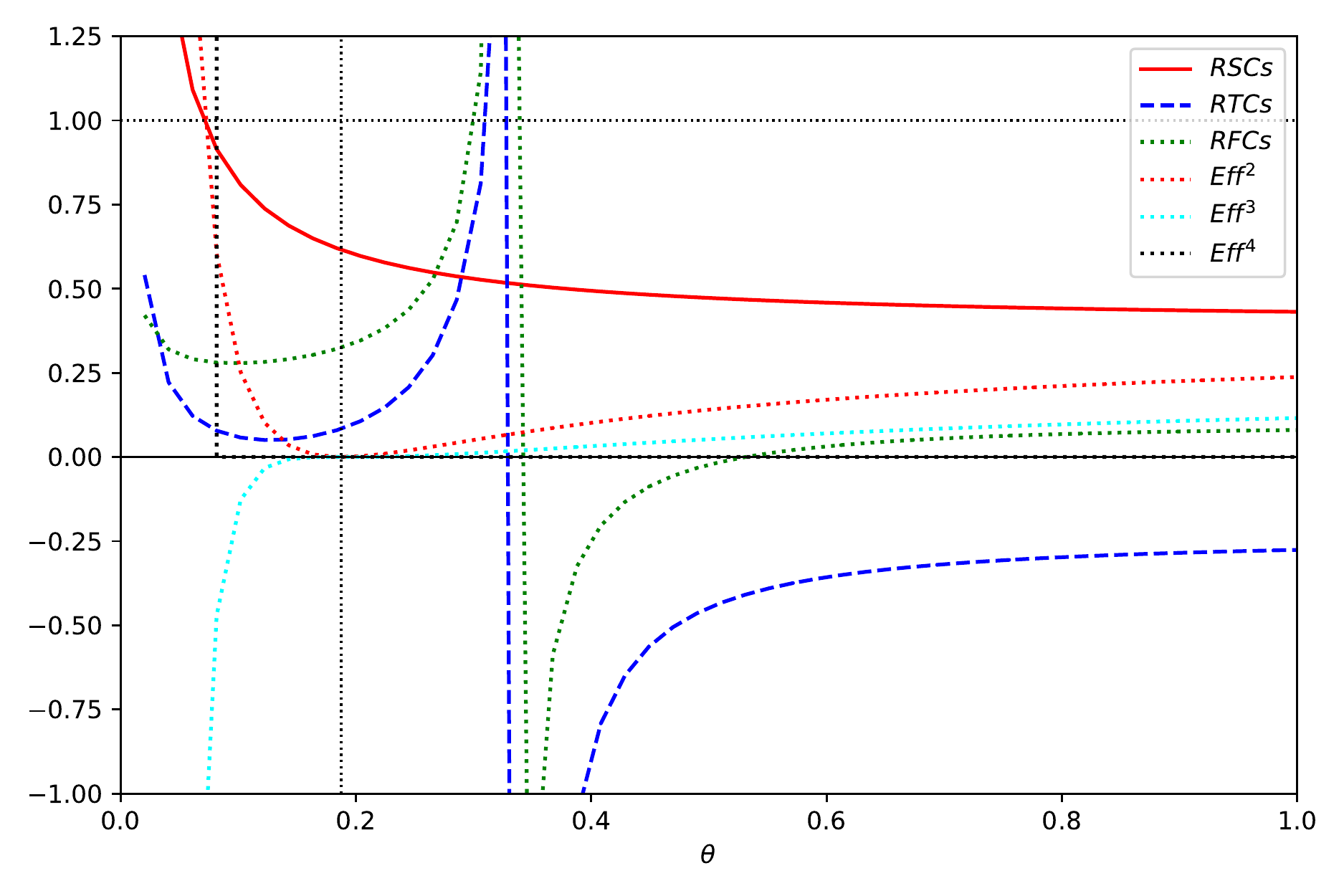}
\caption{Eff refers to efficiency $\langle \eta\rangle$, Eq. (\ref{symmm}). RCSs=ratio of the second cumulants  (red solid line), RTCs=ratio of the third cumulants  (blue dashed line) and RFCs=ratio of the fourth cumulants  (green dotted line) of $W$ and $Q_{M}$ as a function of $\theta$. The parameters are: $\delta=\zeta=0.1$, $\nu_{1}=1$, $\nu_{2}=2$ and $\beta=0.5$. The black vertical dotted line corresponds to $\theta=0.1875$. For $\theta<0.1875$ the system is working as an accelerator and for $\theta>0.1875$ as a heat engine. The red dotted line corresponds to $\langle \eta\rangle^{2}$, cyan dashed line $\langle \eta\rangle^{3}$ and brown dotted line corresponds to $\langle \eta\rangle^{4}$. We see that the ratio of fluctuations is greater than $\langle \eta\rangle^{2}$ and lesser than 1 in the heat engine region as expected. This is still valid even in some regions of the accelerator, in agreement with Eqs. (\ref{UppLo}) and (\ref{UppLo2}). The ratio of the third (fourth) cumulants can be out of $\langle \eta\rangle^{3}\leq .< 1$ ($\langle \eta\rangle^{4}\leq .< 1$).}
\label{BOF1}
\end{figure}

From figure (\ref{0.20.7}), we see that we have either $\llangle W^{3}\rrangle/\llangle W^{2}\rrangle^{3/2}\geq \llangle Q_{M}^{3}\rrangle/\llangle Q_{M}^{2}\rrangle^{3/2}$ or $\llangle W^{3}\rrangle/\llangle W^{2}\rrangle^{3/2}\leq \llangle Q_{M}^{3}\rrangle/\llangle Q_{M}^{2}\rrangle^{3/2}$. Using the already proved analytical lower and upper bound on the ratio of fluctuations, i.e. Eq. (\ref{UppLo}), we obtain from $\llangle W^{3}\rrangle/\llangle W^{2}\rrangle^{3/2}\geq \llangle Q_{M}^{3}\rrangle/\llangle Q_{M}^{2}\rrangle^{3/2}$ that,
\begin{equation}
\frac{\llangle W^{3}\rrangle}{\llangle Q_{M}^{3}\rrangle}\geq \left(\frac{\llangle W^{2}\rrangle}{\llangle Q_{M}^{2}\rrangle}\right)^{3/2}\geq \left(\frac{\llangle W\rrangle}{\llangle Q_{M}\rrangle}\right)^{3}=\langle \eta\rangle^{3}.
\end{equation}
For the case when $\llangle W^{3}\rrangle/\llangle W^{2}\rrangle^{3/2}\leq \llangle Q_{M}^{3}\rrangle/\llangle Q_{M}^{2}\rrangle^{3/2}$ we obtain,
\begin{equation}
\frac{\llangle W^{3}\rrangle}{\llangle Q_{M}^{3}\rrangle}\leq \left(\frac{\llangle W^{2}\rrangle}{\llangle Q_{M}^{2}\rrangle}\right)^{3/2}< 1.
\end{equation}
The same thing can be concluded for the ratio of the fourth cumulants. This suggests that analogously to the ratio of fluctuations, the ratio of the third and fourth cumulants may have a lower and upper bound. However, from the figure (\ref{BOF1}) we see that this is not the case. The reason behind this is purely of quantum origin, since as we showed in Eq. (\ref{3434}) when the adiabatic parameter is zero then the ratio of the cumulants is equal to the $n$-th power of Otto efficiency. However, when $\delta\neq0$ then the ratio of the third (fourth) cumulants is not lower bounded by the third (fourth) power of the Otto efficiency nor it is upper bound by $1$. This shows that the lower and upper on ratio of the second cumulants remains robust against non-adiabaticity but not the ratio of the third and fourth cumulants. This is another central result of the paper.

\begin{figure}[hbtp]
\centering
\includegraphics[scale=0.95]{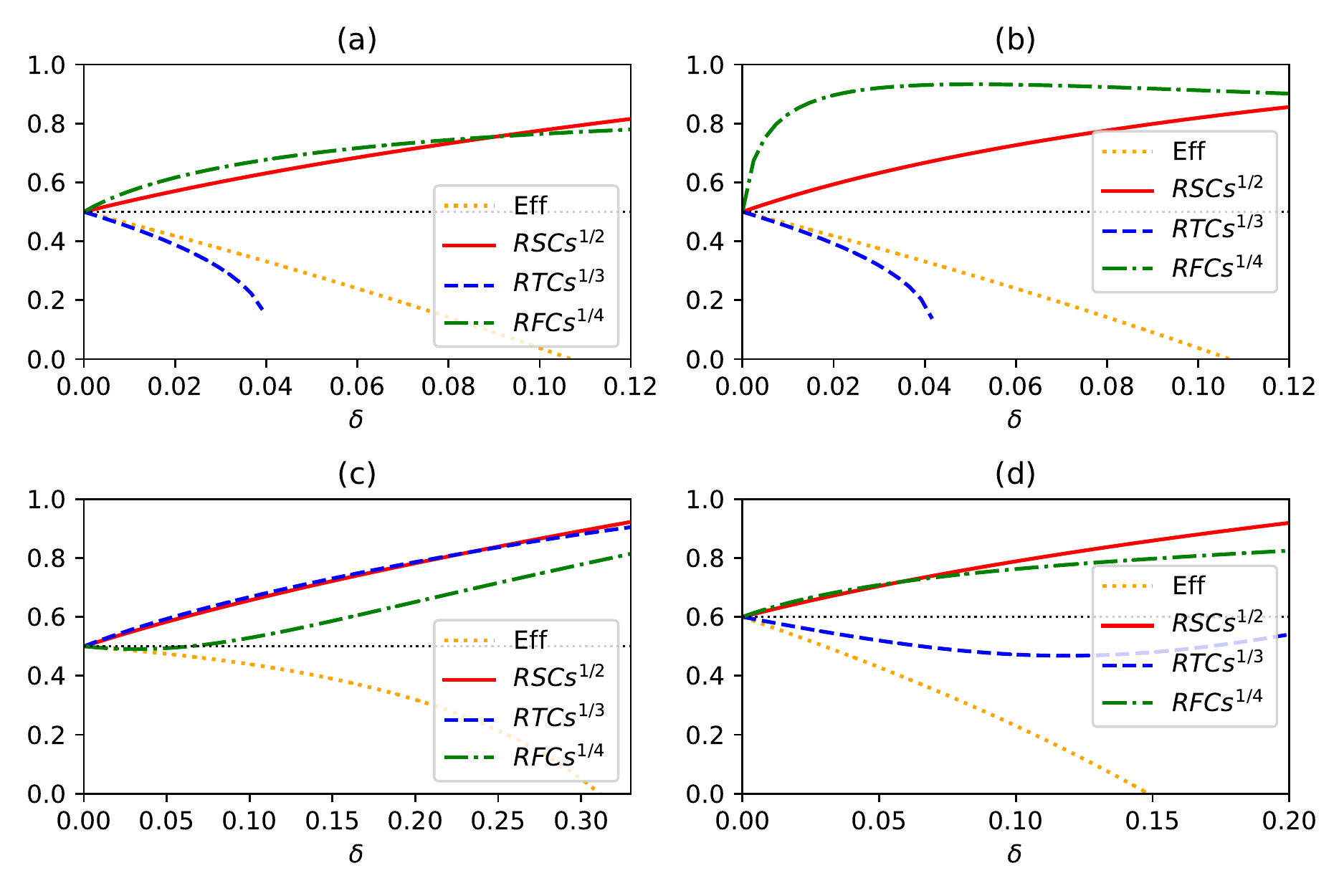}
\caption{Eff and R(S,T,F)Cs as in figure (\ref{BOF1}). (a) $\theta=0.2$, $\nu_{1}=1$, $\nu_{2}=2$ and $\beta=0.5$. (b) $\theta=0.2$, $\nu_{1}=1$, $\nu_{2}=2$ and $\beta=5$. (c) $\theta=0.7$, $\nu_{1}=1$, $\nu_{2}=2$ and $\beta=0.25$. (d) $\theta=0.2$, $\nu_{1}=1$, $\nu_{2}=2.5$ and $\beta=0.5$.
In (a) and (b) we see that the efficiency is upper bounded by the ratio of the second and fourth cumulants. But note that in general, they are above the Otto efficiency, i.e. $0.5$. In Figure (c) the ratio of the fourth cumulants can provide a tighter bound on the Otto efficiency, than the ratio of the second and third cumulants, and even it can be tighter than the Otto efficiency, but only for a very small domain of $\delta$. In (d) we see that all the ratios of the cumulants bound the efficiency. And we see that the ratio of the third cumulants provides a tighter bound that the other two cumulants, and even more, a tighter bound than the Otto efficiency for wide-range values of $\delta$. Unfortunately, this is not universal, i.e. not valid for all heat engine regime parameters.}\label{3443}
\end{figure}

Another strong motivation for considering the ratio of the third and fourth cumulants other than giving us information about the shape of the distribution of thermodynamic quantities such as work and heat, is to investigate if their ratios can provide a bound on the efficiency as it is the case for the second cumulants, Eq. (\ref{UppLo3}). From figure (\ref{3443}) we see that they can sometimes provide a bound on the efficiency which can be tighter than the one provided by the ratio of fluctuations, even sometimes it could be tighter than the Otto efficiency, see Fig. \ref{3443} (d). However, in general, this is parameter dependent and not valid for all heat engine regimes, i.e. not universal as the bound derived from the ratio of fluctuations. This is because these cumulants can have different signs due to nonadiabatic transitions.

An immediate consequence of the results reported in \cite{Gerry2}, is that the $n$-th root of the ratio of the $n$-th cumulants gives a tighter bound on efficiency than the Carnot efficiency. The tighest bound was provided by the ratio of fluctuations. In \cite{BijayKumar}, and in our work, we see that the ratio of second fluctuations also provides a bound on efficiency albeit not being tighter than the one of the Otto. On the other hand, we have seen numerically that sometimes the third and fourth cumulants ratio can provide a tighter bound than that derived from the ratio of fluctuations. Even further, it can be tighter than Otto efficiency. Unfortunately, this is not universal.

We have checked that for the single-spin (quantum Otto cycle) studied by Mohanta et al. \cite{BijayKumar}, the skewness and kurtosis order is arbitrary and that the ratio of the third (fourth) cumulants can violate as well the bounds $\langle \eta\rangle^{3}\leq .\leq\left(1-\beta_{h}/\beta_{c}\right)^{3}$ $\left(\langle \eta\rangle^{4}\leq .\leq\left(1-\beta_{h}/\beta_{c}\right)^{4}\right)$, where $1-\beta_{h}/\beta_{c}$ is the Carnot efficiency. The reason behind this is also the non-adiabaticity. But we note that one can easily show this from our results as well. For example, if we take $\beta_{h}=0$ in Ref. \cite{BijayKumar}, then in this case, the bath becomes completely unital, i.e. it maps an arbitrary state to a mixed state, which is a special case of our results. Then one can easily check our statements under the condition $\theta=1/2$. Already from the figure (\ref{BOF1}) we see that the ratio of the third and fourth cumulants are negative, thus violating the bounds $\langle \eta\rangle^{3}$ and $\langle \eta\rangle^{4}$ which are positive.

\subsection{Numerical analysis of the average work and its relative fluctuations}
Let's now show that considering arbitrary unital channels can have a positive influence on average work as well as its relative fluctuations.

\begin{figure}[hbtp]
\centering
\includegraphics[scale=0.8]{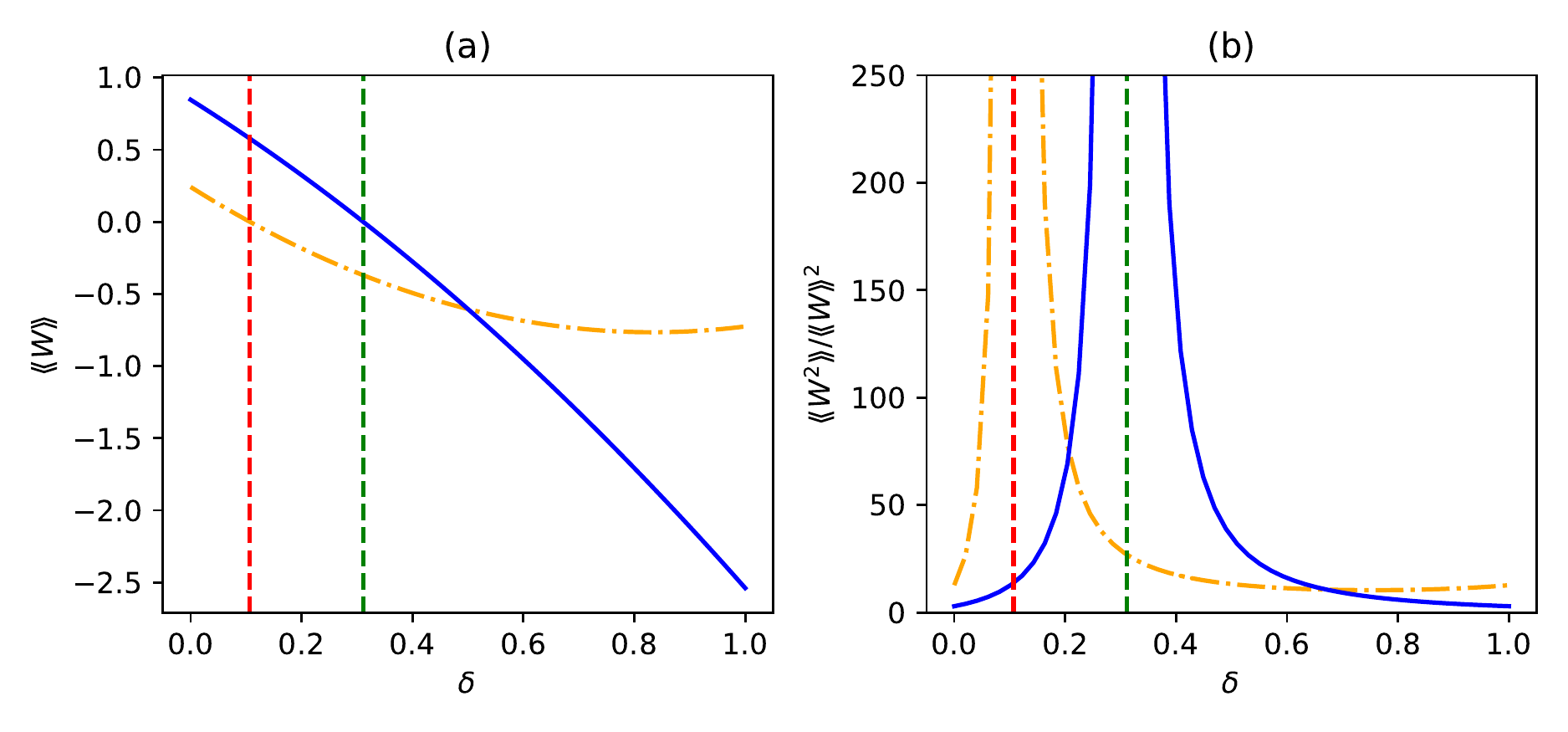}
\caption{Plot of average work (a) and RFs (b), for $\theta=0.2$ (blue solid line) and $\theta=0.7$ (orange dashed-dotted line). The vertical dashed red line corresponds to $0.10685$ and the vertical dashed green line corresponds to $0.31125$.}\label{AVSK0.20.7}
\end{figure}

From figure \ref{AVSK0.20.7}-(a) we see that the average work extracted when $\theta=0.7$ is higher than the case when $\theta=0.2$, and even more the value of the adiabatic parameter $\delta$ above which work becomes negative get increased, that is for $\theta=0.7$ it is $0.31125$ and for $\theta=0.2$ it is $0.10685$. Furthermore, we see that the average work gets lowered as we increase $\delta$ as expected. And we see that the more we increase $\delta$ the more work becomes negative. However, if the bath temperature is negative then this non-adiabaticity increase can be beneficial as we showed already.

For the relative fluctuations of work, figure \ref{AVSK0.20.7}-(b), we see that for $\theta=0.7$ and $\delta\leq0.10685$ the RF is less than the RF for $\theta=0.2$. This shows that increasing $\theta$ not only enhanced the amount of work extracted but also diminishes its relative fluctuations thus increasing its reliability. Of course, increasing $\theta$ increases the variance of $W$, i.e. $\llangle W^{2}\rrangle$(for $\theta=0.7$) $>$ $\llangle W^{2}\rrangle$(for $\theta=0.2)$, but the relative fluctuations as we see get lowered, thus work become more reliable. However, note that for $0.10685 \leq \delta \leq 0.31125$ the RFs of $W$ for $\theta=0.7$ can be lower or higher than the RFs of $W$ for $\theta=0.2$. Fortunately, in this regime, the system is not working as a heat engine for $\theta=0.2$. Further, we see when we increase $\delta$ then the RFs first diverge and then start to decrease again. For $\theta=0.2$ ($\theta=0.7$) the divergence happens for $\delta=0.10685$ ($\delta=0.31125$). Of course, both cumulants are finite, and the divergence is a consequence of the fact that work becomes zero.

Now let's comment on the case when the $\beta$ is negative, i.e. instead of $\beta=0.7$ in figure (\ref{AVSK0.20.7}) we use $\beta=-0.7$. When it is positive the maximum amount of work extracted is when $\delta=0$. It is equal to 0.241747 (0.846115) for $\theta=0.2$ ($\theta=0.7$). The RF is 12.6889 (2.9111) for $\theta=0.2$ ($\theta=0.7$). On the other hand, the maximum amount of work extracted when $\beta$ is negative is when $\theta=1$. It is equal to 0.725241 (2.53834) for $\theta=0.2$ ($\theta=0.7$). The RFs are the same for the case when $\delta=0$. This shows that negative temperatures can help in extracting higher amounts of work with fixed relative fluctuations.

\section{Using the unital map in \cite{Shanhe} in a CS}\label{QOCM}

\subsection{Coherently superposed channels}\label{CS}
\begin{figure}[hbtp]
\centering
\includegraphics[scale=0.140]{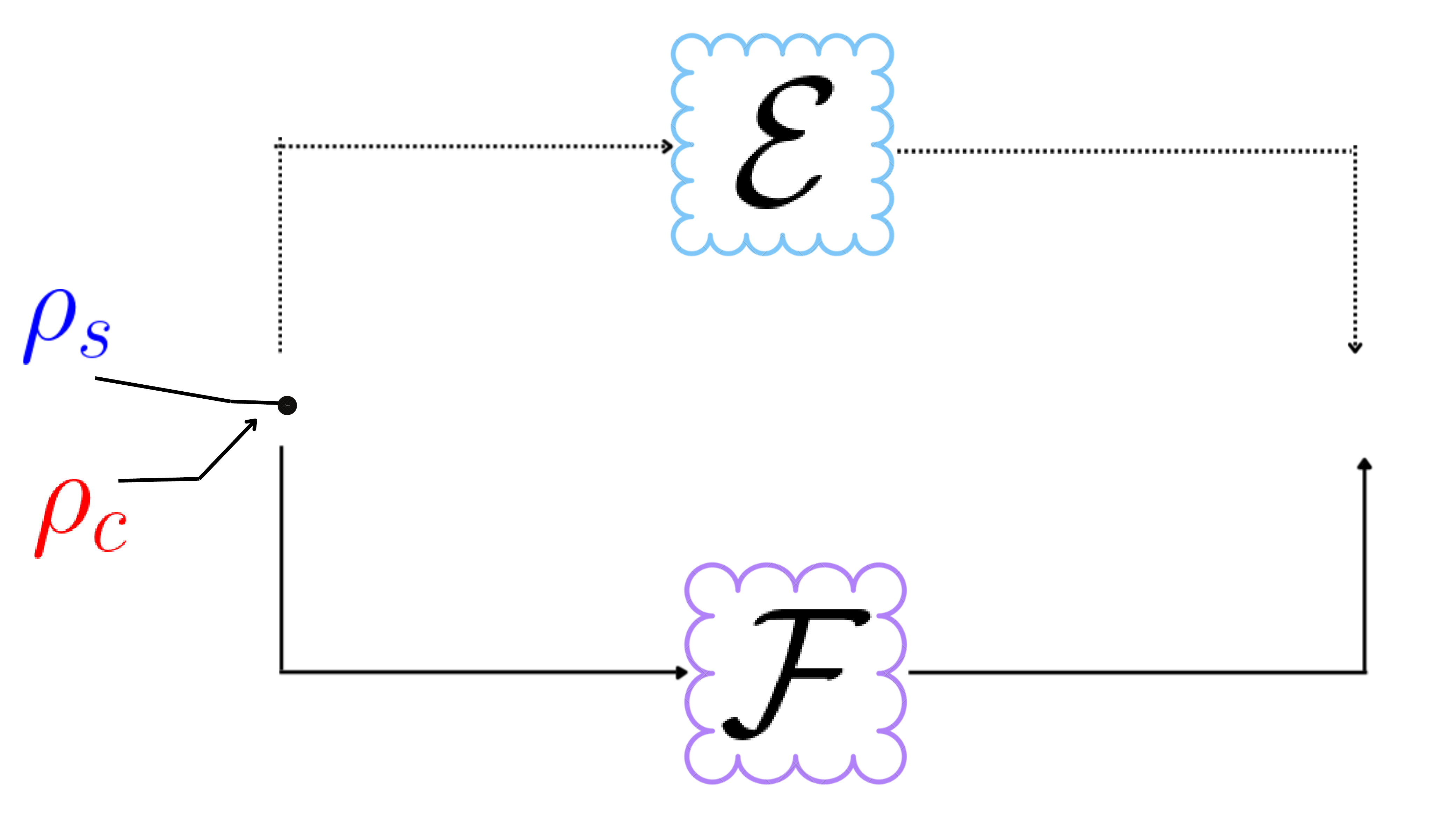}
\caption{This figure shows a CS of the channel $\mathcal{E}$ and the channel $\mathcal{F}$. More precisely, depending on the state of the control qubit $\rho_{c}$ we would act on the state of the system $\rho_{s}$: when the control qubit is in the ground state $|0\rangle_{c}$ it will be affected only by $\mathcal{E}$ and when it is in the excited state $|1\rangle_{c}$ it will be affected only by $\mathcal{F}$. If the system is in a probabilistic mixture of the ground and the excited state, i.e. $p|0\rangle_{cc}\langle0|+(1-p)|1\rangle_{cc}\langle1|$, then, in this case, we have a mixture of the two situations, i.e. $p\mathcal{E}(\rho_{s})+(1-p)\mathcal{F}(\rho_{s})$ with $p$ $\in [0,1]$. However, one should note that this is different from the coherent superposition in the case when the control qubit is in a superposition. More precisely, in the latter case, we have an interference effect between the two quantum channels, which is not possible in the case of a mixture of them.}\label{CSCBB}
\end{figure}

In contrast to ICO \cite{Valiron,BranciardC,Oreshkov1,Mukhopadhyay}, here, we no longer are interested in the order in which the operations are applied. We only care about which operation has been applied on the state of the system $\rho_{s}$. More precisely, suppose that in addition to our system of interest, we have an additional system, that we call a control system, and depending on its state we act on the system of interest. For example, if the control qubit is in the ground state denoted as $|0\rangle_{c}$ we apply the channel $\mathcal{E}$ defined as $\mathcal{E}(.)=\sum_{i}K_{i}(.)K_{i}^{\dagger}$, and if it is in the excited state denoted as $|1\rangle_{c}$ we apply the channel $\mathcal{F}$ defined as $\mathcal{F}(.)=\sum_{j}\tilde{K}_{j}(.)\tilde{K}_{j}^{\dagger}$. $K_{i}$ and $\tilde{K}_{j}$ are known as Kraus operators. Therefore, when the qubit is in an equal superposition of these two states, i.e. $(|1\rangle_{c}+|0\rangle_{c})/2$, we can no longer talk about which operation has been applied (see Fig.(\ref{CSCBB})). And mathematically we have a new big channel-which acts on both the system and the control qubit-with Kraus operators defined as,
\begin{equation}
T_{ij}=K_{i}\otimes|0\rangle_{cc}\langle0|+K_{j}\otimes |1\rangle_{cc}\langle1|\label{Tij}.
\end{equation}
In this work, we only limit ourselves to the case of applying the same channel in a CS, i.e. $K_{i}$ and $\tilde{K}_{j}$ are the same when $i=j$. Furthermore, note that the Kraus operators defined in Eq.(\ref{Tij}) are not normalized. If $i$ and $j$ range from $1$ to $N$, then one can show that,

\begin{equation}
\begin{split}
\sum_{i,j}T_{ij}^{\dagger}T_{ij} & =\sum_{i,j}\left( K_{i}^{\dagger}K_{i}\otimes|0\rangle_{cc}\langle0|+\tilde{K}_{j}^{\dagger}\tilde{K}_{j}\otimes |1\rangle_{cc}\langle1|\right)
\\ &
=\sum_{j}\left( \sum_{i}K_{i}^{\dagger}K_{i}\right)\otimes|0\rangle_{cc}\langle0|+\sum_{i}
\left(\sum_{j}\tilde{K}_{j}^{\dagger}\tilde{K}_{j}\right)\otimes |1\rangle_{cc}\langle1|
\\ &
=N\mathbb{1},
\end{split}
\end{equation}
which is a nontrivial constant, with $\mathbb{1}$ is the identity operator. We can normalize $\sum_{i,j}T_{ij}^{\dagger}T_{ij}$, by dividing it by $N$, thus
$\sum_{ij}T_{ij}^{\dagger}T_{ij}/N=\mathbb{1}$. The new Kraus operators, which are normalized, are defined as follows,
\begin{equation}
T_{ij}'=\frac{T_{ij}}{\sqrt{N}}=\frac{1}{\sqrt{N}}(K_{i}\otimes|0\rangle_{cc}\langle0|+\tilde{K}_{j}\otimes |1\rangle_{cc}\langle1|).\label{NewK}
\end{equation}
Now suppose that the initial state of the control qubit is a given as follows, $\rho_{c}= |\psi_{c}\rangle \langle\psi_{c}|$, where $|\psi_{c}\rangle=\sqrt{\alpha}|0\rangle_{c}+\sqrt{1-\alpha}|1\rangle_{c}$. Therefore the action of $\mathcal{E}$ in a coherently superposed manner is given as follows,
\begin{equation}
S(\rho_{s} \otimes \rho_{c})=\sum_{i,j}T_{ij}'\left(\rho_{s} \otimes \rho_{c}\right) T_{ij}'^{\dagger}.
\end{equation}
Expanding the latter using Eq.(\ref{NewK}), we obtain,
\begin{equation}
\begin{split}
S(\rho_{s} \otimes \rho_{c}) & =\frac{1}{N}\sum_{i,j}\left( K_{i}\rho_{s} K_{i}^{\dagger}\otimes\alpha|0\rangle_{cc}\langle0|
+\tilde{K}_{j}\rho_{s} \tilde{K}_{j}^{\dagger}\otimes(1-\alpha)|1\rangle_{cc}\langle1|
+\sqrt{\alpha(1-\alpha)}(K_{i}\rho_{s} \tilde{K}_{j}^{\dagger}\otimes|0\rangle_{cc}\langle1|
+\tilde{K}_{j}\rho_{s} K_{i}^{\dagger}\otimes|1\rangle_{cc}\langle0|)\right).
\end{split} \label{Eq3}
\end{equation}
The last two terms are the interference between the Kraus operators which would be absent when $\alpha=0$ or $\alpha=1$. After measuring the control qubit in the Fourier basis $\{|+\rangle_{c}=(|1\rangle_{c}+|0\rangle_{c})/\sqrt{2},|-\rangle_{c}=(|1\rangle_{c}-|0\rangle_{c})/\sqrt{2}\}$ and normalizing the state we obtain,
\begin{equation}
\begin{split}
\frac{{}_{c}\langle \pm| S(\rho_{s} \otimes \rho_{c})| \pm \rangle_{c}}{p_{\pm}} & =\frac{1}{2Np_{\pm}}\sum_{i,j}\left(\alpha K_{i}\rho_{s} K_{i}^{\dagger}
+(1-\alpha)K_{j}\rho_{s} \tilde{K}_{j}^{\dagger}\pm\sqrt{\alpha(1-\alpha)}\left( K_{i}\rho_{s} \tilde{K}_{j}^{\dagger}
+\tilde{K}_{j}\rho K_{i}^{\dagger}\right)\right)
\\ &
=\frac{1}{p_{\pm}}\left(\frac{1}{2}\sum_{i}K_{i}\rho_{s} K_{i}^{\dagger}
\pm \frac{\sqrt{\alpha(1-\alpha)}}{N} \sum_{ij}( K_{i} \rho_{s} \tilde{K}_{j}
^{\dagger})\right),
\end{split}\label{ICOCS}
\end{equation}
with $p_{\pm}=\frac{1}{2}\pm\frac{\sqrt{\alpha(1-\alpha)}}{N}\mathrm{Tr}\left( \sum_{i,j} K_{i}\rho \tilde{K}_{j}^{\dagger}\right)$. Note that ICO channels need three important ingredients for the final state of the system to be different from the one when the quantum channels are used in a definite causal order. First, the control qubit is in superposition. Second, the Kraus operators of the channels need to be non-commuting, of course not all of them, and finally, we have to measure the control qubit in the Fourier basis $\{|+\rangle_{c},|-\rangle_{c}\}$, i.e. we have to avoid to project it in the $\{|1\rangle_{c},|0\rangle_{c}\}$ basis. However, CSCs, need only the control qubit to be in a superposition and the measurement to be done in the $\{|+\rangle_{c},|-\rangle_{c}\}$ basis. Furthermore, note that our study is not a comparison between indefinite causal ordered and coherently superposed channels. This is just to explain the difference between them.
\subsection{Applying the unital map in \cite{Shanhe} in a CS}
Our goal in this section is to apply the unital map, which already plays the role of a heat bath, in a CS and investigate the role of coherence on the heats, efficiency and the first two cumulants of work. Differently from the previous section, we consider and limit ourselves to the unital map studied in Refs. \cite{Shanhe}. The Kraus operators of this channel are defined as follows
\begin{equation}
\pi_{1}=|\psi_{1}\rangle\langle\psi_{1}|\hspace{1cm} \ \rm and \hspace{1cm} \ \pi_{2}=|\psi_{2}\rangle\langle\psi_{2}|,\label{P1P2}
\end{equation}
with $\pi_{1}^{\dagger}=\pi_{1}^{T}=\pi_{1}$, $\pi_{2}^{\dagger}=\pi_{2}^{T}=\pi_{2}$ and, $\pi_{1}+\pi_{2}=1$. Here $T$ is the transpose. Note that the expressions of $|\psi_{1}\rangle$ and $|\psi_{2}\rangle$ are not necessary to be specified. When these Kraus operators $\pi_{i}$ characterizing the measurement heat bath are used in ICO \cite{Oreshkov1}, nothing will be changed since they are commuting and satisfy $\pi_{i}\pi_{i}=\pi_{i}$, since they are projectors, however, when used in a CS, the state of the working system along the Otto cycle will be different from the one obtained when $\pi_{i}$ are used in a definite causal order. From equation (\ref{ICOCS}) the state and the Hamiltonian of the system at points $\mathbf{A}$, $\mathbf{B}$, $\mathbf{C}$ and $\mathbf{D}$ of the Otto cycle in figure \ref{b0.55}(b) are given respectively as follows
\begin{widetext}
\begin{equation}
\left\{ \begin{array}{llll}
\rho_{1} \\ H_{1}
\end{array}\right. \rightarrow
\left\{ \begin{array}{llll}
\rho_{2}=U\rho_{1}U^{\dagger} \\ H_{2}
\end{array}\right. \rightarrow
\left\{ \begin{array}{llll}
\rho_{3}^{\pm}=\frac{1}{2p_{\pm}}\rho_{3}\pm\frac{\sqrt{\alpha(1-\alpha)}}{2p_{\pm}}\rho_{2} \\ H_{2}
\end{array}\right. \rightarrow
\left\{ \begin{array}{llll}
\rho_{4}^{\pm}=\frac{1}{2p_{\pm}}\rho_{4}\pm\frac{\sqrt{\alpha(1-\alpha)}}{2p_{\pm}}V\rho_{2}V^{\dagger} \\ H_{1}
\end{array}\right.
,
\end{equation}
\end{widetext}
with, $p_{\pm}=\frac{1}{2}(1\pm\sqrt{\alpha(1-\alpha)})$, $\rho_{1}=e^{-\beta H_{1}}/Z$, $\rho_{3}=\sum_{j}\pi_{j}U\rho_{1}U^{\dagger}\pi_{j}$ and $\rho_{4}=V\left(\sum_{j}\pi_{j}U\rho_{1}U^{\dagger}\pi_{j}\right)V^{\dagger}$.
This is the unmonitored case, i.e. when no projective measurement is applied to the system between the strokes. The $\pm$ refers to which basis the control qubit has been projected.
\subsection{The exact analytical expression of the CF $\chi(\gamma_{W},\gamma_{M})_{F}^{\pm}$}
To differentiate between the already computed CF (eq.\ref{chih}), the new one will be denoted by $\chi(\gamma_{W},\gamma_{M})_{F}^{\pm}$. As we said above the notation $\pm$ is used to distinguish between the case when the control qubit is either projected in the $|+\rangle_{c}$ or the $|-\rangle_{c}$ Fourier basis. Let's now compute the CF. The joint PD of the forward when the unital map (\ref{P1P2}) is used in a CS, is given as follows
\begin{widetext}
\begin{equation}
\begin{split}
P(W,Q_{M})_{F}^{\pm}&=\sum_{n,m,k,l}\frac{e^{-\beta\nu_{n}}}{Z}|{}_{2}\langle m|U|n\rangle_{1}|^{2}\frac{1}{2p_{\pm}}\left(\sum_{j}|{}_{2}\langle k|\pi_{j}|m\rangle_{2}|^{2}\pm\sqrt{\alpha(1-\alpha)}|{}_{2}\langle k|m\rangle_{2}|^{2}\right) |{}_{1}\langle l|V|k\rangle_{2}|^{2}
\\ &
\times \delta(W+(\nu_{m}-\nu_{n}+\nu_{l}-\nu_{k}))\delta(Q_{M}-(\nu_{k}-\nu_{m})).
\end{split}
\end{equation}
This expression can be decomposed into
\begin{equation}
\begin{split}
P(W,Q_{M})_{F}^{\pm}= & \frac{1}{2p_{\pm}}\left( \sum_{n,m,k,l}\frac{e^{-\beta\nu_{n}}}{Z}|{}_{2}\langle m|U|n\rangle_{1}|^{2}\sum_{j}|{}_{2}\langle k|\pi_{j}|m\rangle_{2}|^{2} |{}_{1}\langle l|V|k\rangle_{2}|^{2}\delta(W+(\nu_{m}-\nu_{n}+\nu_{l}-\nu_{k}))\delta(Q_{M}-(\nu_{k}-\nu_{m}))\right) \\ &
\pm\frac{\sqrt{\alpha(1-\alpha)}}{2p_{\pm}}\left(\sum_{n,m,l}\frac{e^{-\beta\nu_{n}}}{Z}|{}_{2}\langle m|U|n\rangle_{1}|^{2} |{}_{2}\langle l|V|m\rangle_{2}|^{2}\delta(W-(\nu_{n}-\nu_{l}))\right) .
\end{split}\label{FKM}
\end{equation}
\end{widetext}
This joint PD is a mixture of a PD of a heat engine when the channel (\ref{P1P2}) is applied in definite causal order and the second PD when the unital channel is the identity, i.e. the trivial unital channel. From (\ref{FKM}), one can straightforwardly show that the compact form of the CF is given as follows
\begin{widetext}
\begin{equation}
\begin{split}
\chi(\gamma_{W},\gamma_{M})_{F}^{\pm} & =\left(\frac{1-\theta}{2p_{\pm}}\pm\frac{\sqrt{\alpha(1-\alpha)}}{2p_{\pm}}\right)\left(1+\left(\frac{\rm 2cos((2\gamma_{W}+i\beta)\nu_{1})}{Z}-1\right)(\delta+\zeta-2\delta\zeta)\right)
\\ &
+\frac{\theta}{2p_{\pm}}\left((1-\delta)\left(\zeta \frac{2\rm cos(2(\gamma_{W}+\gamma_{M})\nu_{2}-i\beta\nu_{1})}{Z}+(1-\zeta) \frac{2\rm cos(2(\gamma_{W}(\nu_{2}-\nu_{1})+\gamma_{M}\nu_{2})-i\beta\nu_{1})}{Z}\right)\right.
\\ &
+\left.\delta\left((1-\zeta) \frac{2\rm cos(2(\gamma_{W}+\gamma_{M})\nu_{2}+i\beta\nu_{1})}{Z}+\zeta \frac{2\rm cos(2((\nu_{1}+\nu_{2})\gamma_{W}+\gamma_{M}\nu_{2})+i\beta\nu_{1})}{Z}\right)\right).\label{result1}
\end{split}
\end{equation}
\end{widetext}
One can check that $\chi(\gamma_{W}=0,\gamma_{M}=0)_{F}^{\pm}=1$. 
\subsection{The first cumulants of work and heats and efficiency enhancement and degradation}
From equation (\ref{result1}), the first cumulants of heats and work can be written as follows
\begin{widetext}
\begin{equation}
\llangle Q_{M}\rrangle_{F}^{\pm}=\frac{2(1-2\delta)\theta\nu_{2}\tanh(\beta\nu_{1})}{1\pm\sqrt{\alpha(1-\alpha)}},\label{QMAL}
\end{equation}
\begin{equation}
\llangle Q_{T}\rrangle^{\pm}=\frac{-2(\theta+\zeta-2\theta\zeta+\delta(1-2\zeta)(1-2\theta))\nu_{1}\tanh(\beta\nu_{1})}{1\pm\sqrt{\alpha(1-\alpha)}}\mp\frac{ 2\sqrt{\alpha(1-\alpha)}(\zeta+\delta-2\delta\zeta)\nu_{1}\tanh(\beta\nu_{1})}{1\pm\sqrt{\alpha(1-\alpha)}} ,\label{QTAL}
\end{equation}
and
\begin{equation}
\begin{split}
\llangle W\rrangle_{F}^{\pm} & =\frac{2(1-2\delta)\theta\nu_{2}\tanh(\beta\nu_{1})}{1\pm\sqrt{\alpha(1-\alpha)}}+\frac{-2(\theta+\zeta-2\theta\zeta+\delta(1-2\zeta)(1-2\theta))\nu_{1}\tanh(\beta\nu_{1})}{1\pm\sqrt{\alpha(1-\alpha)}}
\\ &
\mp\frac{2\sqrt{\alpha(1-\alpha)}(\zeta+\delta-2\delta\zeta)\nu_{1}\tanh(\beta\nu_{1})}{1\pm\sqrt{\alpha(1-\alpha)}}.
\end{split}
\end{equation}
\end{widetext}
As we said before since the considered map is unital, the backward quantities, will follow from those by the correspondence $\delta\leftrightarrow\zeta$. Note that the expression of the second, the third and the fourth cumulants are not reported because they are too long and complicated and not illuminating. Below, we analyze numerically the first two cumulants for work.\par

If we consider $\delta=\zeta$ which corresponds to the symmetric Otto cycle, one can easily show that the efficiency expression is given as follows
\begin{equation}
\langle\eta\rangle^{\pm}=\langle\eta\rangle^{\alpha=0}\mp\frac{\nu_{1}}{\nu_{2}}\frac{\sqrt{\alpha(1-\alpha)}2\delta(1-2\delta)}{(1-2\delta)\theta}.\label{symeta}
\end{equation}
The expression of $\langle \eta\rangle^{\alpha=0}$ is given in equation (\ref{symmm}). For the asymmetric case, we have
\begin{equation}
\langle\eta\rangle^{\pm}=\langle\eta\rangle^{\alpha=0}\mp\frac{\nu_{1}}{\nu_{2}}\frac{\sqrt{\alpha(1-\alpha)}(\delta+\zeta-2\delta\zeta)}{(1-(\delta+\zeta))\theta}.\label{asyeta}
\end{equation}
The expression of $\langle \eta\rangle^{\alpha=0}$ is given in equation (\ref{PropreEff}). The enhancement in efficiency can be easily seen from Eqs.(\ref{symeta}) and (\ref{asyeta}). The enhancement when the control qubit is projected in the $|-\rangle_{c}$ Fourier basis can be understood from the expressions of the heat absorbed $\llangle Q_{M}\rrangle^{-}$ and the heat released $\llangle Q_{T}\rrangle^{-}$ to the cold bath, i.e. Eqs.(\ref{QMAL}) and (\ref{QTAL}), respectively. For $\llangle Q_{M}\rrangle^{-}$ we see that it get increases when we increase $\alpha$ and it reaches the maximal value for $\alpha=1/2$. For the one released to the cold bath $\llangle Q_{T}\rrangle^{-}$, we see that the first term gets increased by the same magnitude of $\llangle Q_{M}\rrangle^{-}$ but we see that there is a second term which is always positive which would decrease the amount of heat released thus efficiency enhancement. In this case, we have work enhancement as well.  When the control qubit is projected in the $|+\rangle_{c}$ Fourire basis, efficiency gets degraded, since in this case, the system will absorb less heat from the hot bath, because of the denominator $1+\sqrt{\alpha(1-\alpha)}$, and the heat released to the cold bath would increase thus work and efficiency decrease.\par

We should mention that the enhancement of the efficiency is possible only in the non-adiabatic regime, i.e. $\delta$ and $\zeta$ being different from zero as equations (\ref{symeta}) and (\ref{asyeta}) show. To see this, one can set $\delta=\zeta=0$, in this case we have,
$\llangle Q_{M}\rrangle^{\pm}=\frac{2\theta\nu_{2}\tanh(\beta\nu_{1})}{1\pm\sqrt{\alpha(1-\alpha)}}$, $\llangle Q_{T}\rrangle^{\pm}=\frac{-2\theta\nu_{1}\tanh(\beta\nu_{1})}{1\pm\sqrt{\alpha(1-\alpha)}}$ and $\llangle W\rrangle^{\pm}=\frac{2\theta(\nu_{2}-\nu_{1})\tanh(\beta\nu_{1})}{1\pm\sqrt{\alpha(1-\alpha)}}$. This shows that $\llangle Q_{M}\rrangle^{\pm}$, $\llangle W\rrangle^{\pm}$ and $\llangle Q_{T}\rrangle^{\pm}$ get increased or decreased by the same magnitude, and therefore efficiency gets reduced to the Otto one, i.e. $1-\nu_{1}/\nu_{2}$.  This analysis shows that coherence can be beneficial in the presence of non-adiabaticity. Since as we showed it can improve the work extracted as well as the efficiency. This means that if one is interested in efficiency and power, i.e. running the cycle in finite time, and not quasistatic as required in the adiabatic regime, coherence can be used advantageously depending on which basis the control qubit has been projected. This is the drawback of this protocol is that it gives an enhancement only probabilistically. More precisley, when we increase $\alpha$ towards 1/2, $p_{-}=(1-\sqrt{\alpha(1-\alpha)})/2$ get decreased towards its lower value 1/4.
\subsection{Heat engine conditions and negativity of $\llangle Q_{T}\rrangle^{\pm}$}
For $\llangle Q_{M}\rrangle^{\pm}>0$ it is still $\delta<1/2$, since coherence affects only the denominator. For $\delta=\zeta$ and for $\llangle W\rrangle^{\pm}>0$, we have to ensure that,
\begin{equation}
\nu_{2}>\frac{\theta+2\delta(1-\delta)(1-2\theta)}{\theta(1-2\delta)}\nu_{1}\pm\frac{2\delta(1-\delta)\sqrt{\alpha(1-\alpha)}}{\theta(1-2\delta)}\nu_{1}.
\end{equation}
This equation shows that when the control qubit is projected in the $|-\rangle_{c}$ Fourier basis, the consumed coherence will be beneficial, since the value of $\nu_{2}$ under which no work can be extracted gets lowered. However, when it is projecred in the $|+\rangle_{c}$ basis, the lowest value of $\nu_{2}$, get increased. That is the bigger we increase $\alpha$ towards $1/2$ the more the minimum of $\nu_{2}$ gets increased. This shows that coherence can be either beneficial or detrimental. When $\delta=\zeta=0$, the positive work condition is not altered by the presence of coherence. Furthermore, in this case as well one can prove that $\llangle Q_{T}\rrangle^{\pm}$ is $\leq0$. We have,
\begin{widetext}
\begin{equation}
\begin{split}
\llangle Q_{T}\rrangle^{\pm} & =\frac{-2(\theta+\zeta-2\theta\zeta+\delta(1-2\zeta)(1-2\theta))\nu_{1}\tanh(\beta\nu_{1})}{1\pm\sqrt{\alpha(1-\alpha)}}\mp\frac{ 2\sqrt{\alpha(1-\alpha)}(\zeta+\delta-2\delta\zeta)\nu_{1}\tanh(\beta\nu_{1})}{1\pm\sqrt{\alpha(1-\alpha)}} 
\\ &
=\frac{-2(\theta+(1-2\theta\pm\sqrt{\alpha(1-\alpha)})(\delta+\zeta-2\delta\zeta))\nu_{1}\tanh(\beta\nu_{1})}{1\pm\sqrt{\alpha(1-\alpha)}}.
\end{split}
\end{equation}
\end{widetext}
Since $0\leq\theta\leq1/2$ and $0\leq\sqrt{\alpha(1-\alpha)}\leq1/2$, we have
$0\leq1-2\theta+\sqrt{\alpha(1-\alpha)}\leq3/2$ and $
-1/2\leq1-2\theta-\sqrt{\alpha(1-\alpha)}\leq1$. Furthemore, since the maximum value of $\delta+\zeta-2\delta\zeta$ is 1, we obtain
\begin{equation*}
\begin{split}
\llangle Q_{T}\rrangle^{\pm}=\frac{-2(1-\theta\pm\sqrt{\alpha(1-\alpha)})\nu_{1}\tanh(\beta\nu_{1})}{1\pm\sqrt{\alpha(1-\alpha)}} 
\end{split}.
\end{equation*}
And since, we have, $1/2\leq1-\theta+\sqrt{\alpha(1-\alpha)}\leq3/2$ and $0\leq1-\theta-\sqrt{\alpha(1-\alpha)}\leq1$, thus we have, $\llangle Q_{T}\rrangle^{\pm}\leq0$, under the assumption that $\beta\geq0$. Until now we have only considered the first cumulants of work and heat. Next, we analyze work and its relative fluctuations. 
\subsection{Numerical analysis of the average work and its relative fluctuations}
\begin{figure}[hbtp]
\centering
\includegraphics[scale=0.95]{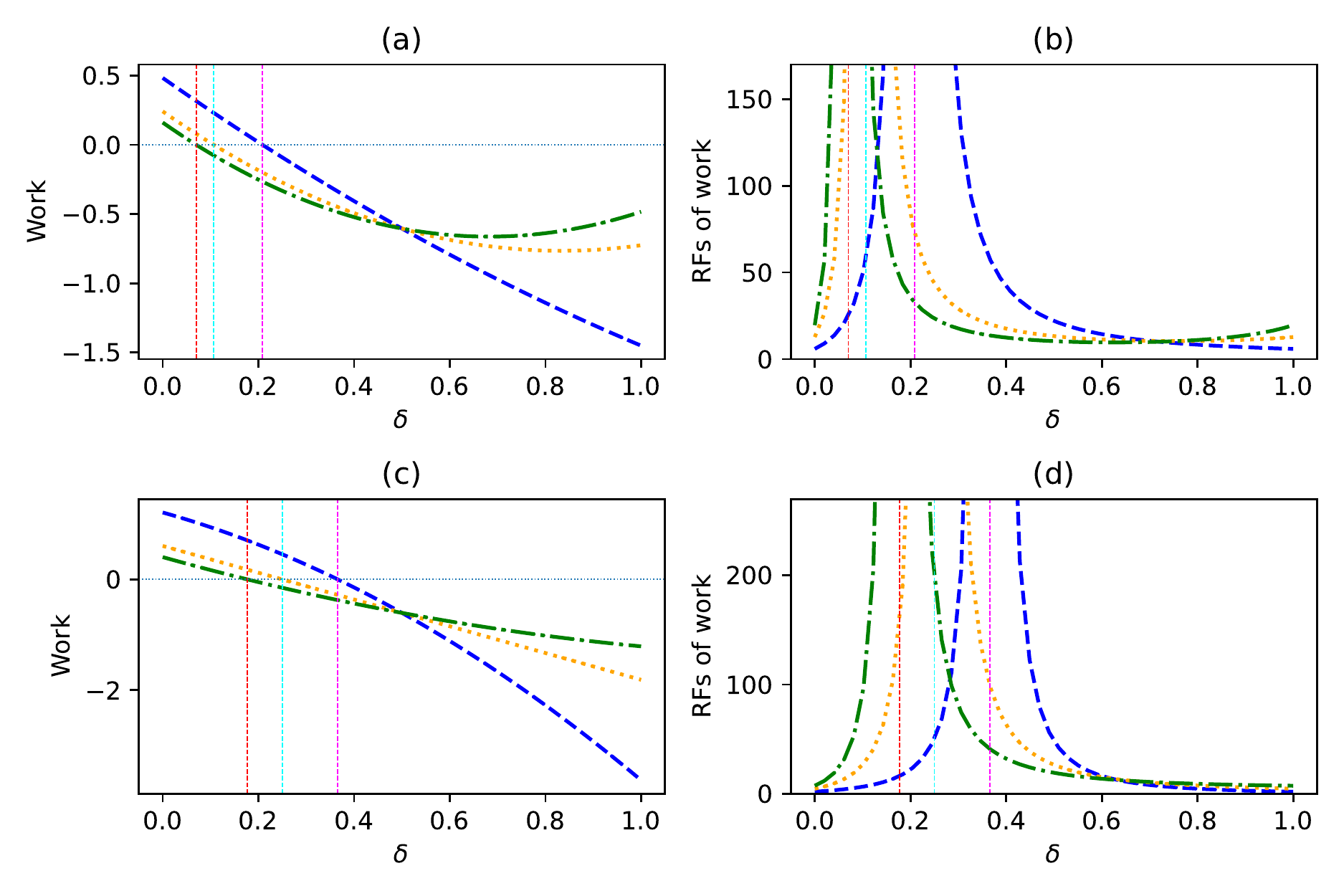}
\caption{The plot of work average $\llangle W\rrangle$ (a) and (c), its RFs $\llangle W^{2}\rrangle/\llangle W\rrangle^{2}$ (b) and (d), as a function of the adiabatic parameter $\delta(=\zeta)$. In all figures: $\beta=0.7$, $\nu_{1}=1$, $\nu_{2}=2$, $\alpha=1/2$ (blue dashed line, when the control qubit is being projected in the $|-\rangle_{c}$ Fourier basis), $\alpha=0$ (orange dotted line, i.e. no coherence) and $\alpha=1/2$ (green dashed-dotted line, when the control qubit is being projected in the $|+\rangle_{c}$ Fourier basis). In (a) and (b): $\theta=0.2$, the red vertical line corresponds to $\delta=0.070290$, the cyan vertical line $\delta=0.10685$ and the magenta vertical line $\delta=0.20871$. In (c) and (d): $\theta=0.5$, the red vertical line corresponds to $\delta=0.17712$, the cyan vertical line $\delta=0.25$ and  the magneta vertical line $\delta=0.36603$.}\label{AVSKP0M0.2P}
\end{figure}

In figure (\ref{AVSKP0M0.2P}) we plot the average work and its relative fluctuations as a function of the adiabatic parameter $\delta$. From Fig.(\ref{AVSKP0M0.2P}) (a) we see that when the control qubit is projected in minus basis (blue dashed line) then work is better than when projected in plus basis (green dashed-dotted line) as well as in the absence of coherence i.e. $\alpha=0$ (orange dotted line). Further, we see that coherence enhances the value of $\delta$ above which work becomes negative. If we consider the temperature being negative, e.g. $\beta=-0.7$, then the maximal possible value of work is reached when $\delta=1$ and when the control qubit is projected in the minus basis. This shows that non-adiabatic transitions can act as a resource when the temperature is negative, in contrast to the case when the temperature is positive where they become detrimental. 

From Fig.(\ref{AVSKP0M0.2P}) (b) we see that the RFs for $\delta\leq0.10685$ and $\alpha=1/2$  (blue dashed line) is always less than the RFs when $\alpha=0$ (orange dotted line) and $\alpha=1/2$ (green dashed-dotted line). If we keep increasing $\delta$ then the RFs (blue dashed line) can be higher or lower than those of green dashed-dotted and orange dotted lines. Furthermore, for example for $\beta=-0.7$ and for $\delta=1$ one would extract the maximal possible amount of work with the same RFs at $\delta=0$. We mention the RFs at $\delta=0$ and $1$ are equal. But note that reversing the sign of $\beta$ does not change the RFs. At the points $0.07029$, $0.10685$ and $0.20871$, the RFs diverge as a consequence of work becoming null.

From these two figures, one can conclude that when the control qubit is measured in the minus basis not only enhances the work average but as well the latter can have lesser relative fluctuations. Of course, the variance can increase when we increase $\delta$, but relative to the square of the average we have a decrease. When the control qubit is measured in the plus basis we see that the average work becomes worse and even further it has increased associated relative fluctuations, thus less reliability. 

In figure (\ref{AVSKP0M0.2P}) (c) and (d) analogously to what we said for the figure (\ref{AVSKP0M0.2P}) (a) and (b), we see that when increasing $\theta$ from 0.2 to 0.5 the average works gets increased. For the RFs, we see that for $\alpha=1/2$ (blue dashed line) and for $\delta\leq0.25$ the RFs are always less than that of $\alpha=0$ (orange dotted line) and $\alpha=1/2$ (green dashed-dotted line).

In the heat engine region, we note that similarly to Fig. (\ref{BOF1}), we found (not shown here for brevity) violations of the bounds $\langle \eta\rangle^{3}\leq.<1$ ($\langle \eta\rangle^{4}\leq.<1$) for the ratio of third (fourth) cumulants, when the control qubit is projected either in the plus or the minus  Fourier basis.
\section{Conclusion}\label{Conclu}
In the first part of this work, a QOHE was considered, where one of the heat baths was replaced by an arbitrary unital map. This is the first time it has been done to our knowledge. Further, we neither specified the driving protocol in the adiabatic stages nor the unital channel which makes our results interesting. First, we have calculated the exact analytical expression of the CF from which all cumulants emerge. By considering an arbitrary unital map we have seen that work and efficiency can be more enhanced. Our results shed light as well on the negative role of projective measurement used to assess the fluctuations of heat and work. We have seen that differently from Ref. \cite{Shanhe} our engine needs the condition $\nu_{2}>\nu_{1}$ to function as a heat engine. We trace this back to the erasure of coherence. More precisely since we have considered an arbitrary unital map that is not completely unital and if projective measurement is not used then the amount of coherence created by $U$ can be transferred to the stroke $\mathbf{C}\rightarrow\mathbf{D}$, which when $V$ can couple the populations and coherences then this coherence can be beneficial. We have shown that the system cannot work as a refrigerator independently of the chosen parameters. We proved mathematically, for a symmetric and asymmetric Otto cycle, that the ratio of the fluctuations of $W$ and $Q_{M}$ is lower and upper-bounded. The same was not true, for the ratio of the third and fourth cumulants. We have seen how the order of relative fluctuations is robust against nonadiabatic transitions, from which it follows Eqs. (\ref{UppLo}) and (\ref{UppLo2}). This result was not possible for the third and fourth cumulants for quantum unital Otto heat engines, highlighting the difference bewteen autonomous continuous quantum thermal machines \cite{Gerry1,Gerry2}, and driven discrete quantum Otto engines.

Note that the exact analytical expression of the CF (eq.\ref{Arbitrary}), is as well one of the main results of the paper. However, we have limited our analysis to only the case when the channels are unital because the time reversal of an arbitrary nonunital channel is not easy to be defined, since the adjoint may not be a valid process as it is for unital channels. Therefore, we cannot take the adjoint of a non-unital channel in the backward cycle as a valid physical channel. This open question, we leave it to the future. In considering negative inverse temperature we have seen that the non-adiabaticity can act as a resource, for efficiency enhancement, for more work extracted with fixed relative fluctuations. It would be interesting to extend these results to e.g. two coupled spins \cite{Abdelkader,Daniel,Das} and investigate how the coupling would affect the results reported here on the cumulants.\par

In the second part concerning the coherent control of channels we have limited ourselves to the channel considered in \cite{Shanhe}. We have shown that efficiency and the work extracted can be enhanced by coherence when the control qubit gets projected in the $|-\rangle_{c}$ Fourier basis. The numerical simulation revealed that work reliability gets enhanced as well. When it is projected in the $|+\rangle_{c}$ Fourier basis we have the inverse of these conclusions. Further, we have shown how non-adiabatic transitions can be a resource when the temperature of the bath is negative.\par 

Our QOHE based on arbitrary unital qubit channels is not just a mathematical curiosity but it can be realized experimentally. We propose the NMR \cite{Peterson, Lisboa, Pedro} and optical \cite{Karni,Shaham} settings as candidates to verify the results and the analysis reported here. We hope our study motivates further interest in the first four cumulants in the field of quantum thermodynamics on quantum thermal machines.

A consequence of the proved realtion (\ref{UppLo}) we have,
\begin{equation}
\llangle Q_{M}^{2}\rrangle\times \langle \eta\rangle^{2}\leq \llangle W^{2}\rrangle < \llangle Q_{M}^{2}\rrangle.
\end{equation}
This shows that the fluctuations of work can never be above those of heat nor they can be equal in the heat engine regime. Further, they are lower bounded by the square of the efficiency and the fluctuations of $Q_{M}$. The lower bound is reached in the quasistatic limit. If this relation was also possible, for quantum unital Otto heat engines for higher cumulants, i.e.
\begin{equation}
\llangle Q_{M}^{n}\rrangle\times \langle \eta\rangle^{n}\leq \llangle W^{n}\rrangle < \llangle Q_{M}^{n}\rrangle,
\end{equation}
as for continuous thermal machines studied in Ref. \cite{Gerry2,Gerry1}, then this would mean the third and fourth cumulants of work $W$ are bounded by those of heat absorbed $Q_{M}$. Furthermore, the cumulants would have the same sign as a consequence of the fact that the lower bound $\langle \eta\rangle^{n}$ and upper bound 1-(analogously to $\langle \eta\rangle^{n}$ and $\langle \eta\rangle_{c}^{n}$ in \cite{Gerry2,Gerry1} for continuous thermal machines, where $\langle \eta\rangle_{c}$ is the Carnot efficiency)-are positive. However, here due to external driving the sign of third (fourth) cumulants of work $W$ and heat $Q_{M}$ can be different.

We believe that quantum Unital Otto heat engines can have great utility. For example in appendix \ref{LZM} we have seen how they can work as a heat engine even without an energy gap change as was mentioned in \cite{Shanhe}, and even more with an efficiency that can go to unity. However, we should note that in the bath stage where it has been replaced by an arbitrary unital channel, then we only focused on the input and output states without caring about how this physical process has happened, and assuming it to be instantaneous. The same thing we have assumed in the thermalization stroke. For the physical implementation of the unital map, we can take a spin and couple it to an environment at zero inverse temperature, i.e. infinite temperature, and then derive its corresponding Linbladian, i.e. the operator that describes the evolution of the state of the system in time. The bath can be either an ensemble of uncoupled spins or a collection of uncoupled harmonic oscillators. We believe that this could bring some usefulness to the power output of the heat engine as well. The authors of Ref. \cite{Misra} have done good work for the derivation of the Linbladian i.e. the generator of the evolution of a single spin (as we considered in our work) immersed in a bath of uncoupled and unpolarized spins. This reference is a good starting point to have a deep understanding of the efficiency and power of quantum unital Otto heat engines and the trade-off between them.

Finally, beyond our results showing a great difference between the bounds on the cumulants of autonomous quantum thermal machines and driven discrete quantum heat engines, our study would open the door to consider other types of quantum noises than unital ones that we have considered here and study their role on important metrics of quantum thermal machines, such as work, efficiency, power and their relative fluctuations.  For example, Farahmand et al.\cite{Zahra}, have shown how quantum noises can enhance heat bath algorithmic cooling over noiseless algorithmic cooling, a result that one would not expect, as it is usually assumed that noises are detrimental especially, from quantum information theory perspective.
\begin{acknowledgments}
We thank, in alphabetical order, Jincan Chen, Rogério J. de Assis, Shanhe Su and Zakaria Mzaouali for useful discussions. We thank a lot Cyril Branciard for pointing out forgetting to sum over the index $j$ in equation (\ref{Eq3}), which was helpful in the derivation of the results of Sec.(\ref{QOCM}). We gratefully thank anonymous reviewers for their constructive remarks and for their insightful comments and suggestions, which have improved the quality of the paper.
\end{acknowledgments}

\renewcommand{\appendixname}{Appendix}
\appendix
\section{The computation of the CF of the forward $\chi(\gamma_{W},\gamma_{M})_{F}$ for an arbitrarily driven two-level system: the hot bath being replaced with an arbitrary unital map}\label{chi}
Here we compute the CF $\chi(\gamma_{W},\gamma_{M})_{F}$, from which all the cumulants of heat and work follow. Our working medium is a two-level system with eigenenergies $\pm\nu_{i}$ and their corresponding eigenvectors $|\pm\rangle_{i}$, with $i=1,2$. But before we compute $\chi(\gamma_{W},\gamma_{M})_{F}$ let's make an important remark. First, when we change one of the baths with an arbitrary qubit channel we do not know if it will play the role of a hot or cold bath. However, when we make the assumption of unitality one can show that the unital channel will play the role of the hot bath, since as we showed in Sec.(\ref{NegativeQc}), the heat exchanged with the bath at inverse temperature $\beta$ is always negative, thus it acts as a sink bath. Hence the name "the hot bath being replaced with arbitrary unital map".

Let's first calculate the expression of $\chi(\gamma_{W},\gamma_{M})_{F}$ for an arbitrary qubit channel $\mathcal{E}$ with Kraus operators $K_{j}$. We have,
\begin{equation}
\begin{split}
\chi(\gamma_{W},\gamma_{M})_{F} & =\sum_{n,m,k,l}\frac{e^{-\beta \nu_{n}}}{Z}|{}_{2}\langle m|U|n\rangle_{1}|^{2}\sum_{j}|{}_{2}\langle k|K_{j}|m\rangle_{2}|^{2} 
|{}_{1}\langle l|V|k\rangle_{2}|^{2}e^{-i\gamma_{W}(\nu_{m}-\nu_{n}+\nu_{l}-\nu_{k})}e^{i\gamma_{M}(\nu_{k}-\nu_{m})}.
\end{split}
\end{equation}
Our main goal is to express $\chi(\gamma_{W},\gamma_{M})_{F}$ in terms of the next transitions probabilities $\delta$, $\theta$ and $\zeta$ that we define as follows,
\begin{equation}
\delta=|{}_{2}\langle+|U|-\rangle_{1}|^{2},
\\
\theta=\sum_{j}|{}_{2}\langle-|K_{j}|+\rangle_{2}|^{2},
\ \  \rm and \ \ \ \
\zeta=|{}_{1}\langle+|V|-\rangle_{2}|^{2}.
\end{equation}
From Table (\ref{Tbale1}) and equation (\ref{Chi}), we get
\begin{equation}
\begin{split}
\chi(\gamma_{W},\gamma_{M})_{F}= & \frac{e^{\beta\nu_{1}}}{Z}\left(
(1-\delta)(h-\theta)(1-\zeta)+(1-\delta)(h-\theta)\zeta e^{-2i\gamma_{W}\nu_{1}}+(1-\delta)(1-h+\theta)\zeta e^{2i\gamma_{W}\nu_{2}}e^{2i\gamma_{M}\nu_{2}}\right.
\\ &
+(1-\delta)(1-h+\theta)(1-\zeta) e^{2i\gamma_{W}(\nu_{2}-\nu_{1})}e^{2i\gamma_{M}\nu_{2}}
+\delta\theta(1-\zeta) e^{-2i\gamma_{W}\nu_{2}}e^{-2i\gamma_{M}\nu_{2}}+\delta\theta\zeta e^{-2i\gamma_{W}(\nu_{1}+\nu_{2})}e^{-2i\gamma_{M}\nu_{2}}
\\ &
+\left.\delta(1-\theta)\zeta+\delta(1-\theta)(1-\zeta) e^{-2i\gamma_{W}\nu_{1}}\right)
\\ &
+\frac{e^{-\beta\nu_{1}}}{Z}\left((1-\delta)(1-\theta)(1-\zeta)+(1-\delta)(1-\theta)\zeta e^{2i\gamma_{W}\nu_{1}}+(1-\delta)\theta\zeta e^{-2i\gamma_{W}\nu_{2}}e^{-2i\gamma_{M}\nu_{2}}\right.
\\ &
+(1-\delta)\theta(1-\zeta) e^{-2i\gamma_{W}(\nu_{2}-\nu_{1})}e^{-2i\gamma_{M}\nu_{2}})+\delta(1-h+\theta)(1-\zeta) e^{2i\gamma_{W}\nu_{2}}e^{2i\gamma_{M}\nu_{2}}+\delta(1-h+\theta)\zeta e^{2i\gamma_{W}(\nu_{1}+\nu_{2})}e^{2i\gamma_{M}\nu_{2}}
\\ &
+\left.\delta(h-\theta)\zeta+\delta(h-\theta)(1-\zeta) e^{2i\gamma_{W}\nu_{1}}\right),
\end{split}
\end{equation}
where $h=\sum_{j}{}_{2}\langle-|K_{j}K_{j}^{\dagger}|-\rangle_{2}$. Rearranging this expression one obtains
\begin{equation}
\begin{split}
\chi(\gamma_{W},\gamma_{M})_{F}= & 
(1-\delta)(1-\zeta)\left(\frac{e^{\beta\nu_{1}}}{Z}(h-\theta)+\frac{e^{-\beta\nu_{1}}}{Z}(1-\theta)\right)+
(1-\delta)\zeta\left(\frac{e^{\beta\nu_{1}}}{Z} e^{-2i\gamma_{W}\nu_{1}}(h-\theta)+\frac{e^{-\beta\nu_{1}}}{Z} e^{2i\gamma_{W}\nu_{1}}(1-\theta)\right)
\\ &
+(1-\delta)\zeta\left(\frac{e^{\beta\nu_{1}}}{Z} (1-h+\theta)e^{2i\gamma_{W}\nu_{2}}e^{2i\gamma_{M}\nu_{2}}+\frac{e^{-\beta\nu_{1}}}{Z}e^{-2i\gamma_{W}\nu_{2}}e^{-2i\gamma_{M}\nu_{2}}\theta\right)
\\ &
+(1-\delta)(1-\zeta)\left(\frac{e^{\beta\nu_{1}}}{Z} (1-h+\theta)e^{2i\gamma_{W}(\nu_{2}-\nu_{1})}e^{2i\gamma_{M}\nu_{2}}+\frac{e^{-\beta\nu_{1}}}{Z}e^{-2i\gamma_{W}(\nu_{2}-\nu_{1})}e^{-2i\gamma_{M}\nu_{2}}\theta\right)
\\ &
+\delta(1-\zeta)\left(\frac{e^{\beta\nu_{1}}}{Z}e^{-2i\gamma_{W}\nu_{2}}e^{-2i\gamma_{M}\nu_{2}}\theta+\frac{e^{-\beta\nu_{1}}}{Z}e^{2i\gamma_{W}\nu_{2}}e^{2i\gamma_{M}\nu_{2}}(1-h+\theta)\right)
\\ &
+\delta\zeta\left(\frac{e^{\beta\nu_{1}}}{Z}e^{-2i\gamma_{W}(\nu_{1}+\nu_{2})}e^{-2i\gamma_{M}\nu_{2}}\theta+\frac{e^{-\beta\nu_{1}}}{Z}e^{2i\gamma_{W}(\nu_{1}+\nu_{2})}e^{2i\gamma_{M}\nu_{2}}(1-h+\theta)\right)
\\ &
+\delta\zeta\left(\frac{e^{\beta\nu_{1}}}{Z}(1-\theta)+\frac{e^{-\beta\nu_{1}}}{Z}(h-\theta)\right)
\\ &
+\delta(1-\zeta)\left(\frac{e^{\beta \nu_{1}}}{Z} e^{-2i\gamma_{W}\nu_{1}}(1-\theta)+\frac{e^{-\beta \nu_{1}}}{Z} e^{2i\gamma_{W}\nu_{1}}(h-\theta)\right).
\end{split}
\end{equation}
It is still a cumbersome equation. This equation can be simplified more under the assumption that the map is unital, i.e. $\sum_{j}K_{j}\mathbb{1}K_{j}^{\dagger}=\mathbb{1}$. Under this condition, we have $h=1$ and after some simple algebra one can show
\begin{equation}
\begin{split}
\chi(\gamma_{W},\gamma_{M})_{F} & =(1-\theta)\left(1+\left(\frac{\rm 2cos((2\gamma_{W}+i\beta)\nu_{1})}{Z}-1\right)(\delta+\zeta-2\delta\zeta)\right)
\\ &
+\theta\left((1-\delta)\left(\zeta \frac{\rm 2cos(2(\gamma_{W}+\gamma_{M})\nu_{2}-i\beta\nu_{1})}{Z}+(1-\zeta) \frac{\rm 2cos(2(\gamma_{W}(\nu_{2}-\nu_{1})+\gamma_{M}\nu_{2})-i\beta\nu_{1})}{Z}\right)\right.
\\ &
+\left.\delta\left((1-\zeta) \frac{\rm 2cos(2(\gamma_{W}+\gamma_{M})\nu_{2}+i\beta\nu_{1})}{Z}+\zeta \frac{\rm 2cos(2((\nu_{1}+\nu_{2})\gamma_{W}+\gamma_{M}\nu_{2})+i\beta\nu_{1})}{Z}\right)\right).
\end{split}
\end{equation}
When $\delta\neq\zeta$, the backward CF $\chi(\gamma_{W},\gamma_{M})_{B}$ follows, directly from $\chi(\gamma_{W},\gamma_{M})_{F}$ by the correspondance $\delta\leftrightarrow\zeta$. Thus, the explicit expression of $\chi(\gamma_{W},\gamma_{M})_{B}$ is
\begin{equation}
\begin{split}
\chi(\gamma_{W},\gamma_{M})_{B} & =(1-\theta)\left(1+\left(\frac{\rm 2cos((2\gamma_{W}+i\beta)\nu_{1})}{Z}-1\right)(\delta+\zeta-2\delta\zeta)\right)
\\ &
+\theta\left((1-\zeta)\left(\delta \frac{\rm 2cos(2(\gamma_{W}+\gamma_{M})\nu_{2}-i\beta\nu_{1})}{Z}+(1-\delta) \frac{\rm 2cos(2(\gamma_{W}(\nu_{2}-\nu_{1})+\gamma_{M}\nu_{2})-i\beta\nu_{1})}{Z}\right)\right.
\\ &
+\left.\zeta\left((1-\delta) \frac{\rm 2cos(2(\gamma_{W}+\gamma_{M})\nu_{2}+i\beta\nu_{1})}{Z}+\delta \frac{\rm 2cos(2((\nu_{1}+\nu_{2})\gamma_{W}+\gamma_{M}\nu_{2})+i\beta\nu_{1})}{Z}\right)\right).
\end{split}
\end{equation}
\begin{widetext}
\begin{table}[h!]
\begin{tabular}{|l|l|l|l|l|l|l|l|}
\hline
Path & Transition from point $\mathbf{A}$ to point $\mathbf{D}$ in Fig.\ref{b0.55}(b) & Probability of the path being followed by the system & $W=-(W_{1}+W_{2})$ & $Q_{M}$  \\
\hline
1 & $|-\rangle_{1}$ $|-\rangle_{2}$ $|-\rangle_{2}$ $|-\rangle_{1}$ & $\frac{e^{\beta \nu_{1}}}{Z}|{}_{2}\langle-|U|-\rangle_{1}|^{2}\sum_{j}|{}_{2}\langle-|K_{j}|-\rangle_{2}|^{2}|{}_{1}\langle-|V|-\rangle_{2}|^{2}$ & 0 & 0   \\
\hline
2 & $|-\rangle_{1}$ $|-\rangle_{2}$ $|-\rangle_{2}$ $|+\rangle_{1}$ & $\frac{e^{\beta \nu_{1}}}{Z}|{}_{2}\langle-|U|-\rangle_{1}|^{2}\sum_{j}|{}_{2}\langle-|K_{j}|-\rangle_{2}|^{2}|{}_{1}\langle+|V|-\rangle_{2}|^{2}$ & $-2\nu_{1}$ & 0   \\
\hline
3 & $|-\rangle_{1}$ $|-\rangle_{2}$ $|+\rangle_{2}$ $|-\rangle_{1}$ & $\frac{e^{\beta \nu_{1}}}{Z}|{}_{2}\langle-|U|-\rangle_{1}|^{2}\sum_{j}|{}_{2}\langle+|K_{j}|-\rangle_{2}|^{2}|{}_{1}\langle-|V|+\rangle_{2}|^{2}$ & $2\nu_{2}$ & $2\nu_{2}$  \\
\hline
4 & $|-\rangle_{1}$ $|-\rangle_{2}$ $|+\rangle_{2}$ $|+\rangle_{1}$ & $\frac{e^{\beta \nu_{1}}}{Z}|{}_{2}\langle-|U|-\rangle_{1}|^{2}\sum_{j}|{}_{2}\langle+|K_{j}|-\rangle_{2}|^{2}|{}_{1}\langle+|V|+\rangle_{2}|^{2}$ & $2(\nu_{2}-\nu_{1})$ &  $2\nu_{2}$   \\
\hline
5 & $|-\rangle_{1}$ $|+\rangle_{2}$ $|-\rangle_{2}$ $|-\rangle_{1}$ & $\frac{e^{\beta \nu_{1}}}{Z}|{}_{2}\langle+|U|-\rangle_{1}|^{2}\sum_{j}|{}_{2}\langle-|K_{j}|+\rangle_{2}|^{2}|{}_{1}\langle-|V|-\rangle_{2}|^{2}$ & $-2\nu_{2}$ & $-2\nu_{2}$  \\
\hline
6 & $|-\rangle_{1}$ $|+\rangle_{2}$ $|-\rangle_{2}$ $|+\rangle_{1}$ & $\frac{e^{\beta \nu_{1}}}{Z}|{}_{2}\langle+|U|-\rangle_{1}|^{2}\sum_{j}|{}_{2}\langle-|K_{j}|+\rangle{}_{2}|^{2}|{}_{1}\langle+|V|-\rangle_{2}|^{2}$ & $-2(\nu_{1}+\nu_{2})$ & $-2\nu_{2}$  \\
\hline
7 & $|-\rangle_{1}$ $|+\rangle_{2}$ $|+\rangle_{2}$ $|-\rangle_{1}$ & $\frac{e^{\beta \nu_{1}}}{Z}|{}_{2}\langle+|U|-\rangle_{1}|^{2}\sum_{j}|{}_{2}\langle+|K_{j}|+\rangle_{2}|^{2}|{}_{1}\langle-|V|+\rangle_{2}|^{2}$ & 0 & 0  \\
\hline
8 & $|-\rangle_{1}$ $|+\rangle_{2}$ $|+\rangle_{2}$ $|+\rangle_{1}$ & $\frac{e^{\beta \nu_{1}}}{Z}|{}_{2}\langle+|U|-\rangle_{1}|^{2}\sum_{j}|{}_{2}\langle+|K_{j}|+\rangle_{2}|^{2}|{}_{1}\langle+|V|+\rangle_{2}|^{2}$ & $-2\nu_{1}$ & 0   \\
\hline
9 &  $|+\rangle_{1}$ $|-\rangle_{2}$ $|-\rangle_{2}$ $|-\rangle_{1}$ & $\frac{e^{-\beta \nu_{1}}}{Z}|{}_{2}\langle-|U|+\rangle_{1}|^{2}\sum_{j}|{}_{2}\langle-|K_{j}|-\rangle{}_{2}|^{2}|{}_{1}\langle-|V|-\rangle_{2}|^{2}$ & $2\nu_{1}$ & 0  \\
\hline
10 & $|+\rangle_{1}$ $|-\rangle_{2}$ $|-\rangle_{2}$ $|+\rangle_{1}$ & $\frac{e^{-\beta \nu_{1}}}{Z}|{}_{2}\langle-|U|+\rangle_{1}|^{2}\sum_{j}|{}_{2}\langle-|K_{j}|-\rangle{}_{2}|^{2}|{}_{1}\langle+|V|-\rangle_{2}|^{2}$ & 0 &  0   \\
\hline
11 & $|+\rangle_{1}$ $|-\rangle_{2}$ $|+\rangle_{2}$ $|-\rangle_{1}$ & $\frac{e^{-\beta \nu_{1}}}{Z}|{}_{2}\langle-|U|+\rangle_{1}|^{2}\sum_{j}|{}_{2}\langle+|K_{j}|-\rangle_{2}|^{2}|{}_{1}\langle-|V|+\rangle_{2}|^{2}$ & $2(\nu_{1}+\nu_{2})$ & $2\nu_{2}$    \\
\hline
12 & $|+\rangle_{1}$ $|-\rangle_{2}$ $|+\rangle_{2}$ $|+\rangle_{1}$ & $\frac{e^{-\beta \nu_{1}}}{Z}|{}_{2}\langle-|U|+\rangle_{1}|^{2}\sum_{j}|{}_{2}\langle+|K_{j}|-\rangle_{2}|^{2}|{}_{1}\langle+|V|+\rangle_{2}|^{2}$ & $2\nu_{2}$ & $2\nu_{2}$  \\
\hline
13 & $|+\rangle_{1}$ $|+\rangle_{2}$ $|-\rangle_{2}$ $|-\rangle_{1}$ & $\frac{e^{-\beta \nu_{1}}}{Z}|{}_{2}\langle+|U|+\rangle_{1}|^{2}\sum_{j}|{}_{2}\langle-|K_{j}|+\rangle_{2}|^{2}|{}_{1}\langle-|V|-\rangle_{2}|^{2}$ & $2(\nu_{1}-\nu_{2})$ & $-2\nu_{2}$   \\
\hline
14 & $|+\rangle_{1}$ $|+\rangle_{2}$ $|-\rangle_{2}$ $|+\rangle_{1}$ & $\frac{e^{-\beta \nu_{1}}}{Z}|{}_{2}\langle+|U|+\rangle_{1}|^{2}\sum_{j}|{}_{2}\langle-|K_{j}|+\rangle{}_{2}|^{2}|{}_{2}\langle+|V|-\rangle_{1}|^{2}$ & $-2\nu_{2}$ & $-2\nu_{2}$  \\
\hline
15 & $|+\rangle_{1}$ $|+\rangle_{2}$ $|+\rangle_{2}$ $|-\rangle_{1}$ & $\frac{e^{-\beta \nu_{1}}}{Z}|{}_{2}\langle+|U|+\rangle_{1}|^{2}\sum_{j}|{}_{2}\langle+|K_{j}|+\rangle{}_{2}|^{2}|{}_{1}\langle-|V|+\rangle_{2}|^{2}$ & $2\nu_{1}$ & 0  \\
\hline
16 & $|+\rangle_{1}$ $|+\rangle_{2}$ $|+\rangle_{2}$ $|+\rangle_{1}$ & $\frac{e^{-\beta \nu_{1}}}{Z}|{}_{2}\langle+|U|+\rangle_{1}|^{2}\sum_{j}|{}_{2}\langle+|K_{j}|+\rangle_{2}|^{2}|{}_{1}\langle+|V|+\rangle_{2}|^{2}$ & 0 & 0   \\
\hline
\end{tabular}
\caption{Paths followed by a two-level system and their associated probabilities with the hot bath being replaced by an arbitrary unital channel with Kraus operators $K_{j}$. $W$ and $Q_{M}$ are the stochastic work and the heat exchanged with the hot bath corresponding to each path. Furthermore, note that even though the system can work sometimes as a refrigerator for some paths, on average the system cannot, and this is forbidden by the second law of thermodynamics. In addition to this one should note that when both the compression and expansion adiabatic strokes are quasistatically then we have only four possible paths that will be taken by the two-level system: 1, 4, 13 and 16. However, in the non-adiabatic case, all the paths are possible.} \label{Tbale1}
\end{table}
\end{widetext}
\section{Proof of $0\leq \theta\leq1/2$ for the unital map considered in Ref.\cite{Shanhe}} \label{Proof}
The unital map considered in Ref.\cite{Shanhe} has a number of Kraus operators equal to two. The latters are given as $\pi_{1}=|\psi_{1}\rangle\langle\psi_{1}|$ and $\pi_{2}=|\psi_{2}\rangle\langle\psi_{2}|$. Thus, we see that they are hermitian and satisfy
\begin{equation}
\pi_{1}+\pi_{2}=1.
\end{equation}
After a simple algebra one can show that $\theta\left(=\sum_{i}{}_{2}\langle-|\pi_{i}|+\rangle_{22}\langle+|\pi_{i}|-\rangle_{2}\right)$, is given as follows,
\begin{equation}
\theta=|_{2}\langle-|\psi_{1}\rangle|^{2}|\langle\psi_{1}|+\rangle_{2}|^{2}+|_{2}\langle-|\psi_{2}\rangle|^{2}|\langle\psi_{2}|+\rangle_{2}|^{2}.
\end{equation}
And by defining $p_{1}=|_{2}\langle-|\psi_{1}\rangle|^{2}$ and $p_{2}=|_{2}\langle-|\psi_{2}\rangle|^{2}$, one can find that,
\begin{equation}
\theta=p_{1}(1-p_{1})+p_{2}(1-p_{2}).
\end{equation}
This follows from the microreversibility principle, i.e.
\begin{equation}
\begin{split}
|_{2}\langle+|\psi_{1}\rangle|^{2} & =|\langle\psi_{1}|+\rangle_{2}|^{2}
\\ &
=\langle\psi_{1}|+\rangle_{22}\langle+|\psi_{1}\rangle
\\ &
=\langle\psi_{1}|(\mathbb{1}-|-\rangle_{22}\langle-|)|\psi_{1}\rangle
\\ &
=1-|_{2}\langle-|\psi_{1}\rangle|^{2}
\\ &
=1-p_{1}.
\end{split}
\end{equation}
In the same manner one can show easily, $|_{2}\langle+|\psi_{2}\rangle|^{2}=1-p_{2}$. Furthermore, one can find as well that, $p_{2}=1-p_{1}$. Thus, we obtain,
\begin{equation}
\theta=2p_{1}(1-p_{1}).
\end{equation} 
From the fact that, $0\leq p_{1},p_{2}\leq1$, we conclude that,
\begin{equation}
0\leq \theta\leq1/2.
\end{equation}
This is because the maximum value of $p_{1}(1-p_{1})$ is 1/4, which correspond to the case of $p_{1}=1/2$. This concludes the proof.

\section{Nature of the energy provided by the unital channel to the working substance}\label{WHL}
Here we give a detailed analysis of the nature of energy provided by the unital channel to the working substance. First The Hamiltonian of the system given by $H_{2}=\nu_{2}\left(|+\rangle_{22}\langle+|-|-\rangle_{22}\langle-|\right)$ stays fixed in the stroke $B\rightarrow C$. When the adiabatic parameter $\delta=0$, in this case, the state before applying the unital channel is just the thermal state given in the basis $\{|+\rangle_{2},|-\rangle_{2}\}$ as,
\begin{equation}
\rho_{th}=\begin{pmatrix}
1-p & 0 \\
0 & p \\
\end{pmatrix}
\end{equation}
with $p=\frac{e^{\beta\nu_{1}}}{e^{-\beta\nu_{1}}+e^{\beta\nu_{1}}}$. When $p_{1}+p_{2}=1$, in this case, the two Pauli operators just swap the populations. Therefore, we have
\begin{equation}
\mathcal{E}\left( \rho_{th}\right)=p_{1}\sigma_{1}\rho_{th}\sigma_{1}+p_{2}\sigma_{2}\rho_{th}\sigma_{2}=\left( p_{1}+p_{2}\right)\rho_{th}'=\rho_{th}'.
\end{equation} 
with,
\begin{equation}
\rho_{th}'=\begin{pmatrix}
p & 0 \\
0 & 1-p \\
\end{pmatrix}.
\end{equation}
Since the entropy of $\rho_{th}$ and $\rho_{th}'$ are the same, thus the energy exchanged in this stroke i.e.
\begin{equation}
\mathrm{Tr}
\left[ \left(\mathcal{E} \left( \rho_{th}\right)-\rho_{th}\right) H_{2}\right]=2\nu_{2}
(1-2p).
\end{equation}
is work and not heat. We can write any one qubit density matrix be it pure or mixed as
\begin{equation}
\rho=\frac{1}{2}\left( \mathbb{1}+v_{x} \sigma_{x} +v_{y} \sigma_{y} +v_{z} \sigma_{z} \right)=\begin{pmatrix}
(1+v_{z})/2 & (v_{x}-i v_{y})/2 \\
(v_{x}+i v_{y})/2 & (1-v_{z})/2 \\
\end{pmatrix}.
\end{equation}
When $v_{x}^{2}+v_{y}^{2}+v_{z}^{2}=1$ the state $\rho$ is pure, and when $v_{x}^{2}+v_{y}^{2}+v_{z}^{2}<1$ the state $\rho$ is the mixed state. After applying $\mathcal{E}$ on the state $\rho$ the latter becomes,
\begin{equation}
\mathcal{E}(\rho)=\begin{pmatrix}
(1-v_{z})/2 & (p_{1}-p_{2})(v_{x}+i v_{y})/2 \\
(p_{1}-p_{2})(v_{x}-i v_{y})/2  & (1+v_{z})/2 \\
\end{pmatrix}
\end{equation}
One can check that the eigenvalues of $\rho$ and $\mathcal{E}(\rho)$ are the same only when either $p_{1}$ or $p_{2}$ is zero. In this case,
the unital channel is just characterized by one unitary operator either $\sigma_{1}$ or $\sigma_{2}$, and therefore entropy stays conserved. Therefore, if either $p_{1}$ or $p_{2}$ is null then the exchanged energy is work, since this transformation is entropy-conserving. And when both of them are non-null then there is entropy change.
\section{A comparison between the monitored and unmonitored Landau-Zener model}\label{LZM}
\begin{figure}[hbtp]
\centering
\includegraphics[scale=0.97]{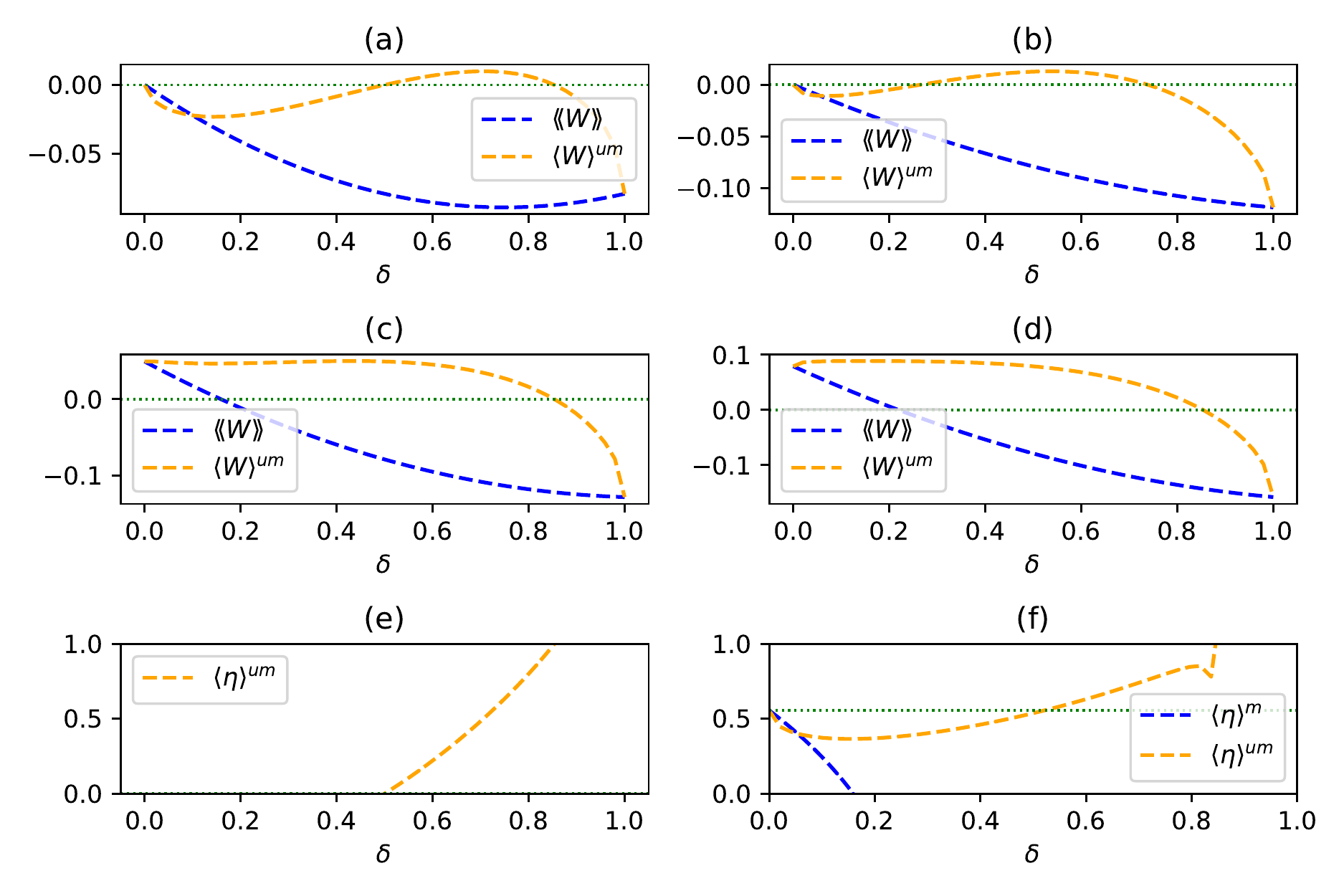}
\caption{
In all plots we have $\nu_{1}=0.4$ and $\beta=0.5$.
(a) $\nu_{2}=0.4$, $\alpha=\pi/4$, $\phi=0$ and $\chi=0$. (b)  $\nu_{2}=0.4=$, $\alpha=\pi/3$, $\phi=0.1$ and $\chi=0.1$. (c)  $\nu_{2}=0.9$, $\alpha=\pi/4$, $\phi=0$ and $\chi=0$. (d) $\nu_{2}=1.2$, $\alpha=\pi/3$, $\phi=0.1$ and $\chi=0.1$. (e) $\nu_{2}=0.4=$, $\alpha=\pi/4$, $\phi=0$ and $\chi=0$. (f) $\nu_{2}=0.9$, $\alpha=\pi/3$, $\phi=0.1$ and $\chi=0.1$. First we see that when $\nu_{2}=\nu_{1}$, Figs. (a) and (b), only the unmonitored heat engine can still work as a heat engine in agreement with our results and Ref. \cite{Shanhe}. We see that when $\nu_{2}>\nu_{1}$, Figs. (c) and (d), both of them can work as a heat engine, however, the unmonitored one can work as a heat engine in a big domain of $\delta$, which is not the case for the monitored one. Furthermore, note that the unmonitored one work can still be positive, even when $\delta\geq 1/2$, something that was not possible in the previous results in the literature, see e.g. \cite{Andrea} when we have an Otto cycle based on two completely thermalizing baths. In (e) only the average efficiency of the unmonitored case is positive and we see that it can even be equal to unity. In (f) we see that the monitored one can be higher than the unmonitored one only for very small values of theta. However, it could not exceed the efficiency, a result that we have proved already. On the other hand, the efficiency of the unmonitored engine could exceed the Otto efficiency and can reach values up to unity as we see in Figure (e).}\label{WMWUM}
\end{figure}

For a two-level system where the protocol driving is linear in time one can show from the results of Refs. \cite{Zener1,Zener2,Zener3,JuzarSong} on the Landau-Zener model and from the Supplemental Material of Ref. \cite{Denzler}, that the unitary operator $U$ that governs the evolution in the expansion stroke $A\rightarrow B$ is given as follows,
\begin{equation}
U=\sqrt{1-\delta}\left(e^{-i\phi}|+\rangle_{21}\langle+|+e^{i\phi}|-\rangle_{21}\langle-|\right)+\sqrt{\delta}\left(|-\rangle_{21}\langle+|-|+\rangle_{21}\langle-|\right).\label{Unit}
\end{equation}
$\phi$ is a phase. Limiting ourselves to the case when the cycle is symmetric i.e. $\delta=\zeta$, the unitary operator characterizing the compression stroke $C\rightarrow D$ is given by,
\begin{equation}
V=CU^{\dagger}C=\sqrt{1-\delta}\left(e^{-i\phi}|+\rangle_{12}\langle+|+e^{i\phi}|-\rangle_{12}\langle-|\right)-\sqrt{\delta}\left(|-\rangle_{12}\langle+|-|+\rangle_{12}\langle-|\right).
\end{equation}
$C$ here is the complex conjugation operator. Note that for the considered comparison we do not need to specify the expressions of $\delta$ and $\phi$. For the unital channel fueling the engine we consider the one studied by Shanhe et al. in Ref. \cite{Shanhe}. The quantum measurement channel is defined by $|\pi_{1}\rangle\langle \pi_{1}|$ and $|\pi_{2}\rangle\langle \pi_{2}|$, where
\begin{equation}
|\pi_{1}\rangle=e^{-i\chi}\sin\left(\alpha/2\right)|+\rangle_{2}-\cos\left(\alpha/2\right)|-\rangle_{2},\label{Pi1}
\end{equation}
and,
\begin{equation}
|\pi_{2}\rangle=\cos\left(\alpha/2\right)|+\rangle_{2}+e^{i\chi}\sin\left(\alpha/2\right)|-\rangle_{2}.\label{Pi2}
\end{equation}
With $0\leq \alpha \leq\pi$ and $0\leq \chi \leq 2\pi$. One can show that for the monitored Otto cycle we have, 
\begin{equation}
\theta=\sin\left(\alpha\right)^{2}/2\left(\leq 1/2\right).\label{awili}
\end{equation}
We see that for $\alpha=0$ we have $\theta=0$, thus $\llangle Q_{M}\rrangle=0$, since the Kraus operators commute with the Hamiltonian $H_{2}$. Note that for the considered unmonitored case we did not consider arbitrary unital channel, but we limit ourselves to the unital channel considered by Shanhe et al. \cite{Shanhe}, since in this case, it becomes difficult to compute the average energies, especially $E_{4}$. The average quantities for the monitored case are given as follows,
\begin{widetext}
\begin{equation}
\llangle Q_{M}\rrangle=(1-2\delta)\sin^{2}\left(\alpha\right)\nu_{2}\tanh(\beta\nu_{1}),
\end{equation}
\begin{equation}
\llangle Q_{T}\rrangle=-2(\sin^{2}\left(\alpha\right)/2+2\delta(1-\sin^{2}\left(\alpha\right))(1-\delta))\nu_{1}\tanh(\beta\nu_{1}),
\end{equation}
and,
\begin{equation}
\begin{split}
\llangle W\rrangle & =2((1-2\delta)\sin^{2}\left(\alpha\right)\nu_{2}/2-(\sin^{2}\left(\alpha\right)/2+(1-\sin^{2}\left(\alpha\right))2\delta(1-\delta))\nu_{1})\tanh(\beta\nu_{1}).
\end{split}
\end{equation}
\end{widetext}
Wen there is no monitoring we have,
\begin{widetext}
\begin{equation}
\langle Q_{M}\rangle^{um}=\nu_{2}\sin(\alpha)(2\sqrt{\delta(1-\delta)}\cos(\alpha)\cos(\phi+\chi)+\sin(\alpha)-2\delta\sin(\alpha)) \tanh(\beta\nu_{1}).\label{QUM}
\end{equation}
\end{widetext}
One can check easily that when $\delta=0$ we have, $\langle Q_{M}\rangle^{um}=\llangle Q_{M}\rrangle$. The expressions of $\langle Q_{T}\rangle^{um}$ and $\langle W\rangle^{um}$ are long so we do not report their analytical expressions here. See Figure (\ref{WMWUM}) for a comparison between the work extracted and the efficiency of the monitored and unmonitored Landau-Zener model.

Now after showing how monitoring can influence work and efficiency negatively, let's compare our results with some interesting results in the literature. In Ref. \cite{Andrea} Solfanelli et al. have done a detailed study on the operation regimes of engine/refrigerator/accelerator and heater. Translated into our notation, it was shown that for $\delta\geq1/2$, the system cannot work as a heat engine. Here we see that quantum unital Otto engines have different properties such as the capability of working as a useful machine for $\delta\geq1/2$ and that they are not limited by the Otto efficiency, i.e. $\eta_{Otto}=1-\nu_{1}/\nu_{2}$. This makes this type of engine interesting to be studied.

For example, in the work of Camati et al. \cite{Camati}, the authors have considered a quantum Otto heat engine with some similarities to our quantum unital Otto engine. They have considered an Otto cycle where the hot bath is not completely thermalizing. This has been shown to not erase all the coherence created in the first unitary stroke and therefore can interfere either constructively or destructively with the coherence created in the second unitary stroke. They have found that this phenomenon can have either a positive or a negative influence on thermodynamic metrics such as power and efficiency, depending on the parameters of the cycle. The similarity between our work and theirs is in the sense that when there is no monitoring, some coherence can be transferred from stroke $A\rightarrow B$ to stroke $C\rightarrow D$. However, in our work, we are considering monitored quantum Otto heat engines based on arbitrary unital channels. Furthermore, in \cite{Camati} it has been found that, even though considering a non-thermalizing environment, the efficiency was still to be limited by the Otto. For the case of monitored quantum unital Otto heat engines, we already have proved (\ref{PropE}), that they are also limited by the Otto bound. On the other hand for the non-monitored cycle, we see from figures: \ref{WMWUM} (e) and \ref{WMWUM} (f), that the efficiency can exceed the Otto efficiency and even go to unity. And in addition to this, we see that from figures: \ref{WMWUM} (a) and \ref{WMWUM} (b), that work can still be extracted even when $\nu_{2}=\nu_{1}$, an interesting result that has already been  shown by the authors of Ref. \cite{Shanhe}.

As a final remark we note that our results on the cumulants of monitored quantum unital Otto heat engines do not depend on the decomposition given in Eq.(\ref{PauliU}). The results only depend on the assumption of unitality. Furthermore, we already have shown that under the decomposition Eq. (\ref{PauliU}), the populations and coherences cannot be coupled, thus coherence created in the first stroke, i.e. $A\rightarrow B$, cannot affect $\langle Q_{M}\rangle^{um}$, but some of it can be transferred to the other unitary stroke, i.e. $C\rightarrow D$. For this, we have considered the Kraus operators (\ref{Pi1}) and (\ref{Pi2}) which in general are not diagonal in the basis of the Hamiltonian $H_{2}$, thus can couple the populations and coherences. This means that the expression of $\langle Q_{M}\rangle^{um}$ in Eq. (\ref{QUM}), is in general affected by the coherences created by the unitary operator $U$, given in Eq. (\ref{Unit}).

\end{document}